%% file: aanda_PJ_mar22.tex
\documentclass[]{aa}  

\usepackage{graphicx}
\usepackage{natbib}
\usepackage{color,soul}
\usepackage{txfonts}
\usepackage{mathtools}
\usepackage{hyperref}
\bibpunct{(}{)}{;}{a}{}{,} 

\newcommand{\apo}{APOKASC}

\newcommand{\Msun}{M$_\odot$}

\begin{document}

   \title{Cannibals in the thick disk II}

   \subtitle{Radial-velocity monitoring of the young $\alpha$-rich stars }

   \author{Jofr\'e, P.
          \inst{1}
          \and
          Jorissen, A.\inst{2}
          \and
            Aguilera-G\'omez, C.\inst{1}
          \and
          Van Eck, S.\inst{2}
          \and
          Tayar, J.\inst{3}
          \and
          Pinsonneault, M.\inst{4}
          \and
          Zinn, J.\inst{4}
          \and
          Goriely, S.\inst{2}
          \and
          Van Winckel, H.\inst{5}
          }

   \institute{N\'ucleo Milenio ERIS \& N\'ucleo de Astronom\'\i a, Facultad de Ingenier\'\i a y Ciencias,
Universidad Diego Portales,
Santiago de Chile\\
              \email{paula.jofre@mail.udp.cl}
         \and
            Institut d'Astronomie et d'Astrophysique, Universit\'e Libre de Bruxelles, Campus Plaine C.P. 226, Boulevard du Triomphe,\\ B-1050 Bruxelles, Belgium
\and
Department of Astronomy, University of Florida, Bryant Space Science Center, Stadium Road, Gainesville, FL 32611, USA
\and
Department of Astronomy, The Ohio State University, Columbus, OH 43210, USA
\and
Institute of Astronomy, KU Leuven, Celestijnenlaan 200D, B-3001 Leuven, Belgium\\
             }

   \date{Received ; accepted}

 
  \abstract
   {Determining ages of stars for reconstructing the history of the Milky Way remains one of the most difficult tasks in astrophysics. This involves knowing when it is possible to relate the stellar mass with its age and when it  is not. The {young $\alpha-$rich} (YAR)  stars present such {a} case in which we are still not sure about their ages because they are relatively massive, implying young ages,  but their abundances are $\alpha-$enhanced, which implies old ages. }
   {We report the results from new observations from a long-term radial-velocity-monitoring campaign complemented with high-resolution spectroscopy, as well as new astrometry and seismology of a sample of 41 red giants from the third version of APOKASC, which includes YAR stars. {The aim is to better characterize the YAR stars in terms of binarity, mass, abundance trends, and kinematic properties.}
   }
   {The radial velocities of HERMES, APOGEE, and Gaia were combined to 
   {determine} the binary 
   fraction among YAR stars.
   In combination with 
   their mass estimate, evolutionary status, chemical composition, and kinematic properties, 
   it allowed us to better constrain the nature of these objects.
   } 
   {We found that  stars  with $\mathrm{M} < 1 \mathrm{M}_\odot$ were all single, whereas stars with $\mathrm{M} > 1 \mathrm{M}_\odot$  could be either single or binary. This is in agreement with theoretical predictions of  population synthesis 
   models. Studying their [C/N], [C/Fe], and [N/Fe], trends with mass, it became clear that many YAR stars do not follow the  APOKASC stars, favoring the scenario that most of them are the product of mass transfer. 
    Our sample further includes two likely undermassive stars, that is to say of such as low mass that they cannot have reached the red clump within the age of the Universe, unless their low mass is the signature of mass loss in previous evolutionary phases. These stars do not show signatures of currently being binaries. Both YAR and undermassive stars might show some anomalous APOGEE abundances  for the elements N, Na, P, K, and Cr; although, higher-resolution optical spectroscopy might be needed to confirm these findings.}
   {Considering the significant fraction of stars that are formed in pairs and the variety of ways that makes mass transfer possible, the diversity in properties in terms of binarity, and chemistry of the YAR and undermassive stars studied here implies that most of these objects are likely not young. }

   \keywords{ }

   \maketitle
%

\section{Introduction}

The  young $\alpha-$rich (YAR) stars were first reported by \cite{Chiappini-2015} and \cite{Martig-2015} when performing studies that 
combined
spectroscopy, and asteroseismology. From spectroscopy it is possible to study stellar populations of different metallicities and $\alpha-$element abundances, hence   {belonging to different Galactic components} \citep[e.g.,][]{Hayden-2015}, and from asteroseismology it is possible to study the distributions of masses,  and thus ages, of stellar populations \citep[e.g.,][]{Miglio-2013}. 

The YAR stars were identified as stars that coexist with thick disk stars. The latter stars are defined as 
objects with enhanced [$\alpha$/Fe] abundance ratios that formed 8-10 Gyr ago \citep[e.g.,][]{Chiappini97}.  While they have similar abundance patterns as thick disk stars, YAR stars are inferred to be significantly younger \citep{Martig-2015}. Since the vast majority of the thick disk stars are very old \citep{Fuhrmann98,Haywood-2016, silva-aguirre-2018, Miglio-2021}, the YAR stars might be seen as just  outliers. 

However, understanding these few outliers has great implications on Galactic archaeology and on stellar evolution theory in general. On the one hand, current models of Galaxy formation and evolution do not predict young stars in the thick disk \citep{Haywood-2016, Buck21}. Hence, if some thick disk stars are truly young, we need to significantly modify our theory of how the thick disk of the Milky Way has assembled.  Alternatively, considering that {dating} stars is still one of the most challenging tasks in astrophysics, the YAR stars may not pose a problem to existing
{Galactic evolutionary} models if their ages are misestimated. This may plausibly be the case if YAR stars are in fact the result of binary stellar evolution, resulting in the inferred asteroseismic masses being larger than 
{their initial mass}.
This would cause the inferred ages to be systematically too low.

{Distinguishing between these two possibilities is important,}
hence understanding the nature of these few outliers, the YAR stars, has become a central topic in this field in the past few years.  This is because YAR stars seem to be present in most samples of old stellar populations in the Galaxy, as revealed by a number 
of recent studies  \citep[e.g.,][]{silva-aguirre-2018,Das-202, Britosilva-2021, Zinn-2021, Matsuno-2021}.

Studies on YAR stars can be divided in two avenues. The Galactic evolution avenue, for example, compares the kinematic and chemical properties of the YAR stars with those of other Galactic stellar populations, 
or the stellar evolution avenue, for example looking for evidence of binary evolution and mass transfer in existing YAR stars. None of these avenues have concluded so far that YAR stars are any different from single thick disk stars 
and therefore the nature of YAR stars remains a mystery.

More specifically, regarding the Galactic evolution avenue, after the first reports of the existence of YAR stars by \cite{Martig-2015} and \cite{Chiappini-2015}, whose abundances were derived from APOGEE (hence infrared) spectra,  \cite{Yong-2016} performed a follow-up analysis in the optical of four YAR stars and determined that the stars were essentially "normal", that is,  there was neither an indication of s-process enhancement due to pollution from asymptotic red giant (AGB) stars, nor some other chemical anomaly that would suggest that YAR stars had a different origin compared to the thick-disk population.  \cite{Matsuno-2018} continued that work by performing a careful analysis of YAR stars alongside a control sample of thin disk stars. They could not find any signature of YAR stars being distinct from that of other thin disk stars, hinting 
at a binary evolution process behind their formation. 

By using larger samples of asteroseismic data from K2 combined with APOGEE and carefully determined ages considering opacities consistent with $\alpha$ enhancement in the models as well as corrections of the scaling relations in asteroseismology, \cite{Warfield21} were able to draw age distributions of $\alpha-$rich stars at different galactic heights. They also found the presence of YAR stars in the K2 data. They 
{derived a different age distribution},
finding more intermediate-age stars at $z>1$~kpc  than at the Galactic plane, which is the area covered by APOKASC.  They suggested that such stars could also be the result of radial migration or star formation episodes happening in clumpy bursts throughout the Galaxy. Since their ages are more uncertain compared to Kepler-derived ages because K2 data hve a shorter cadence, a direct comparison of the age distributions between K2 and Kepler fields is still not possible. Therefore, confirming the existence of these intermediate-age stars at large $z$ still requires the analysis of larger samples of stars. These intermediate-age stars are  older than the YAR stars by two Gyr. 

Using Gaia data and taking advantage of new machine-learning tools to derive ages for large samples of stars, \cite{sun-2020} and \cite{Zhang-2021} have been able to identify YAR stars in the LAMOST dataset and to study them statistically. They have compared the overall behavior of the YAR population with that of the thin and thick disks, finding that the YAR stars indeed have kinematic and chemical distributions that mimic very well those of the thick disk. 
Interestingly, \cite{Zhang-2021} found that while the YAR stars have essentially the same chemical distributions as the thick disk in $\alpha-$ and iron-peak elements, they tend to be enriched in the s-process element barium {as well as, for a fraction of them, in carbon and nitrogen (C+N)}. They attributed this offset to the fact that mass accretion from {RGB or} AGB companions has probably been more frequent for YAR stars than for normal low-mass thick-disk stars. 

Moreover, following the second avenue with the aim of finding indications of mass transfer due to binary evolution to account for the existence of YAR stars, \cite{Yong-2016}
studied the spectral energy distributions of four YAR stars
and found out that 
three out of four had some infrared excess. In parallel,  \citet[][hereafter Paper~I]{Jofre-2016} performed a radial-velocity (RV) monitoring campaign to evaluate whether YAR stars have a higher binary frequency than normal stars. 
Unfortunately, this earlier study relied on small-number statistics and a time span that was not sufficient to cover long-period binaries to
derive any 
strong conclusion on the binary frequencies. 

Paper~I was then complemented by the theoretical analysis of \cite{Izzard-2018}, who demonstrated, using population-synthesis models which included binary systems, that it is perfectly possible to create YAR stars through a binary channel. The number of created such stars strongly depends upon the properties of the binary system (orbital separation, eccentricity, and mass ratio). Many of these interactions led the binary to merge {producing a single star at the end. This makes it difficult to interpret the nature of YAR stars by only looking at current binary fractions such as those found in Paper~I } or the number of YAR stars compared to low-mass stars found in the large surveys.  

The fact that population-synthesis models   {including binary systems} can explain the formation of YAR stars has motivated a further search for spectral signatures 
in YAR stars that would   {hint at the importance of binarity.}
This was investigated by \cite{Hekker-2019}, who performed a careful reanalysis of C, N, and O abundances from APOGEE spectra of YAR stars with the aim to identify signatures of extra-mixing. While some YAR stars showed anomalous N/C ratios, others followed the same trends as the rest of the thick-disk stars. \cite{Hekker-2019} could therefore not clearly conclude whether these stars are the products of binary evolution or whether they are truly young.

Considering that there is still no consensus on the nature of the YAR stars,  a follow-up of Paper~I might provide new insights. Here we update the binary frequency of the targets from Paper~I, thanks to new RV measurements {as well as 
Gaia observations}, and extend the sample with 15 new targets aiming at improving the accuracy of the 
binary statistics. Moreover, the stellar properties (mass, metallicity, $\alpha$ abundances...) of all targets have been reassessed thanks to the new data releases of APOGEE and the subsequent new analyses of asteroseismic data. All this motivates the revision of the results of Paper~I, in addition {to providing a more robust diagnostic of the possible presence of long-period binaries in the sample.} 

The paper is organized as follows. Section~\ref{sect:data} describes the data used in this work.  Section~\ref{sect:results} presents our results about the binary frequency of the different monitored samples. In Sect.~\ref{sect:nature} we explore the properties of the stars given their binary or nonbinary nature, masses, chemistry, and kinematics and in Sect.~\ref{sect:discussion}, we 
{discuss the possible formation scenario for the YAR stars}
and discuss individual cases. We conclude our analysis in Sect.~\ref{sect:conclusion}. 

\begin{figure}[t]
   \centering
   \includegraphics[scale=0.6]{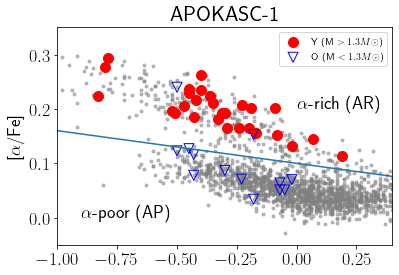}
      \includegraphics[scale=0.6]{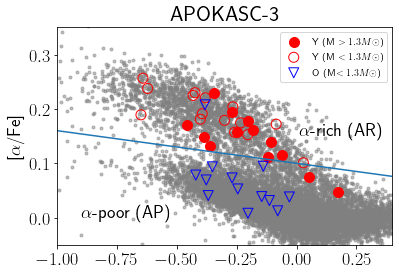}
         \includegraphics[scale=0.6]{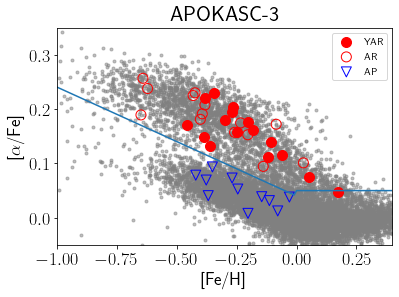}
   \caption{Tinsley-Wallerstein diagrams for the stars in our sample alongside with the rest of the APOKASC catalogue version 1 (\textit{top panel}) or version 3 (\textit{middle and bottom panels}). The blue lines indicate 
   different thresholds between  $\alpha$-rich ("AR") and $\alpha$-poor ("AP") stars:
   {Eq.~\ref{eq:alpha} for the middle panel and Eq.~\ref{eq:alpha2} for the bottom panel, see Sect.~\ref{sect:data}}. Filled symbols represent stars whose mass is above 1.3~M$_\odot$ and open symbols represent stars with masses below that limit. The mass is taken from APOKASC-1 (\textit{top panel}) or APOKASC-3 (\textit{middle and bottom panels}). The bottom panel indicates our final classification as YAR, $\alpha-$rich (AR) or $\alpha-$poor (AP; see text).}
              \label{fig:tinsley13}%
    \end{figure}

\input{table_ids_jul2022}

\begin{table*}[t]
\caption{\label{Tab:starsbyclass}
Stars classified according to Eq.~\ref{eq:alpha2} for their chemistry and according to a threshold of 1.3~M$_\odot$ for their mass.}
   \centering
    \begin{tabular}{|c|c|l|}
    \hline
    class & $N$ & stars \\
    \hline
    YAR & 16 &  
 Y1, Y2, Y3, Y4, Y5, Y6, Y7, Y8, Y9, Y10, Y11, Y12, Y13, Y16, Y18, Y28  \\
    AR & 14 &  O3,  O4, Y14, Y15, Y17, Y19, Y20, Y21, Y22, Y23, Y24, Y25, Y26, Y27 \\
    AP & 11 & O1, O2, O5,  O6, O7, O8, O9, O10, O11,  O12, O13\\
    \hline
    \end{tabular}
    \end{table*}

\section{Data}\label{sect:data}

\subsection{Chemical abundances and masses}

{The initial sample of 26 stars from Paper~I contained 13 YAR stars reported by \cite{Martig-2015}. It also included  a control sample of 13 stars with similar atmospheric parameters but with masses below 1.2~$\mathrm{M}_{\odot}$, as inferred from the \apo\ catalogue, version 1 \citep[][hereafter APOKASC-1]{Pinsonneault-2014}, which is a joint project between APOGEE and {\it Kepler}. In Table~\ref{Tab:sample}, the two groups are labeled "Y" and "O", respectively, in accordance with the denomination used in Paper~I.  This initial sample of 26 stars has been supplemented in 2017 by 15 supposedly YAR {(Y)} stars, as judged from APOKASC-1.} The stars are plotted in Fig.~\ref{fig:tinsley13} with different symbols.

More precisely, to select the new stars for monitoring, we defined at the time the Y stars as having masses in excess of 1.3~$\mathrm{M}_{\odot}$ 
and belonging to the thick disk (thus with $[\alpha/\mathrm{Fe}] > [\alpha/\mathrm{Fe}]_\mathrm{thresh}$), with [$\alpha/\mathrm{Fe}]_\mathrm{thresh}$ defined according to:
\begin{equation}\label{eq:alpha}
[\alpha/\mathrm{Fe}]_\mathrm{thresh} =  -0.06 \times [\mathrm{Fe/H}] + 0.1.
\end{equation}
 This relation was taken from \cite{Masseron-2017} and follows the analysis of \cite{Izzard-2018}, who also used the same mass threshold for selecting the overmassive stars, under the argument that individual stars with 
 masses {exceeding this limit} cannot exist in the old thick disk anymore. Additionally, for this study we considered a cut in magnitude, selecting only stars with $Ks < 11$ (with the only exception of O4) to be observable by the HERMES spectrograph \citep{Raskin-2011}. 

{The initial and complementary samples of} stars are listed in Table~\ref{Tab:sample}. In addition to listing different IDs of the stars, we list the magnitudes and the atmospheric parameters as taken from APOGEE DR16. Although the DR17 is now available \citep{2022ApJS..259...35A}, the most recent mass determinations were based on DR16; {therefore DR16 is used throughout this study}
for consistency.

{In the top panel of Fig.~\ref{fig:tinsley13}}, the $\alpha$ abundances and metallicities are from APOGEE DR14 \citep{Holtzman-2018}, which were the values we had at the time the first radial-velocity observations were obtained (see below). We note that the latest published APOKASC catalogue is version 2 \citep[][hereafter APOKASC-2]{Pinsonneault-2018}.  We nevertheless chose to make use of APOKASC-3 (Pinsonneault et al., in prep.) rather than APOKASC-2 to avoid having to revise our results soon after publication.

\subsubsection{Revised classification because of updates in published data}

We stress that the original nomenclature Y1 - Y28 and O1 - O13, which has been kept here only for backward compatibility with Paper~I, does not correspond to any physical reality any longer, as revealed by the middle panel of Fig.~\ref{fig:tinsley13} built from APOGEE DR16 \citep{Ahumada-2020} and APOKASC-3. It shows that some Y stars have $\alpha$ abundances below the threshold of Eq.~\ref{eq:alpha}. Moreover, for several stars previously classified as Y, new masses are now too small to maintain that classification (these stars are represented as open circles in the middle and bottom panels of Fig.~\ref{fig:tinsley13}). 

A comparison and discussion of the differences between these catalogs can be found in Appendix~\ref{mass_comparison}.  Since the distribution of our sample stars among the mass and [$\alpha$/Fe] categories is so different from their original assignment (in particular, many candidates tagged as Y from APOKASC-1  -- red filled symbols in the upper panel of Fig.~\ref{fig:tinsley13} -- do not belong any longer to that category in the APOKASC-3 catalogue), a  new classification is necessary, which we now describe.

We first need a new definition of  $\alpha-$rich (AR) and $\alpha-$poor (AP) stars which separates better the two sequences in the Tinsley-Wallerstein diagram with APOGEE DR16 values. To do so, we adopt the criterion of \cite{Miglio-2021}, namely 
\begin{equation}\label{eq:alpha2}
   [\alpha/\mathrm{Fe}]_\mathrm{thresh} = 
\begin{cases}
    -0.2 \times [\mathrm{Fe/H}] + 0.04  ,& \text{if }   \mathrm{[Fe/H]} <0\\
    0.04,              & \text{if }   \mathrm{[Fe/H]} >0.
\end{cases}
\end{equation}

 We split the sample into 
AR stars (with  $[\alpha/\mathrm{Fe}] > [\alpha/\mathrm{Fe}]_\mathrm{thresh}$) and AP stars (with $[\alpha/\mathrm{Fe}]$ below the threshold) as displayed in the bottom panel of Fig.~\ref{fig:tinsley13} with red and blue symbols, respectively.  Then, among AR stars, those with a mass in excess of 1.3~M$_\odot$ (according to APOKASC-3) are flagged 'YAR', and are plotted as red filled circles. Table~\ref{Tab:starsbyclass} lists the stars according to this final classification.

\begin{figure*}[t]
\centering
    \includegraphics[scale=0.45]{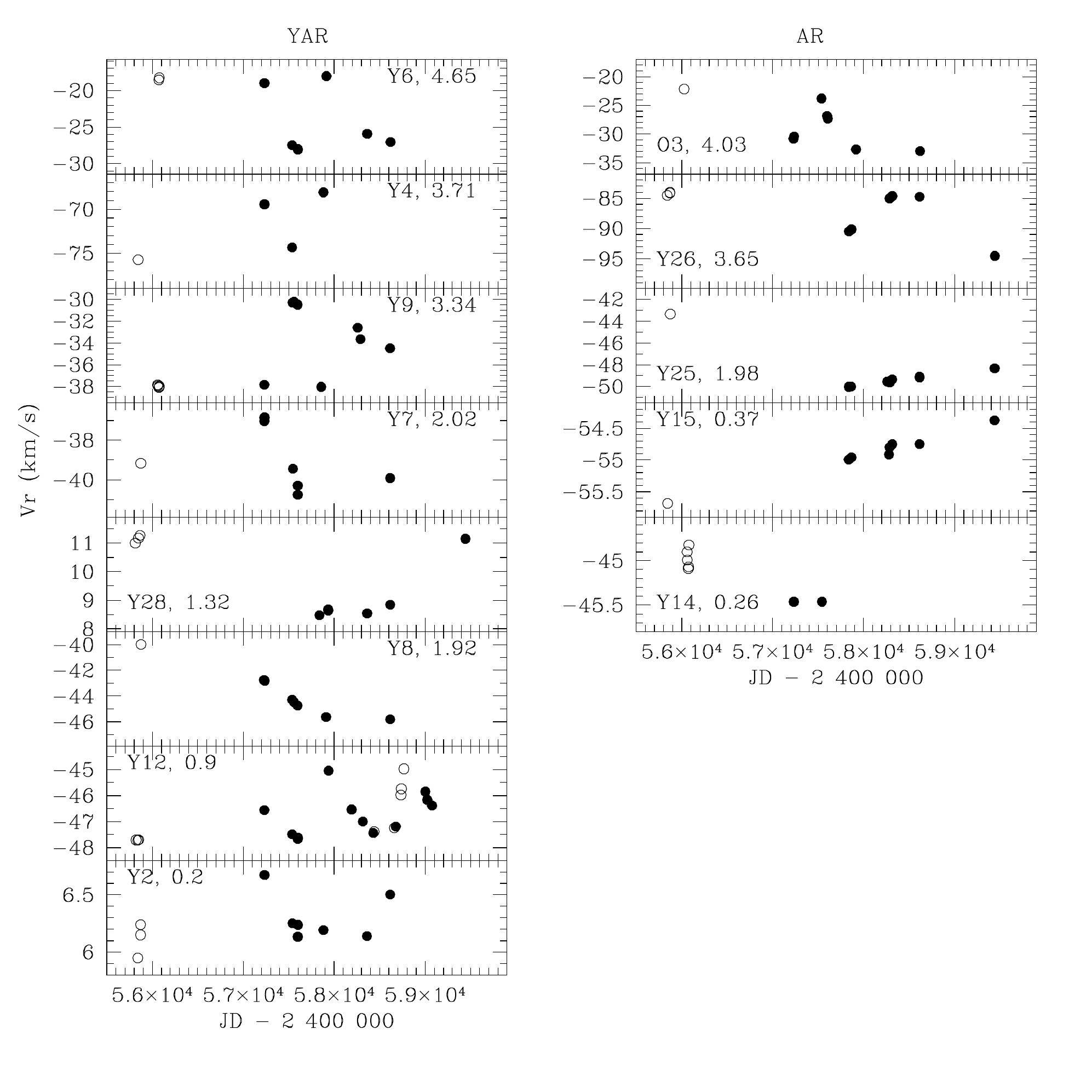}
    \includegraphics[scale=0.45]{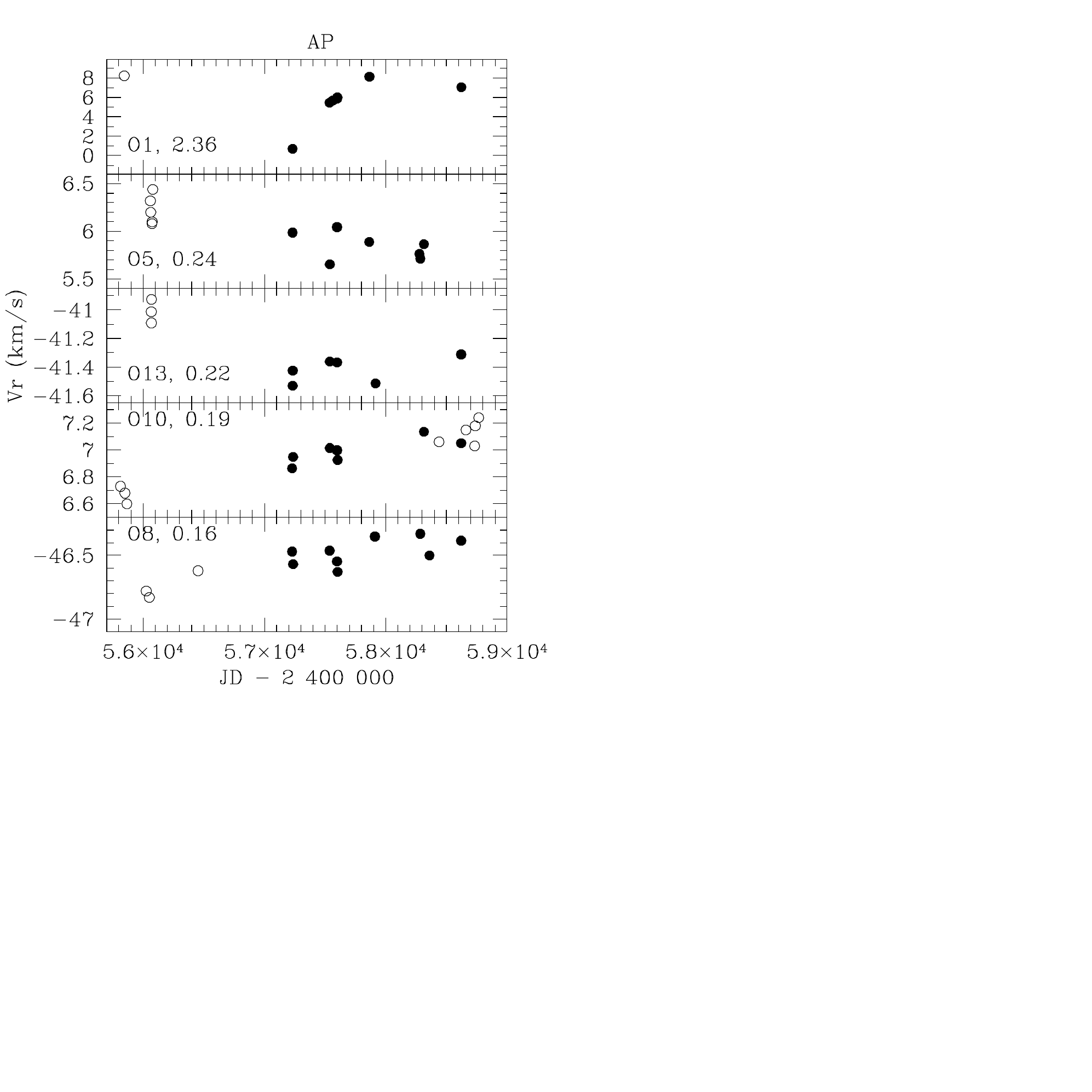}\vspace{-0.7cm}\\
    \mbox{}\hspace{-9cm}
    \includegraphics[scale=0.45]{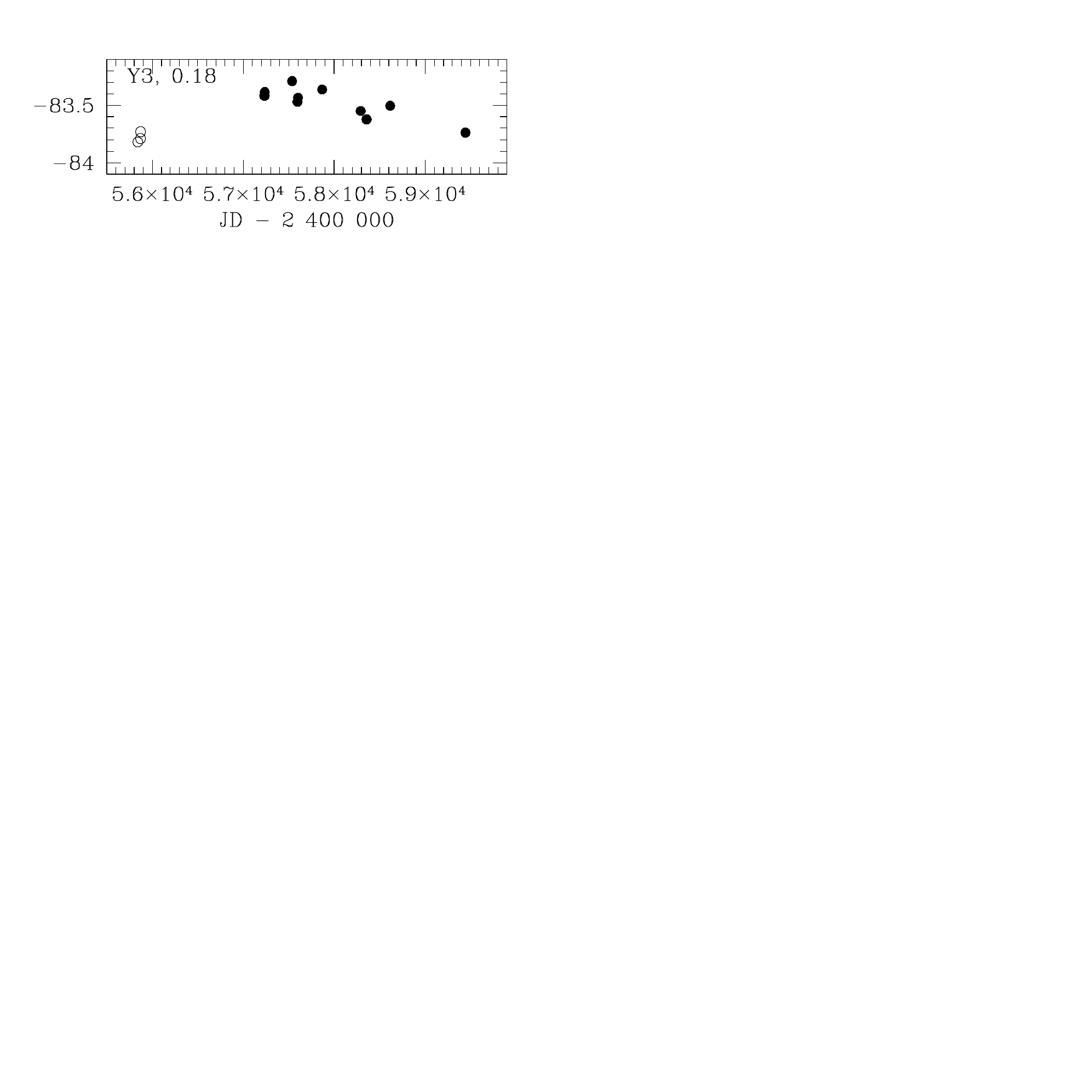}
\vspace{-5cm}
\caption{
{Heliocentric radial velocities plotted as a function of Julian date}
for  stars flagged as binaries (filled dots: HERMES; open dots: APOGEE). Each column collects stars of a given category (YAR, AR or AP, respectively). The labels in each panel give the star number and the HERMES+APOGEE RV standard deviation $\sigma(RV)$ (in km~s$^{-1}$). Stars are ordered in terms of decreasing $\sigma(RV)$ values (HERMES+APOGEE).
}
\label{Fig:binaries}
\end{figure*}
 
There is some degree of arbitrariness in the definition of YAR stars, especially considering the adopted mass threshold and the uncertainties on the masses. Some AR stars fall very close to the here adopted 1.3~M$_\odot$ mass threshold for YAR stars,  and could well classify as YAR considering the mass uncertainties.  The impact of the arbitrariness in the definition of YAR stars on the conclusion of our analysis is extensively discussed in Sect.~\ref{sect:masses_apo}.

\subsection{Kinematic properties}\label{sect:data_kin}
 To interpret our results, we consider the space motions of the stars from Gaia eDR3 \citep{Gaia-EDR3}. In particular, we cross-matched the entire APOKASC-3 sample with the Value Added Catalog {\tt AstroNN} to have information about the Galactic orbits. That catalogue uses distances derived with a neural network trained directly on the APOGEE spectra of stars with known parallaxes \citep{jofre15} as described in \cite{Leung-19}. The determination of dynamical properties, such as total velocities, as well as  actions, angles, eccentricities and energies of the orbits are described in \cite{mackereth-18}. {}That work assumes the solar Galactic radius and height above the midplane to be 
$R_\odot = 8~$kpc, $z_\odot = 25~$pc and a circular velocity of  220~km/s. It further assumes a solar motion of  $[U, V, W]_\odot = [11.1,
12.24, 7.25]~$km/s \citep{schoenrich}. \cite{mackereth-18} have implemented the {\tt galpy} package \citep{Bovy-2015} to derive the dynamical properties using the gravitational potential  {\tt MWPotential2014}.

  {For most of our target stars, the Gaia RUWE ('reduced unit-weight error') parameter, which traces stars with large uncertainties on their astrometric data \citep[when RUWE~$ > 1.4$]{Lindegren2018}, remains below that threshold (Table~\ref{Tab:Gaia_DR2_DR3}), indicating that the astrometric proper motions used for the kinematical computations are reliable. The binary stars O3 and Y25 are the only exceptions, however, with RUWE values of 4.05 and 1.83, respectively.}

\begin{figure*}[t]
\centering
    \includegraphics[scale=0.45]{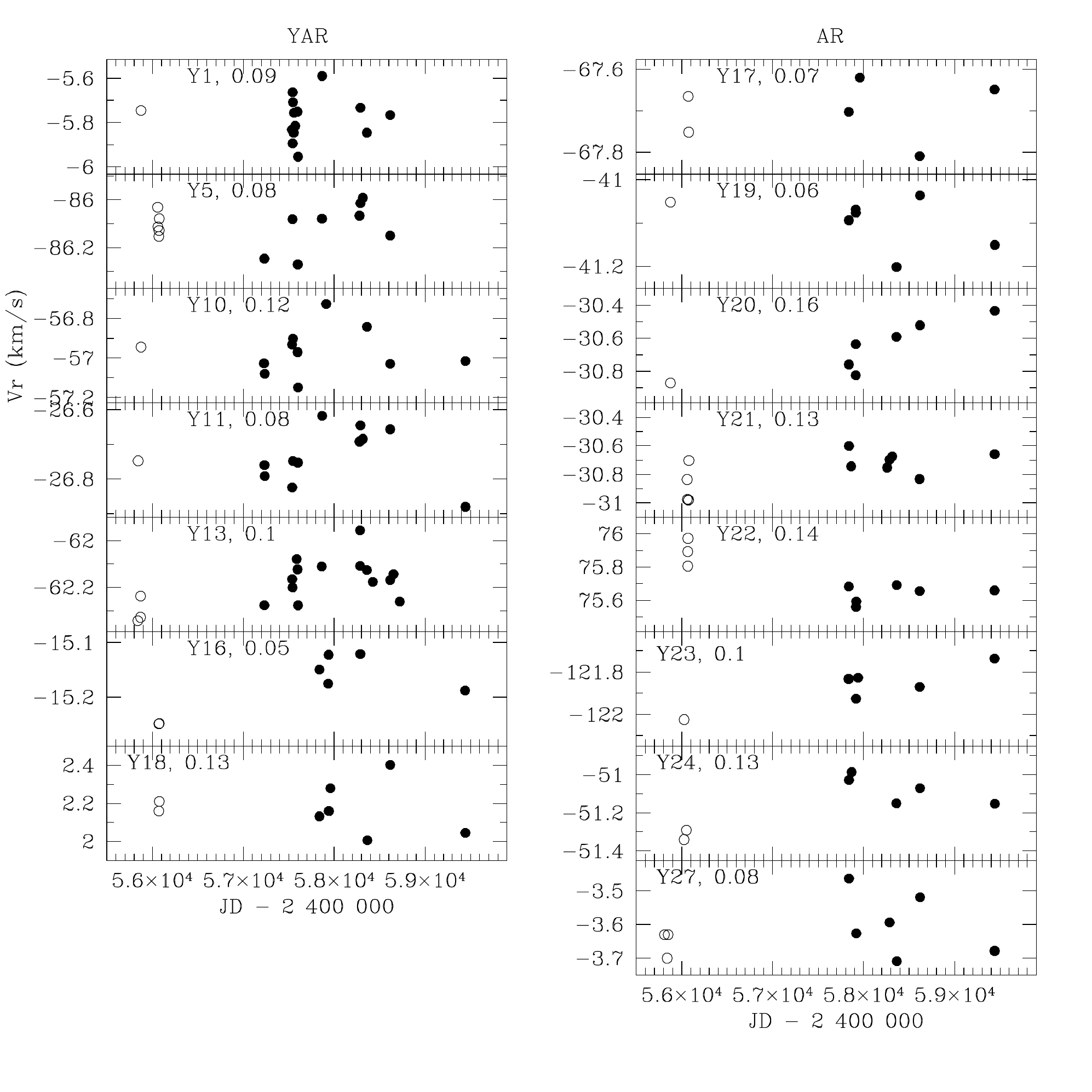}
    \includegraphics[scale=0.45]{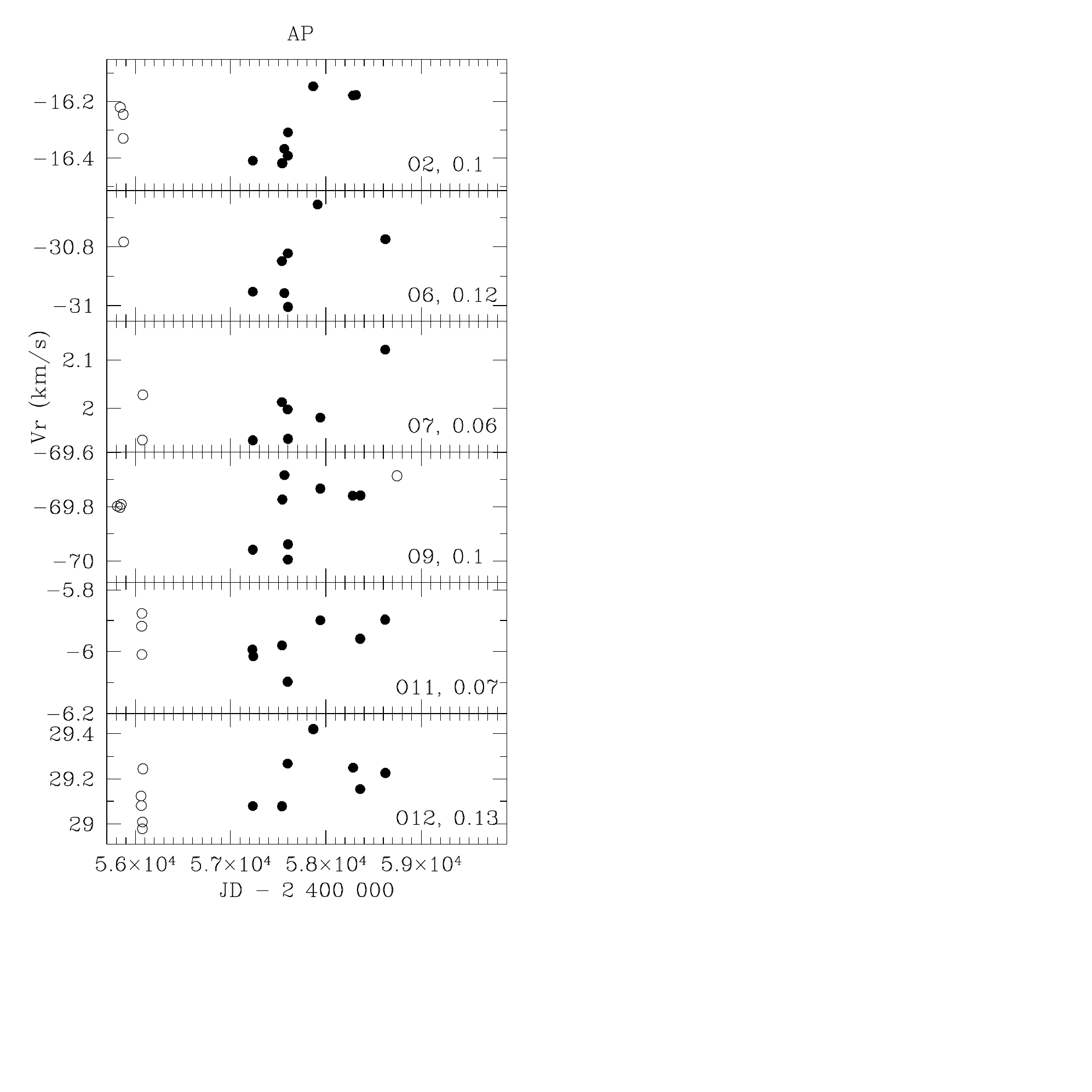}
    \vspace{-0.7cm}\\
    \includegraphics[scale=0.45]{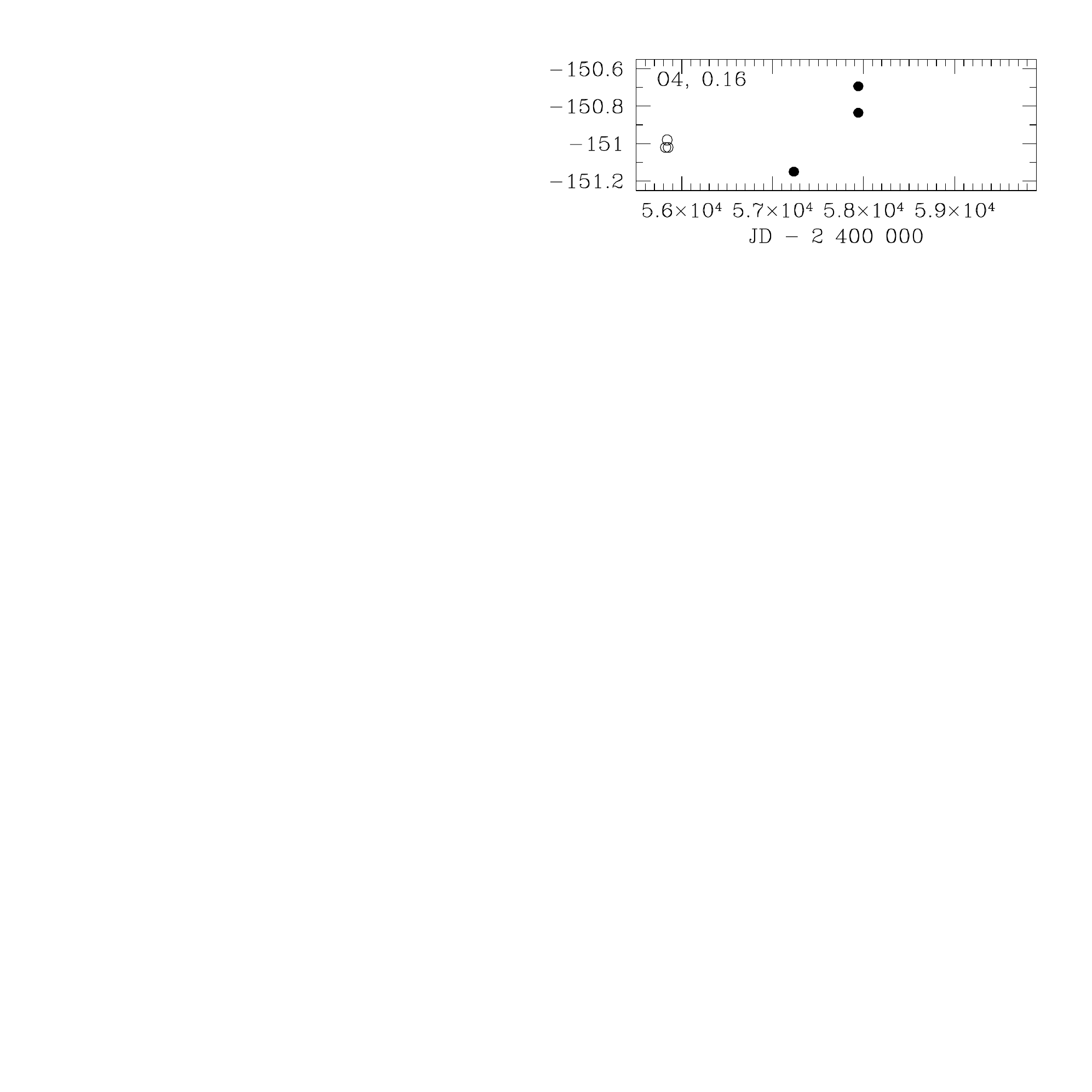}\hspace{\fill}
    \vspace{-6.7cm}\\
\caption{Same as Fig.~\ref{Fig:binaries}, but for  stars not flagged as binaries.}
\label{Fig:constant}
\end{figure*}

\subsection{Radial velocities} 

The RV data were obtained with the HERMES spectrograph \citep{Raskin-2011} mounted on the 1.2~m Mercator telescope,  at the Roque de Los Muchachos Observatory,  La Palma, Canary Islands.  The HERMES spectrograph covers the optical wavelength range from $380$ to $900~\mathrm{nm}$ with a spectral resolution of about 86~000. RVs were derived by cross-correlating the stellar spectrum with a mask covering the wavelength range $480 - 650~\mathrm{nm}$ and mimicking the spectrum of Arcturus (K1.5~III). The restricted wavelength span is to avoid both telluric lines at the red end and the crowded and poorly exposed blue end of the spectrum. \\
\indent The exposure times were calculated according to the brightness of the star in order to achieve a signal-to-noise ratio (S/N) of about 15   {per pixel around 550~nm.} 
This is normally sufficient to obtain a well-defined cross-correlation function (CCF) with its minimum defined within a few $\mathrm{m\; s^{-1}}$. The actual uncertainty on the RV is however larger than the formal uncertainty on the minimum obtained by the Gaussian fit of the CCF, as this actual uncertainty should take into account the long-term RV stability of the spectrograph, on the order of 55~m~s$^{-1}$ \citep[see][where more details about the RV acquisition may be found; also Sect.~\protect\ref{Sect:RV} below]{Jorissen-2016}.

The HERMES observations for the initial sample (Y1 - Y13 and O1 - O13) cover the 
time span July 2015 till May 2019 (about $\sim 1400$~d)  
and March 2017 
till August 2021 (about $\sim 1600$~d) for the new sample (Y15 - Y28). Due to an encoding problem, the  star Y14 was only observed twice over a 1-yr time span. 

 The individual RVs for all target stars are listed in Table~\ref{Tab:RV}. 
Table~\ref{28stars} summarizes the RV results, which will be further discussed in the following sections. We complement our data with measurements obtained by APOGEE and Gaia.

\section{Binarity diagnostics}\label{sect:results}
\label{Sect:RV}

In order to assess whether a star is a binary we use the criteria from Paper~I applied on data from HERMES and APOGEE.

Figure~\ref{Fig:binaries} presents the RV curves for all stars  flagged as spectroscopic binaries (SB), whereas Fig.~\ref{Fig:constant} displays the RVs of stars flagged as constant. Each column of the figures collects  binaries from a given category (YAR, AR, or AP listed in Tab~\ref{Tab:starsbyclass}). Filled circles correspond to HERMES observations while open circles correspond to APOGEE observations. In this section we explain how we conclude on the binary nature of the stars.

\subsection{HERMES radial velocities}

Assuming that the uncertainty on measurement RV$_i$ is $\sigma_i$, the $\chi^2$ value based on all $N$ observations is
\begin{equation}
\label{Eq:chi2}
\chi^2 = \sum_{i=1}^N  \frac{\left(RV_i- \overline{RV}\right)^2}{\sigma_i^2} \, ,
\end{equation}
\noindent where $\overline{RV}$ is the mean RV value computed from the set of  $N$ observations. 
Since these $\chi^2$ distributions have different degrees of freedom depending on the number of observations available for the different stars, 
we use instead as binarity diagnostics the quantity $F2$  related to the reduced $\chi^2$ \citep{Wilson-1931}, and defined as:
\begin{equation}
F2 = \left(\frac{9\nu}{2}\right)^{1/2}\left[\left(\frac{\chi^2}{\nu}\right)^{1/3}+\frac{2}{9\nu}-1\right],
\label{Eq:F2}
\end{equation}
where $\nu = N-1$ is the number of degrees of freedom of the $\chi^2$ variable. The transformation of ($\chi^2, \nu$) to $F2$ eliminates the inconvenience of having the distribution depending on the additional variable $\nu$, which is not the same over the whole sample of stars. $F2$ follows a normal distribution with zero mean and unit standard deviation, provided that the $\sigma_i$ are correctly estimated.   {As in Paper I, we adopt $\sigma_i = 0.09$~km~s$^{-1}$ for all HERMES measurements in order} 
{to avoid a deficit of stars in the left wing of the $F2$ distribution}
(displayed in the top panel of Fig.~\ref{fig:mass_prob}) even though the stability of radial-velocity standards would rather call for 0.055~km~s$^{-1}$  \citep{Jorissen-2016}. There might be a slight excess of binary stars on the right wing of the distribution (in the range $1 \le F2 \le 4$; top panel of Fig.~\ref{fig:mass_prob}), but we consider that the statistical evidence is not strong enough to flag them as binaries. Among those, Y20 and Y23 may be considered as possible binaries based on the RV trend seen on Fig.~\ref{Fig:constant}.  
Finally, the probability $Prob$ of a star being a SB was calculated from the $\chi^2$ distribution with $\nu$ degrees of freedom.  
Stars are flagged as SB when $F2 \ge 3$, which is roughly equivalent to $Prob \ge 0.9990$. 

These values, together with the mean $\overline{RV}$ and its standard deviation, are listed in the first set of columns of Table~\ref{28stars}.

\subsection{APOGEE radial velocities}

APOGEE data from DR17 \citep{2022ApJS..259...35A} for the samples under discussion  were obtained between JD 
2455811 and  2456452 (i.e., September 7, 2011 to June 8, 2013, with a few more measurements from JD 2458439 to 2458766, i.e., November 16, 2018 to October 10, 2019, specifically for O9, O10, and Y12). In the case of O10, these measurements (represented as open circles in Fig.~\ref{Fig:binaries}) are simultaneous to existing HERMES data, and reveal a zero-point offset $RV_{\rm HERMES} - RV_{\rm APOGEE}$ of $-0.40$~km~s$^{-1}$. Therefore, this zero-point offset has been applied to all APOGEE velocities listed in Table~\ref{Tab:RV}. These zero-point-corrected APOGEE RVs were then combined to the HERMES RVs to compute the overall standard deviation and $F2$ value (Eq.~\ref{Eq:F2}), as listed in Table~\ref{28stars}, with $F2 > 3$ used as well as binarity diagnostic.   

We note that, although the formal error on the individual APOGEE RVs is about 0.02 - 0.03~km~s$^{-1}$, the standard deviation of the APOGEE RVs for stars not flagged as SB by the other data sets is about 0.06 - 0.07~km~s$^{-1}$ (see for instance O2, O7, O9, O11... in Table~\ref{28stars}). Therefore, an uncertainty of 0.09~km~s$^{-1}$ has been adopted for each individual APOGEE RV when computing $\chi^2$ and the associated $F2$.   {This uncertainty is consistent with improving the symmetry of the $F2$ distribution}. APOGEE data are especially important to qualify O8 and O10 as binaries (see Fig.~\ref{Fig:binaries}), 
{because they reveal a RV trend that was not clear from} 
HERMES data {alone}.

\subsection{Gaia radial velocities}
\label{Sect:Gaia_DR2}

{Before Gaia Data Release 3 became available in June 2022, an extensive analysis of the RV data provided by  Gaia Data Release 2 \citep[DR2;][]{GaiaDR2,  Katz-2019} had been performed. Gaia DR2 RVs  span} the range JD 2456863.5 to 2457531.5 (2014 July 25 to 2016 May 23), just prior to the HERMES RV monitoring.   {The absence of temporal overlap between HERMES and Gaia DR2 offers an advantage to detect SB from this comparison, and this advantage would be somewhat diluted by using DR3 RVs instead.} 
Although the individual Gaia RV data will not be available until Gaia DR4, the average RV provided by Gaia DR2 turns out to be useful.
The expected uncertainty $\epsilon_{\rm RV,DR2}$ on the Gaia DR2 average velocity $\overline{RV}_{\mathrm {DR2}}$ is computed from the number of transits $N$ and Gaia RVS magnitude $G_{\rm RVS}$ using the data from Fig.~18 of \citet{Katz-2019}, namely:
\begin{eqnarray}
\epsilon_{\rm RV,DR2}  &=&  (-0.429 + 1.019\; G_{\rm RVS} - 4.456\;10^{-1} \;G_{\rm RVS}^2 \nonumber \\
&& +\; 8.542\; 10^{-2} \; G_{\rm RVS}^3 - 7.629\; 10^{-3}\; G_{\rm RVS}^4 \nonumber \\
&&+\; 2.626\; 10^{-4}\; G_{\rm RVS}^5)\nonumber \\
&& \times\; (1.809 -0.179\; N + 1.38\;10^{-2}\; N^2\nonumber  \\
&&-\; 6.06\; 10^{-4}\; N^3 + 1.345 \;10^{-5}\; N^4\nonumber  \\
&&-\; 1.16\; 10^{-7}\; N^5), 
\end{eqnarray}
where the first polynomial in powers of $G_{\rm RVS}$ corresponds to $N = 8$  \citep[as given in][]{Jorissen-2020} and the polynomial in powers of $N$ applies a scaling factor. The quantities $\overline{\mathrm{RV}}_{\mathrm {DR2}}$, $\sigma_{\rm RV,DR2}$, and $\epsilon_{\rm RV,DR2}$ are listed in Table~\ref{28stars}. A star is flagged as a binary if $\sigma_{\rm RV,DR2} > 3\;\epsilon_{\rm RV,DR2}$, or if $\Delta RV \equiv |RV_{\mathrm{HERMES}} - RV_{\mathrm{DR2}}| > 3\; \epsilon_{\rm RV,DR2}$. The Gaia DR2 RV data do not bring up new SB detections, but confirms those flagged as such by the HERMES or APOGEE data. 
  {For the sake of completeness, the average Gaia DR2 and DR3 RVs are compared in Table~\ref{Tab:Gaia_DR2_DR3}. With the Gaia DR3 RVs comes a rigorous statistical analysis of the same kind as that performed with the HERMES RVs, namely the publication of an $F2$-like variable (named 'renormalized-gof' in Table~\ref{Tab:Gaia_DR2_DR3}) and the associated $p$-value (named 'Prob' in Table~\ref{Tab:Gaia_DR2_DR3}). Star Y21 is the only one to appear as a new binary from this analysis. However, as Gaia DR3 is the only data set to flag it as binary, we still consider it as uncertain, and will not add it in the binary statistics of our analysis.} 

\begin{figure}[t]
   \centering
   \includegraphics[scale=0.45]{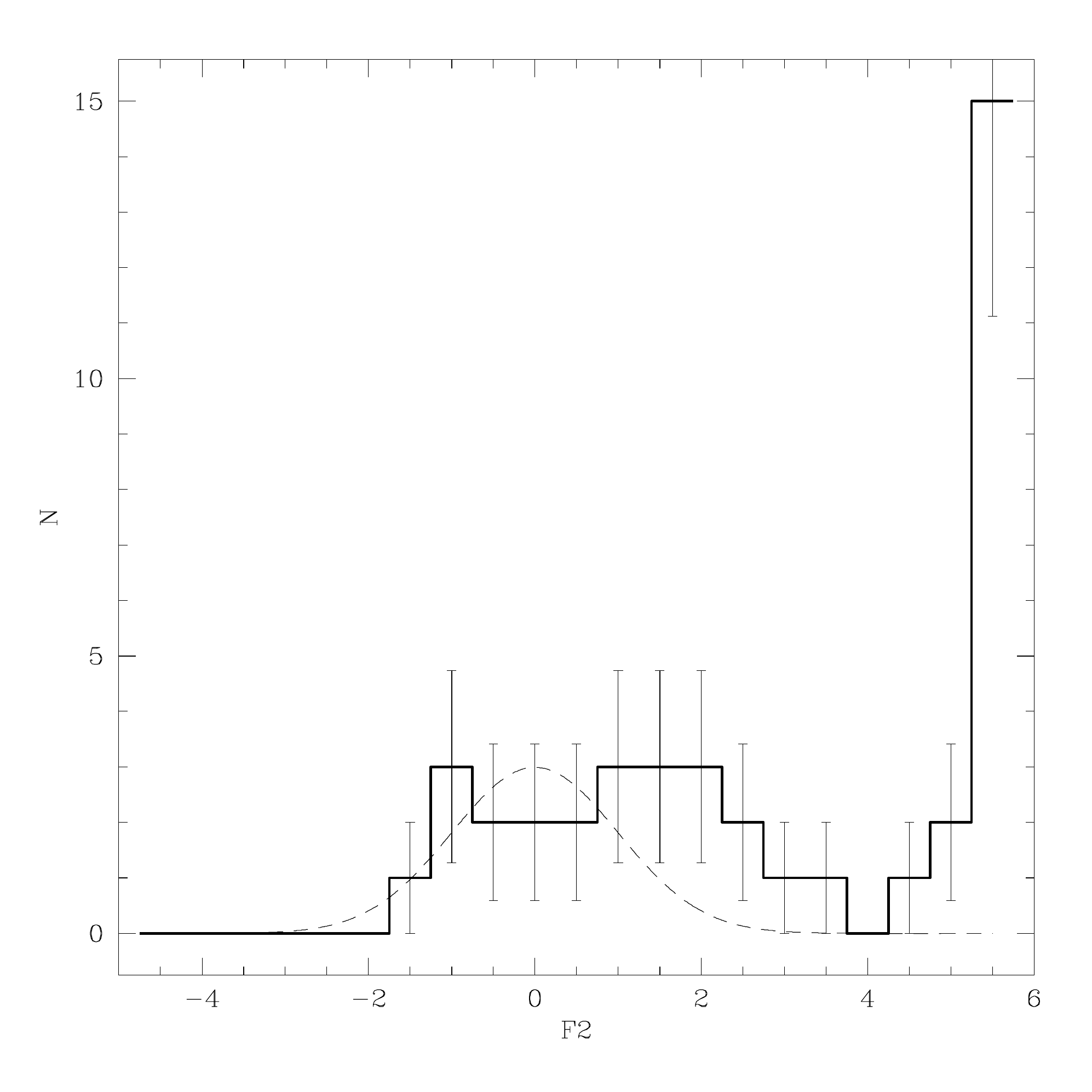}
   \includegraphics[scale=0.45]{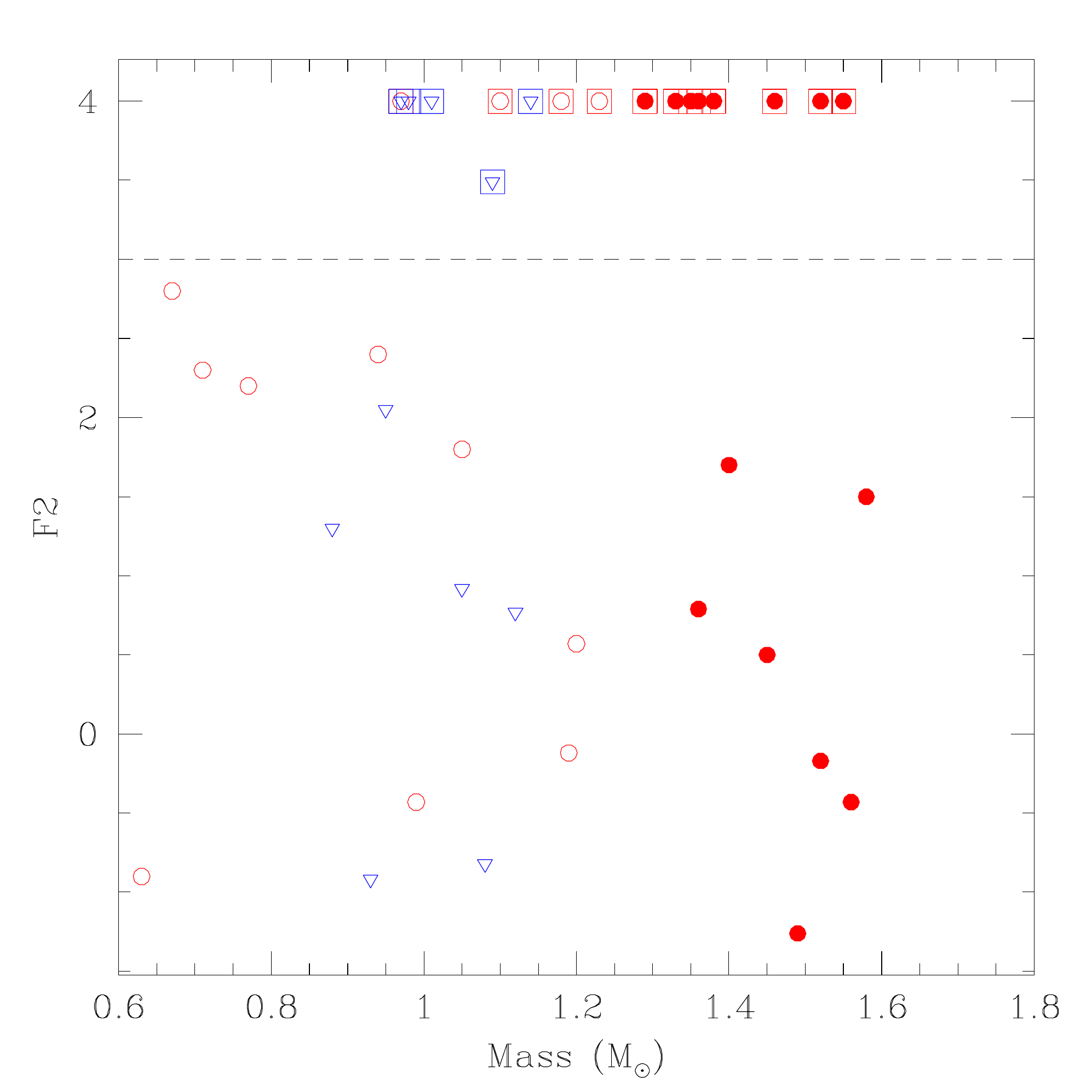}
   \caption{Distribution of $F2$. 
   \textit{Top panel:}   {Distribution for the full sample, adopting 0.09~km~s$^{-1}$ as the typical error on HERMES radial-velocity measurements.}
   \textit{Bottom panel:} Stellar mass vs $F2$ (see Eq.~\ref{Eq:F2}) for APOGEE and HERMES data combined. The threshold for binarity has been set at  $F2 > 3$. 
   The symbols follow the classification of Fig.~\ref{fig:tinsley13} (when the star is considered as a binary, its symbol is enclosed within a square). Stars with $F2 \ge 4$ are displayed at $F2 = 4$. }
              \label{fig:mass_prob}%
    \end{figure}

    \begin{figure}[t]
   \centering
       \includegraphics[scale=0.465]{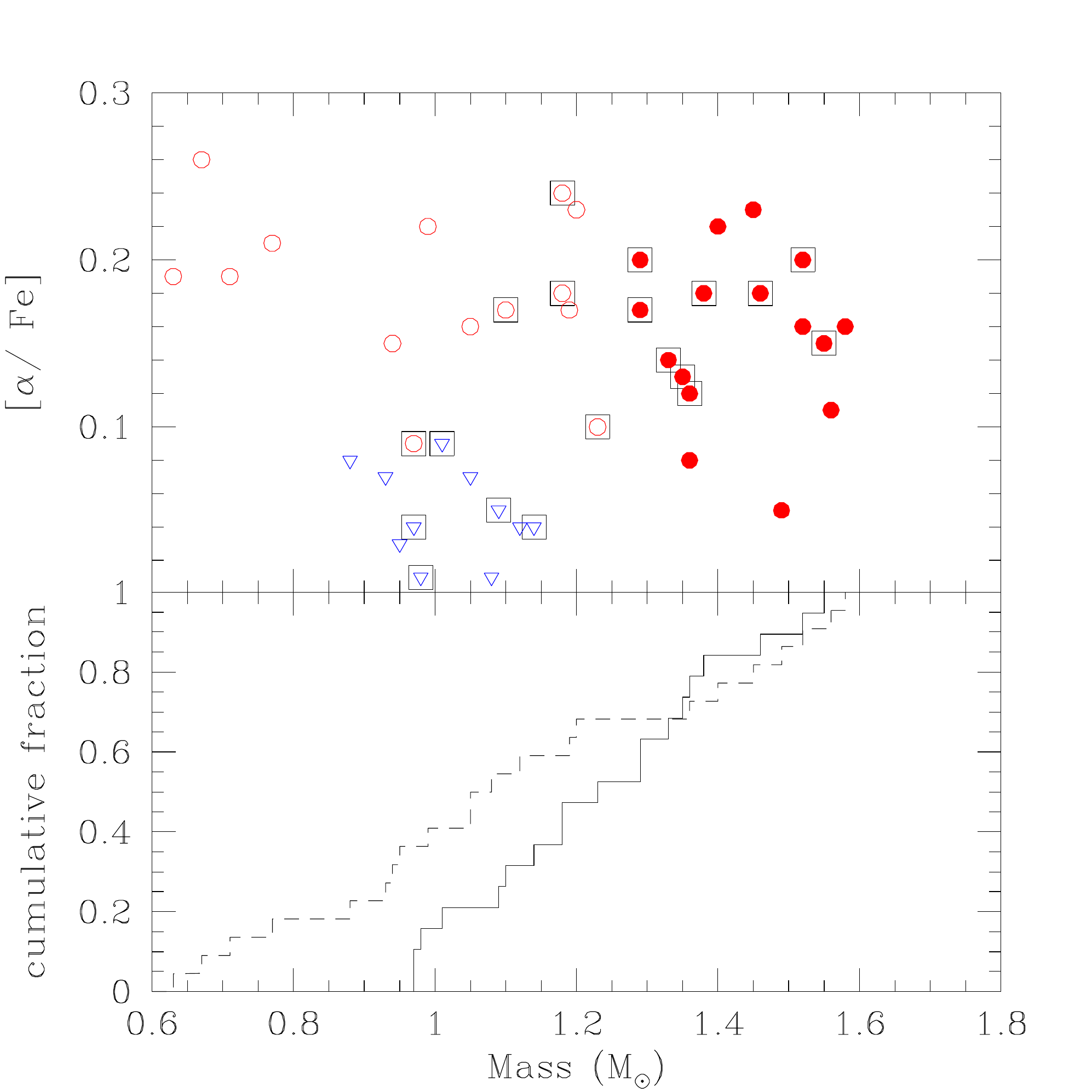}
       \includegraphics[scale=0.45]{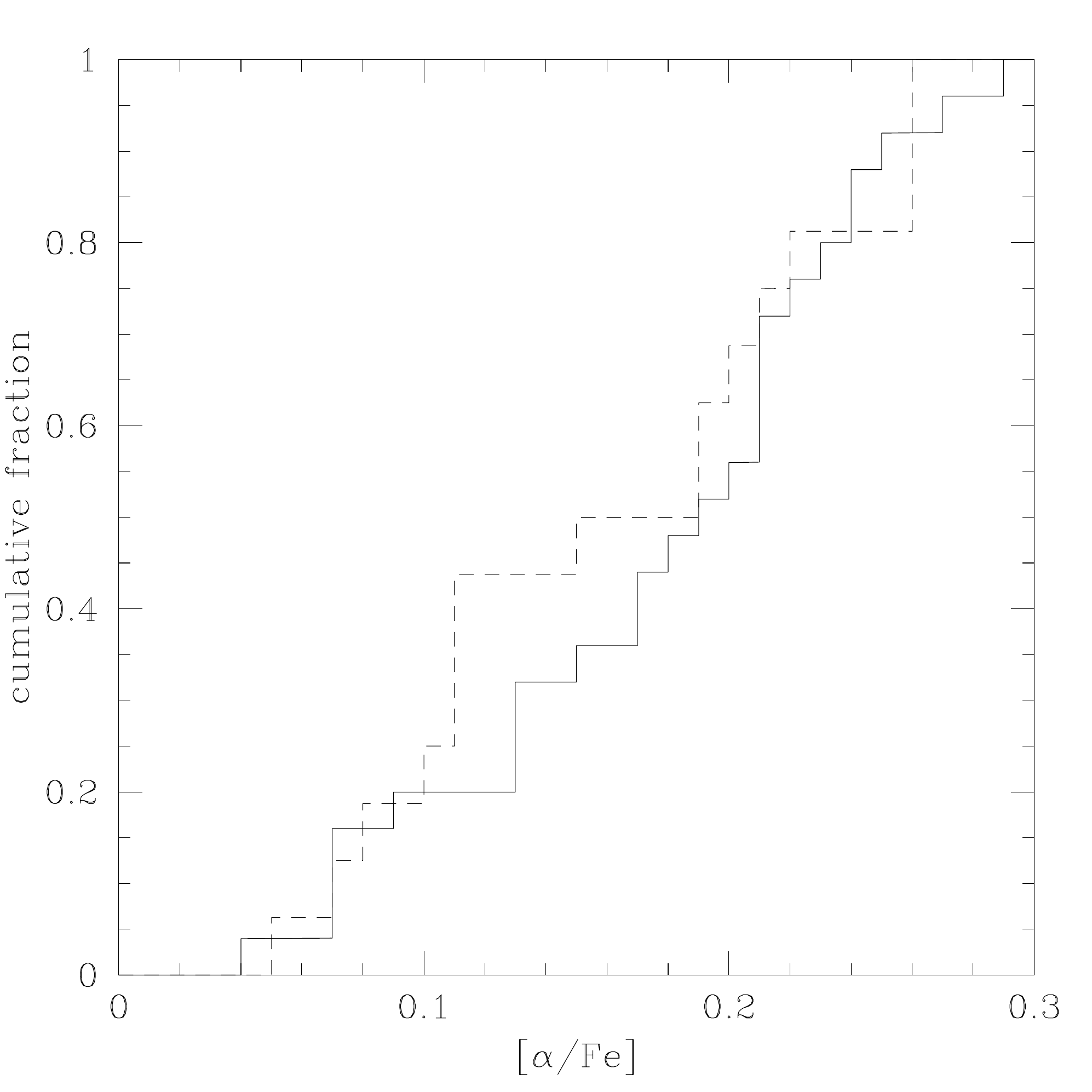}
   \caption{\textit{Top panel}: Stellar mass versus [$\alpha/$Fe] abundance ratio. Symbols are as in Figs.~\ref{fig:tinsley13} and \ref{fig:mass_prob}.  \textit{Middle panel}: The cumulative frequency distributions of SBs (solid line) and nonSBs (dashed line). \textit{Bottom panel}: Same as middle panel, but for [$\alpha$/Fe]. }
              \label{Fig:mass_metal}%
    \end{figure}

   \begin{figure}[t]
   \centering
       \includegraphics[scale=0.465]{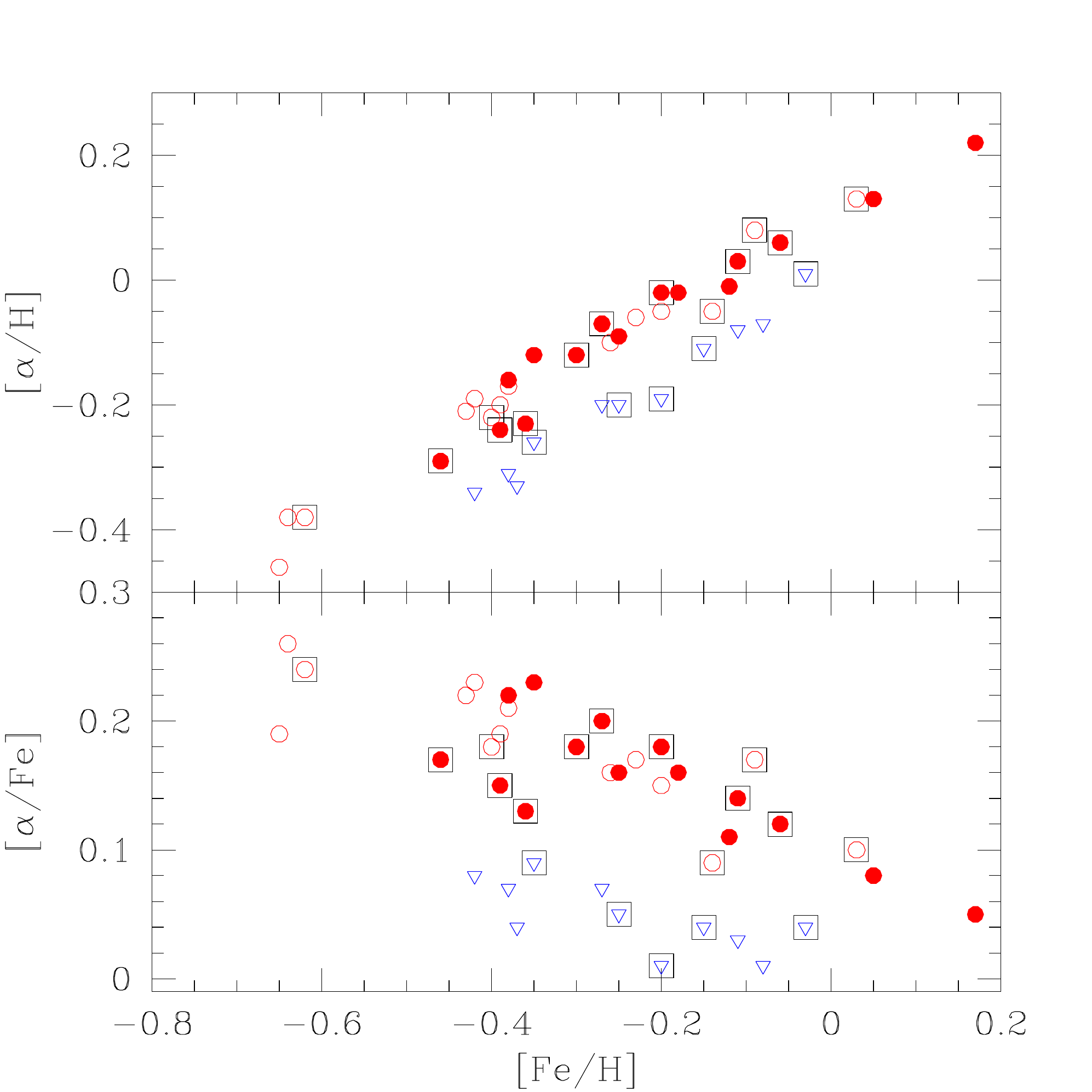}
   \caption{\textit{Top panel}: Binary and nonbinary stars in the [$\alpha/$H] versus [Fe/H] diagram. Symbols are as in Figs.~\ref{fig:tinsley13} and \ref{fig:mass_prob}.   \textit{Bottom panel}: Same as top panel, but  for [$\alpha$/Fe] vs [Fe/H]. }
              \label{Fig:alpha_FeH}%
    \end{figure}

The Gaia DR3 nonsingle star (NSS) data \citep{Arenou2022} further  confirm the results obtained above. It provides SB1 orbits for stars O3, Y4, Y6, and Y9 (as listed in Table~\ref{Tab:orbital_elements}), and detect a first-degree RV trend for Y8, Y25, and Y26 and second-degree RV trend for O1 (see Table~\ref{Tab:Gaia_DR2_DR3}). The RUWE parameter listed in Table~\ref{Tab:Gaia_DR2_DR3} does not allow us to expand the sample of SBs already detected by other means.

\subsection{Binarity as a function of stellar properties}
\label{Sect:binarity-vs-stellar_properties}

In summary, when a star is flagged as binary from either HERMES, APOGEE or Gaia, we classify it as SB, as indicated by the last column of Table~\ref{28stars}. Comparing the present results with those of Paper~I for the stars in common (stars 1-13), {the binarity diagnostic is confirmed for many stars:  O1, O3, O8, O10, Y4, Y6, Y7, Y8, Y9, and Y12 are flagged as SBs here and in Paper~I}. 
However, there appears to be a few differences.  They concern stars  Y2, Y3, O5, and O12 which are newly flagged SBs (thanks to the larger number of data points and the more extended time span). 
In addition, Y1 was classified as a binary in Paper~I but here we show it is constant. The reason was one measurement in Paper~I corresponding to the radial velocity of another target, increasing the scatter of the RV associated to Y1 and hence the probability of Y1 being a binary.

In the lower panel of Fig.~\ref{fig:mass_prob}, the stellar mass is plotted as a function of the $F2$ index, with the star being considered as a binary when $F2 > 3$ (either for HERMES and APOGEE data as a whole, or considered separately; see Table~\ref{28stars}) with its corresponding symbol being enclosed within a square. 
{Figure~\ref{fig:mass_prob} reveals that at the low-mass end ($M < 0.95$~\Msun), the stars have a preference for low $F2$ values, meaning we do not find any binaries for such low masses. At the high mass end however,  the stars span a large range of $F2$ values, implying that they can be either  binaries or nonbinaries. } 

Figure~\ref{Fig:mass_metal}
displays the mass of the stars as a function of the [$\alpha$/Fe] abundance ratio, with symbols as in Fig.~\ref{fig:mass_prob}. 
The middle panel in that figure shows the cumulative frequency of binaries (solid line) and single stars (dashed line) as a function of mass. There is a   {clear} tendency for the binaries to become more numerous than single stars in the upper mass range (1.4 to 1.6~M$_\odot$),   {and to be lacking at the lower-mass end ($< 0.95$~M$_\odot$). The null hypothesis that the two samples (19 binaries, 22 nonbinaries) are extracted from the same parent population in term of its mass distribution cannot be rejected  ($p$-value of 24\%; Table~\ref{Tab:SBfrequency})}.

The lower panel of Fig.~\ref{Fig:mass_metal} displays the cumulative frequency of binaries (solid line) and single stars (dashed line) as a function of $[\alpha$/Fe]. The largest frequency difference amounts to 0.25 and is reached at $[\alpha/\mathrm{Fe}] \sim 0.12$, where there are more single stars than binary stars, but this difference is not large enough to be any significant in the framework of a Kolmogorov-Smirnov test comparing samples with sizes {19} (SB) and {22} (non SB; see Table~ \ref{Tab:SBfrequency}). 

Figure~\ref{Fig:alpha_FeH} shows the distribution of binary stars in the ([$\alpha$/H], [Fe/H]) and ([$\alpha$/Fe], [Fe/H]) diagrams, to be compared with Fig.~6 of \citet{mazzola2020}. But as we show in a quantitative manner in Table~\ref{Tab:SBfrequency}, our sample does not confirm in a statistically significant way the  prevalence of binaries found by  \citet{mazzola2020} among the AP stars.

\subsubsection{The role of the evolutionary stage}\label{sect:masses_apo}

It is well known that while mass is the main driver of time scales in stellar evolution, metallicity, and $\alpha-$abundances also play some role \citep[and references therein]{Warfield21}.  It is also  known that stars along the red giant branch (RGB) and in the red clump (RC) have indistinguishable spectra \citep{masseron-17b} even though RC stars are more evolved than RGB stars. 
Thus a straight cut at $M = 1.3$~M$_\odot$ (or any other value) in mass for the entire sample might not be representative of an age limit in a sample of stars with different metallicities,  $\alpha-$abundances,  and evolutionary stages. 

\begin{figure}[t]
   \centering
     \includegraphics[scale=0.40]{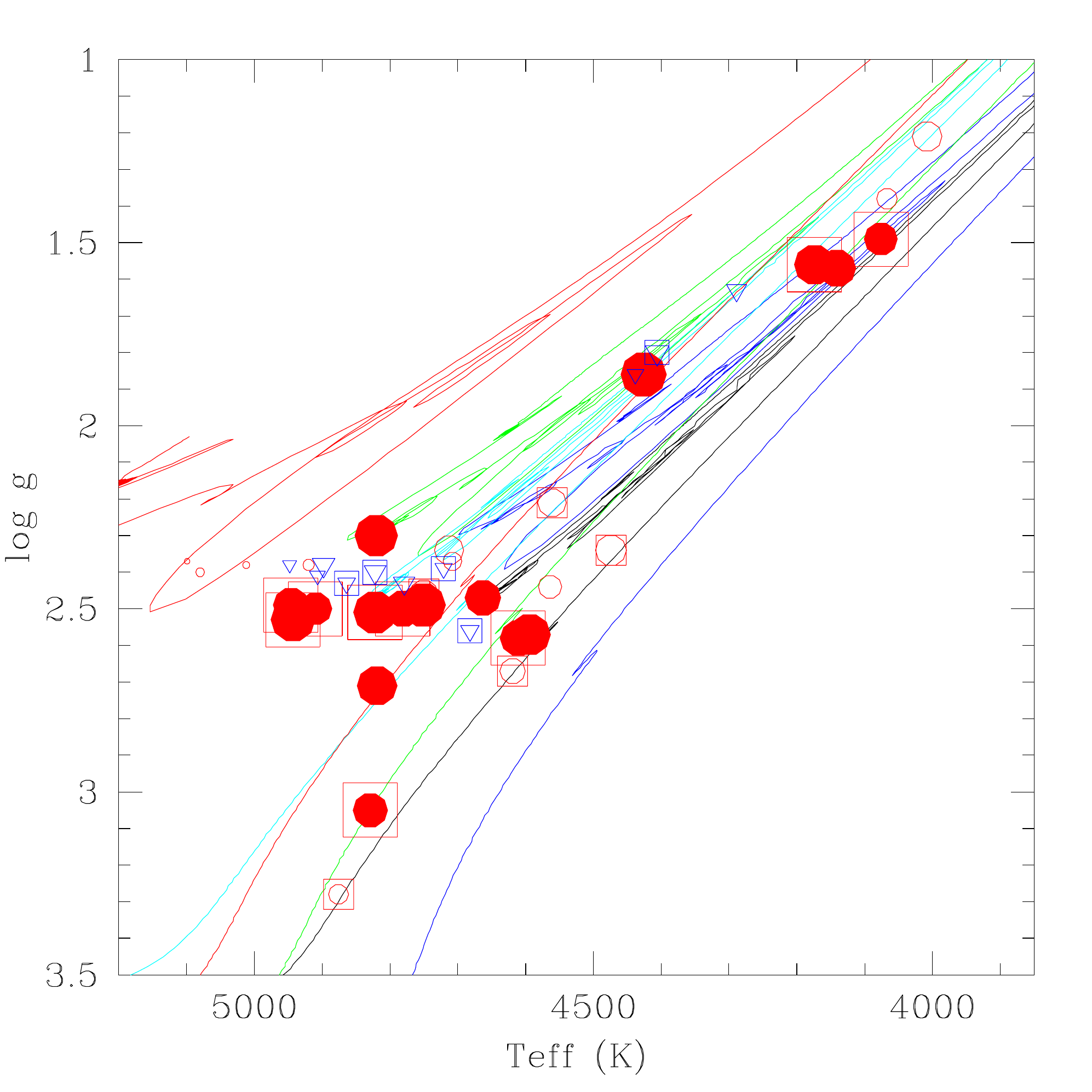}
   \caption{Kiel diagram $(T_{\rm eff}, \log g)$ for the sample stars. The symbol   {radius} is proportional to the mass. Evolutionary tracks from STAREVOL (see text) have been overplotted for stars with [Fe/H] = 0, and masses of 1~\Msun\  (blue curve), 1.5 \Msun\   (black curve). Also plotted are tracks for [Fe/H] = -0.25, 1~\Msun\  (green curve) and 1.5~\Msun\  (cyan curve), and [Fe/H] = -0.5, 0.9 \Msun\   (red curve). }
            \label{fig:Kiel}%
    \end{figure}

Figure~\ref{fig:Kiel} presents the Kiel diagram ($T_{\rm eff}, \log g$)  of the sample stars, alongside with STAREVOL stellar evolutionary tracks for different masses and metallicities \citep{Siess-2000,Siess-2006,Escorza-2017}, which are represented with different colors. The figure overplots the AR, YAR, and AP, showing how they are indeed at a variety of evolutionary stages.  The symbols follow our classification of YAR, AR, and AP using the red circles and blue triangles as before. The symbol   {radius} is proportional to the stellar mass and encapsulated are the binaries.   Most of the YAR stars fall as expected close to the black and cyan tracks corresponding to 1.5~\Msun\ stars with [Fe/H] = 0 or $-0.25$, respectively. On the other hand, 
{we note the presence of} some low-mass stars 
lying at the blue end of the horizontal branch. Their current mass (around 0.7~\Msun) and metallicity (around [Fe/H]~$= -0.5$) is compatible with the evolutionary track of a star of initial mass 0.9~\Msun\ and [Fe/H]~$= -0.5$, which reaches the horizontal branch within the age of the universe.

\begin{figure*}[t]
\centering
    \includegraphics[scale=0.5]{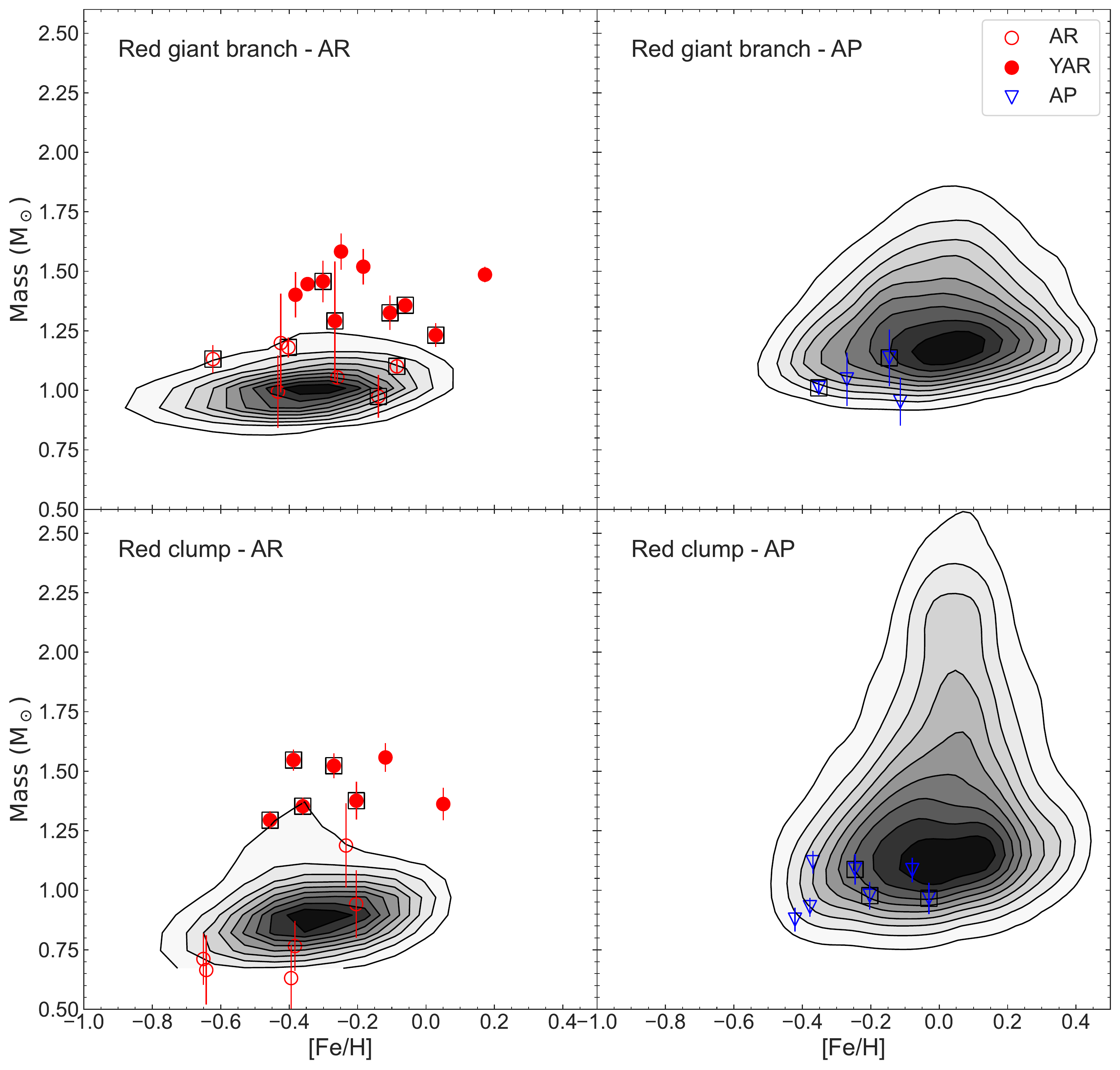}
\caption{Metallicities and masses of our stars alongside with the density distribution of APOKASC-3 for RGB stars  ({\it upper panel}) and RC stars ({\it lower panel}). The $\alpha-$rich stars are plotted in the left-hand panels, and the $\alpha-$poor stars are plotted in the right-hand panels. The chemical separation uses Eq.~\ref{eq:alpha2}. Contours indicate the percentages 10, 20, 30\%, and so on of the APOKASC-3 distributions. The outer contour corresponds to 90\% of the stars being inside the distribution.
}
\label{Fig:APO3mass}
\end{figure*}

In Fig.~\ref{Fig:APO3mass} we plot the metallicity and mass of our stars following the usual symbols for YAR, AR, and AP stars. In the background the overall density distributions of the entire APOKASC-3 catalogue are displayed. Thanks to the Kepler data, we know the evolutionary stage of the stars (as listed in the last column of Table~\ref{Tab:sample}), and separate them in RGB and RC stars in the upper and lower panels of Fig.~\ref{Fig:APO3mass}, respectively. Thanks to the APOGEE data, we can as well divide the stars of the background according to their [$\alpha$/Fe] ratio according to Eq.~\ref{eq:alpha2} (left-hand panels: AR stars;  right-hand panel: AP stars). 

The first interesting aspect to notice in Fig.~\ref{Fig:APO3mass}  is the difference in the overall APOKASC-3 distributions, especially at solar metallicities. Whilst it is rare to find RGB stars with masses above 1.75~M$_\odot$, the RC stars can have masses up to 
2.5~M$_\odot$.

The second interesting aspect is the difference in the mass distributions for stars in the same evolutionary stage but different $\alpha-$abundances. The contours show that $\alpha-$enhanced stars have overall a narrower mass distribution than stars with  solar $\alpha-$abundances. This is expected, since $\alpha-$enhanced stars in the solar neighbourhood are likely belonging to the thick disk, which has a narrow and old  age {distribution}, so its stars have preferentially lower masses \citep{masseron-2015, Miglio-2021}. The contour of AP stars, on the other hand, are likely belonging to the thin disk, which is forming stars for a long time until the present day, allowing for stars with a wider range of masses to exist today.  

The third interesting aspect to notice in Fig.~\ref{Fig:APO3mass} is that not all of our AR stars (open red circles) lie within the 80\% contour of APOKASC-3.  
Because the mass distributions in APOKASC-3 are different for RGB and RC stars, we see here how assigning a star the AR category with a simple cut in mass does not reflect clean samples of stars that are truly different from each other.   For example, the most metal-poor AR star (Y15), despite having a mass below the threshold of  1.3~M$_\odot$, is still too massive for the allowed mass distribution at that metallicity for RGB stars in APOKASC-3. Still, our selection process has selected all the outlying stars in the mass distribution of both RC and RGB for YAR stars, but has rejected some AR stars that are still overmassive.  Comparing the binary frequencies of YAR and AR stars might lead to uncertain conclusions about the nature of YAR and AR stars (see further discussion in the Appendix~\ref{Sect:binaries}).

Finally,  not accounting for a threshold in the lower mass range implies including AR stars whose masses are too low to be on the red giant branch today. Given their masses and age of the Universe, they should still be on the main sequence, but they have evolved to the red giant phase. While some stars might be on the horizontal branch (see Fig.~\ref{fig:Kiel}), some have even lower masses. Very recently, \cite{Li-2022} has reported the existence of such stars in APOKASC, and we discuss them more below and in the next sections.

\section{Kinematics and chemistry}
\label{sect:nature}

Despite the fact that our sample might still be too small to draw any strong conclusion about the binary frequencies between the groups defined in the previous sections (see Appendix~\ref{Sect:binaries}),  this sample has great potential to study in detail the evolutionary history of each star, given that spectroscopic data (time domain for radial-velocity variations and high-resolution high-signal-to-noise for abundance analyses), kinematic data (for inferring their Galactic dynamics) and seismologic data (for inferring their evolutionary stage) are available.  It is thus interesting to study the stars further, by putting them in the context of the entire APOKASC catalogue.

\subsection{New stellar classifications}\label{sect:new_groups}
Understanding that the evolutionary stage of a star plays a role in the mass threshold for it to be seen as overmassive, we group the stars differently. 
 We define bulk stars as those falling within the 80\% contour drawn from the full APOKASC sample in Fig.~\ref{Fig:APO3mass} (e.g, the second outer contour in the figure). 
 Those stars falling outside and above this region 
are defined as ``overmassive" while those that are outside and below that region 
are called ``undermassive". 

From Fig.~\ref{Fig:APO3mass}, we may conclude that all our initial classification of YAR stars are indeed overmassive, but this new category includes some AR stars as well (one in the RC and three along the RGB). This yields a total of 20 overmassive stars. In addition, two 
{red clump}
AR stars are undermassive.

\subsection{Kinematic and chemical abundance distributions}

In this section, we study if the overmassive and undermassive stars share the kinematical and chemical properties of standard stellar populations, namely thin and thick disk. 
 From here on we focus solely on the properties of the overmassive and undermassive stars and compare them with the entire APOKASC-3 as a control sample.
 
\begin{figure}[t]
   \centering
   \includegraphics[scale=0.28]{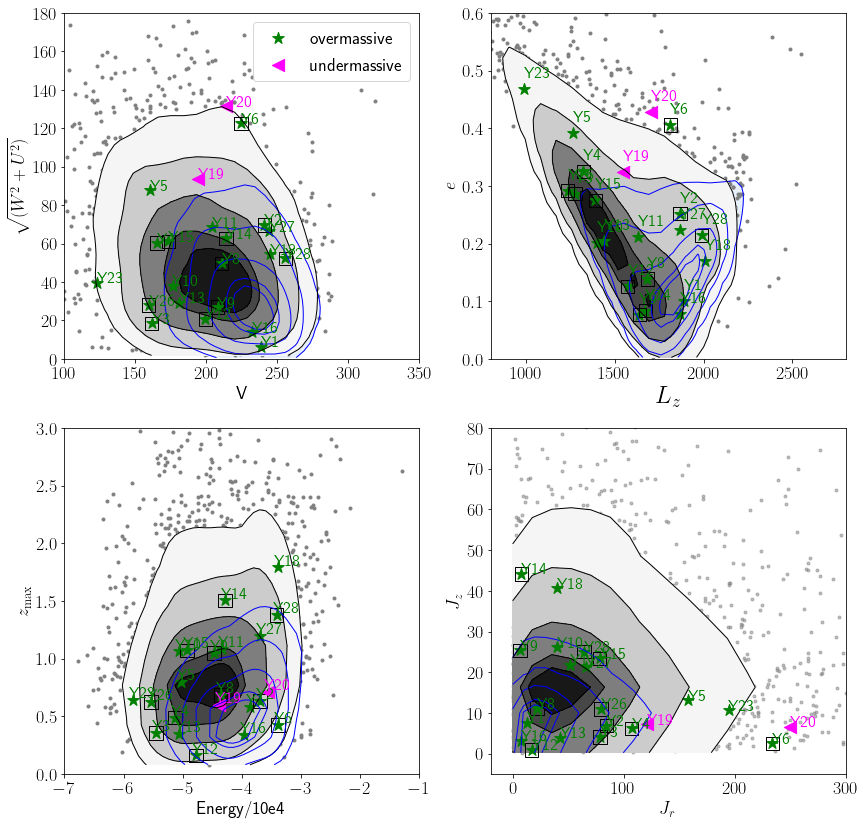}
   \caption{Dynamical properties of the overmassive and undermassive stars. In the background we display in gray the $\alpha-$rich population from APOKASC-3 as defined by Eq.~\ref{eq:alpha2} and in blue we display the contours of the $\alpha-$poor population, for reference. {Green
stars represent the overmassive stars, and magenta triangles the undermassive stars. Objects encapsulated in squares are the binaries.}}         \label{fig:toomre}%
    \end{figure}

\subsubsection{Kinematics}

{In this section, we compare the kinematical  properties of overmassive and undermassive stars (computed as described in Sect.~\ref{sect:data_kin}) with those  of the thick disk, following }
 \citet{Zhang-2021} and \citet{Warfield21}.  Different dynamical 
{diagnostics} are plotted in Fig.~\ref{fig:toomre}.  The top left panel shows the classical Toomre diagram.
The x-axis shows the rotational velocities ($V$) and the y-axis shows the combination of the velocity toward the Galactic center ($U$) and the velocity toward the Galactic poles ($W$).  The thin disk {stars} 
are usually considered
as {having} circular orbits in the Galactic plane, rotating with velocities consistent with the rotation of the Galaxy at $V \sim 220$~km/s \citep[e.g.][]{bensby03}. Hence, thin disk stars have small $U$ and $W$. This is indeed what 
{is observed for the blue contours in Fig.~\ref{fig:toomre}, which 
confirms the assignment of $\alpha$-poor stars }
to the thin disk. 

 Thick disk stars rotate slightly slower than the thin disk, that means they have normally $V < 220$ km/s \citep{bensby03, Soubiran2003}. Since they form part of a thicker disk, their velocities toward the Galactic poles are larger than those of the thin disk stars, making them able to reach higher distances from the Galactic plane. 
 In the Toomre diagram, thick disk stars are expected to be
 shifted toward lower values of $V$ with respect to thin disk stars, and to span a larger range of $U$ and $W$ compared to the thin disk stars.
 This is indeed the tendency shown by 
 the gray contours in Fig.~\ref{fig:toomre}. 
 We  therefore interpret $\alpha-$rich stars as belonging to the thick disk.  We do not see a notable difference 
 between 
 the over/undermassive stars.

\begin{figure*}[t]
   \centering
   \includegraphics[scale=0.35]{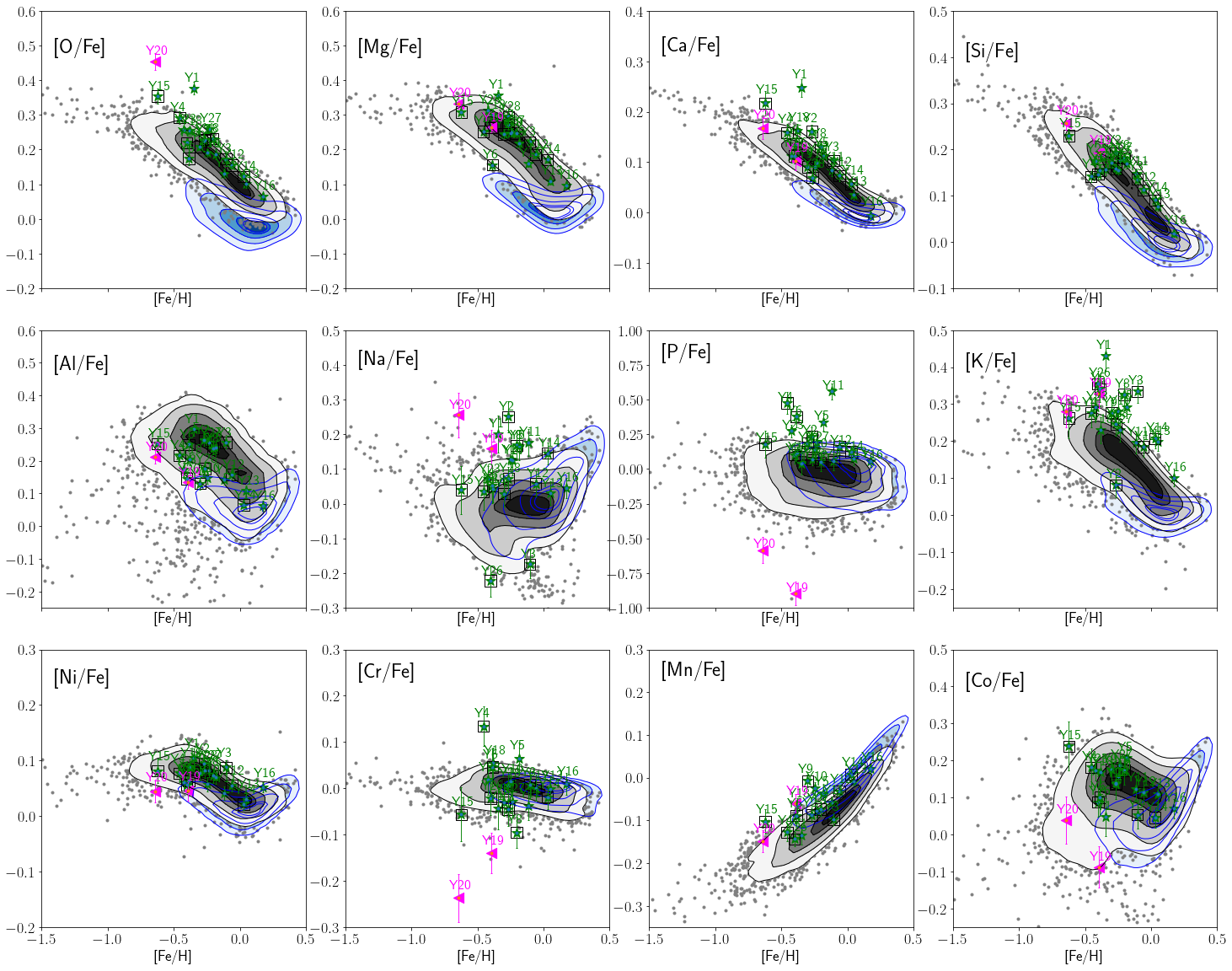}
   \caption{Abundance ratios for APOKASC-3 stars (contours and gray dots) and for our sample stars. Symbols and colors are as in Fig.~\ref{fig:toomre}. }
            \label{fig:abus}%
    \end{figure*}
    
To dig  into possible differences between over- and undermassive stars and standard thick disk stars further, we study the distributions of the angular momentum $L_z$ and eccentricities $e$ of the stellar orbits (top right panel of Fig.~\ref{fig:toomre}). The thick-disk stars tend to have lower angular momenta with respect to thin-disk stars, because of their lower $V$. Their orbits span a larger range in eccentricities than the thin disk. 
The rest of the stars follow the dynamical behavior of the  disk, except perhaps Y20 and Y6.  

In the lower left panel of Fig.~\ref{fig:toomre} we plot the energy and the maximum heights of the orbits. We see how our stars are consistent with disk stars.  
The lower right panel of Fig.~\ref{fig:toomre} shows the radial and vertical actions, denoted as $J_r$ and $J_z$, respectively. We see how the thin-disk stars have small actions overall, but the thick disk stars span a wider range.  Our stars fall well within the contours of the thin and thick disks. 
Y20 and Y6  have relatively higher radial actions compared to the disk. 

We conclude that most of our stars have normal thick disk kinematics. The overmassive stars are in general consistent with thick disk kinematics. This is consistent with previous kinematic studies of young $\alpha-$rich stars \citep[e.g.][]{Zhang-2021}. undermassive stars also have kinematics that are consistent with the thick disk.  We do not find a notable difference among binaries and nonbinaries in their overall kinematics. 

\subsubsection{Chemical abundances}

In Fig.~\ref{fig:abus} we show the same populations as in Fig.~\ref{fig:toomre} but now the metallicity is plotted against different abundance ratios. {This is customary to evaluate whether stars have been chemically enriched by a common nucleosynthetic channel and can be attributed to a common stellar population  \citep[e.g.][]{hawkins2015, jofre2019, buder19, kobayashi2020}.} We consider the abundances reported in APOGEE DR16 (see Sect.~\ref{sect:data}). The idea here is to see from the chemical point of view whether the overmassive and undermassive stars belong to the {stellar populations of the thick and thin }disks as inferred from their kinematics or whether they present instead some anomalies in their chemical imprint. The contours and symbols follow Fig.~\ref{fig:toomre}, namely the gray contours represent the $\alpha-$rich (thick disk) population and the blue contours the  $\alpha-$poor (thin disk) population. Green stars represent the overmassive stars, and magenta triangles the undermassive stars. 

When focusing on the panels with $\alpha-$capture elements (O, Mg, Ca, and Si), we see that the overmassive stars follow the  thick disk. We note however that Y1 and Y15 are more O-rich and Ca-rich than the  thick disk.  As the undermassive stars are concerned, we note that their abundances tend to be slightly lower than the thick disk, except for O where Y20 has high O whereas Y19 has no reported O abundance.

\begin{figure*}[t]
   \centering
     \includegraphics[scale=0.40]{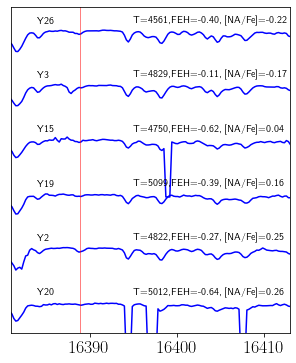}
       \includegraphics[scale=0.4]{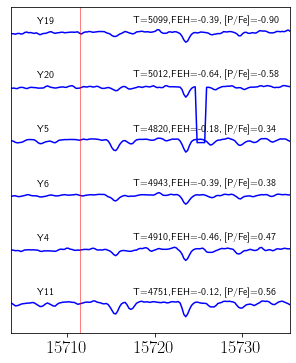}
         \includegraphics[scale=0.4]{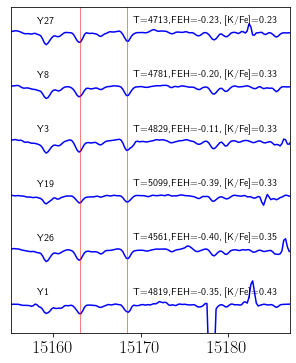}
           \includegraphics[scale=0.4]{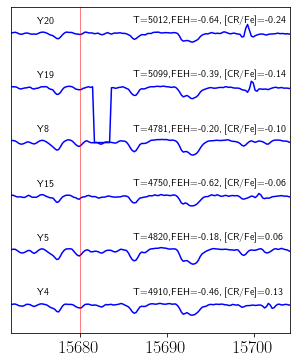}
   \caption{Line profiles of Na~I ($\lambda$ 16388.9 \AA), P~I ($\lambda$ 15711.5 \AA), K~I ($\lambda$ 15163.1 and 15168.4 \AA) and Cr~I ($\lambda$ 15680.1 \AA) 
   for a selection of stars sorted by decreasing abundance. The line under consideration is indicated with a vertical red line. For each star, parameters are given for reference, including the abundance of the considered element.} \label{fig:profiles}%
    \end{figure*}

Aluminum, sodium, phosphorus, and potassium also trace different star-formation environments. Like the $\alpha-$capture elements, they are produced in massive stars and released to the interstellar medium (ISM) through type II supernova, but their yields depend on metallicity. Our overmassive and undermassive stars follow the distributions of the thick disk for Al, although for Na there is more scatter and some stars fall off the overall disk distribution. We note that the sodium abundance is derived from only two weak lines in APOGEE spectra, and therefore is not derived with particularly high precision \citep{apogee-dr16}.  In the left panel of Fig.~\ref{fig:profiles} we plot the spectra around the Na~I $\lambda 16388.9$~\AA\  line, which is indicated by the red vertical mark.  We show six stars, which are sorted by increasing 
Na abundance. The temperature and metallicity of the stars are indicated alongside with the [Na/Fe] abundance, showing how these parameters have an effect in the shape of the Na line. In all cases, however, the Na line is very weak, even for the [Na/Fe]-enhanced cases, which means that these abundances should be
taken with care. Sodium can furthermore be altered during the evolution of stars, slightly increasing in the atmosphere of red giants through dredge-up mechanisms, by an amount that might correlate with mass \citep[see][and discussion therein]{Smiljanic09}. 

Other interesting abundances are those of P and K. Although P is one of the most uncertain elements in APOGEE \citep{apogee-dr16}, because the lines are weak and can suffer contamination from telluric lines \citep[see also][]{Hawkins-2016}, we have a number of overmassive and undermassive stars that are clearly outside the disk populations.  Potassium, on the other hand, is an element whose abundance is measured from strong lines which results in a precision that is comparable to other elements \citep{jonsson-dr12-14, apogee-dr16}.  
We can see that our overmassive and undermassive stars have a tendency not to follow the distribution of the thin and the thick disk, with overmassive stars being systematically P and K rich while the undermassive stars are P-poor. 

The middle panels of Fig.~\ref{fig:profiles} show the line profile of P and K for stars covering the wide range in abundances as seen in the panels of Fig.~\ref{fig:abus}. 
The reason why 
the P abundances are so uncertain appears clearly from these spectra: it is almost impossible to detect the $\lambda15711.5~$\AA\ line 
and it is very hard to tell if the line 
is not 
blended with Fe \citep[see][]{jonsson-dr12-14, Hawkins-2016}. Indeed, P abundances are not reported in DR17, which means that these measurements, particularly those for Y19 and Y20 which are very low, should be taken with care. Determining upper limits might be more suitable in this case. The profiles of the K lines at 15163.1 and 15168.4\AA, on the other hand, show that the measurements of these abundances 
should rely on more solid grounds. 

\begin{figure*}[t]
   \centering
   \includegraphics[scale=0.35]{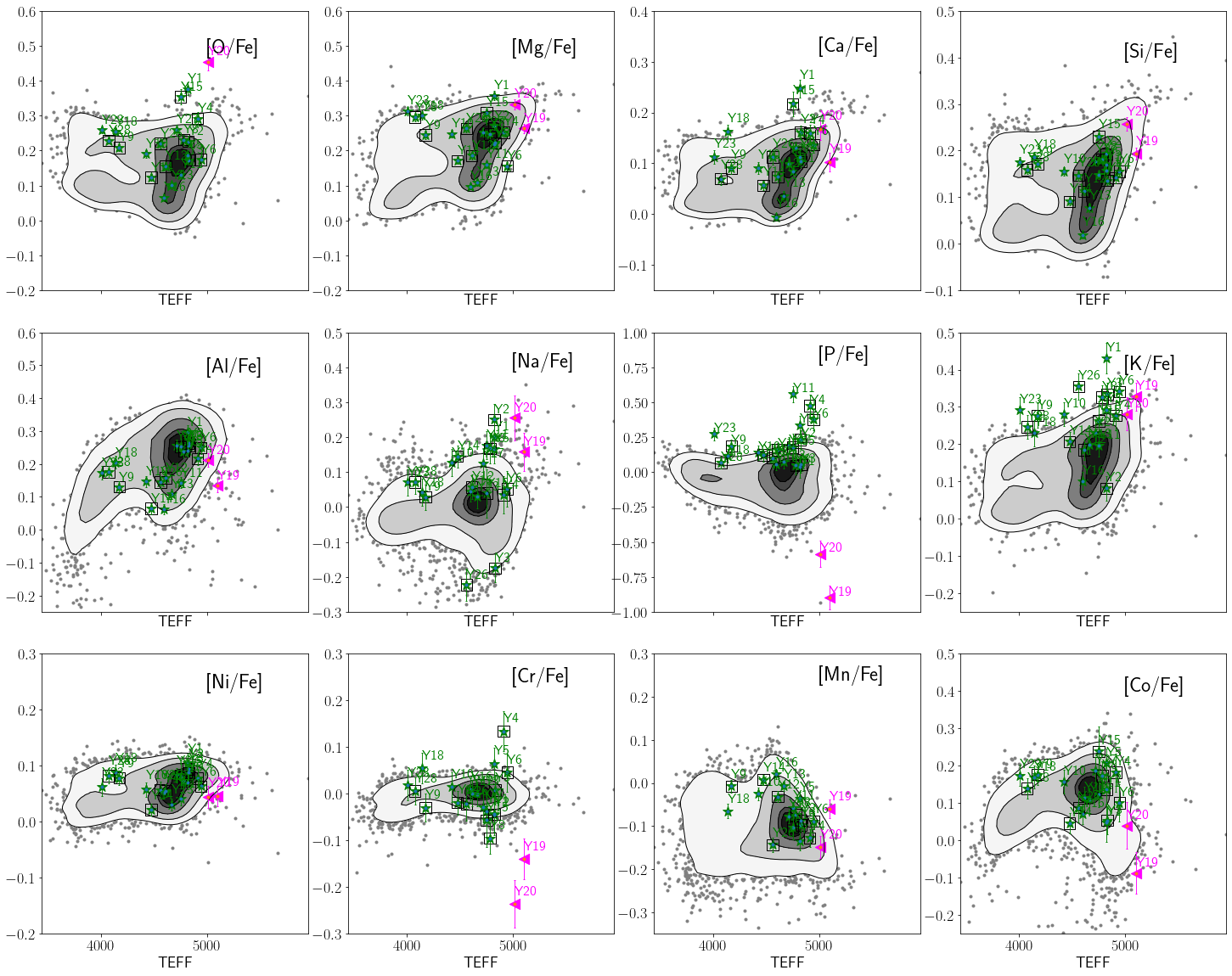}
   \caption{Abundance ratios for APOKASC-3 in contours and our sample. Symbols and colors follow the same definition as in Fig.~\ref{fig:toomre}. }
            \label{fig:abusteff}%
    \end{figure*}

Nickel, chromium, manganese, and copper are  tracers of star-formation histories through their SN Ia production, and our stars follow the trends of the disk. However, the undermassive stars have lower [Ni/Fe] and [Cr/Fe] ratios than the disk populations. There are some overmassive stars with low Cr abundances as well. The right-hand panel of Fig.~\ref{fig:profiles} shows the Cr~I line at 15680.1\AA, which was the only Cr line used by \cite{Hawkins-2016} to determine the Cr abundance. We can see how the line becomes weak for low Cr abundances, therefore we 
conclude that the low Cr abundance reported for some stars is likely real. {In fact, we show in Fig.~\ref{fig:abusteff} below that it cannot be attributed to a temperature effect.} 
We note however that this Cr line is blended \citep[see Fig.~10 of ][]{Hawkins-2016}.

We finally notice that for some but not all abundance ratios there is a systematic difference between the over- and undermassive stars. The undermassive stars have systematically lower abundances for Ca, Al, P, K, Ni, Cr, and Co.  It is possible that this  is due to a systematic effect related to stellar parameters, since undermassive stars are systematically hotter than 
 overmassive
stars, undermassive stars being 
likely blue horizontal-branch stars (see Fig.~\ref{fig:Kiel}). In Fig.~\ref{fig:abusteff} we show the same panels as in Fig.~\ref{fig:abus} but as a function of effective temperature $T_{\mathrm{eff}}$. The contours show that the abundances of some elements {are correlated to} 
$T_{\mathrm{eff}}$. This might account for the systematic difference for some overmassive and undermassive stars in Ca, Al, K, Ni, and Co, but does not 
explain the difference in P and Cr, 
since a few over- and undermassive stars do not follow the abundance trend with temperature and clearly stand out.

\subsubsection{Conclusion about stellar populations}

From Figs.~\ref{fig:toomre} and \ref{fig:abus} we might deduce that our overmassive and undermassive stars seem to belong mainly to the thick disk. 
We do not see a significant difference between binaries and constant stars in their overall chemical or kinematic patterns.

We detect a systematic difference between overmassive and undermassive stars in several abundance ratios such as  P, Ni, and Cr.  Some such differences can be attributed to a systematic effect,  a detection limit or blends. We can however not exclude the possibility that these differences 
are caused by an alteration in the atmospheres of stars that might contain interesting information about their nature. 
{A few stars stand out as far as} 
Na, P, K, Cr 
are concerned, having systematically higher or lower abundances than the rest of the stars. 
We stress that after the visual inspection the spectra around the best Na and P lines for these stars in APOGEE, they can still be too weak or blended to measure any abundance with confidence.
In fact, these abundances are flagged as uncertain in APOGEE DR7. To confirm the anomalies of these abundances, other wavelength domains or higher resolution than APOGEE would be needed.

\begin{figure*}[t]
   \centering
     \includegraphics[scale=0.35]{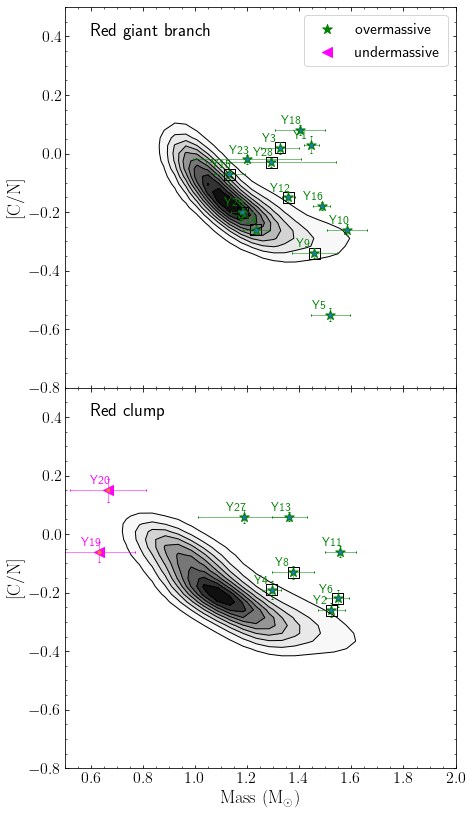}
        \includegraphics[scale=0.35]{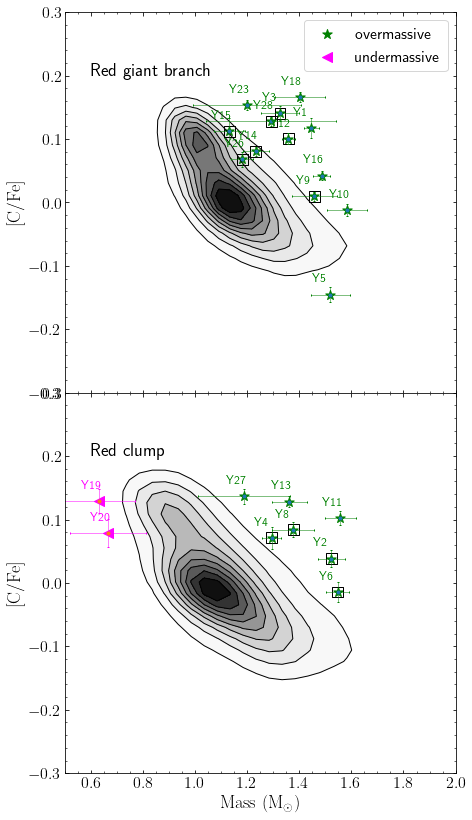}
           \includegraphics[scale=0.35]{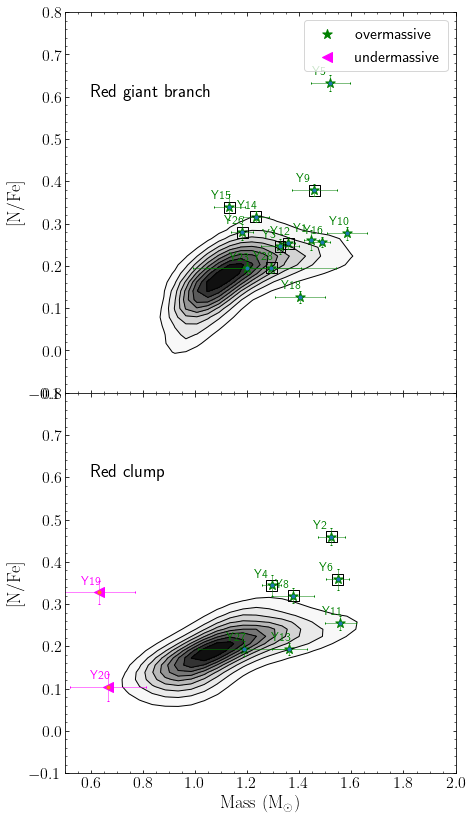}
   \caption{Left: [C/N]-mass relation for the stars in our sample alongside with the $\alpha-$rich APOKASC-3 distribution. The upper panel shows the stars classified as RGB and the lower panel shows the RC stars. The symbols follow our new classification based on the contours of Fig.~\ref{Fig:APO3mass}.  Binary stars are enclosed with squares. Middle and right panels: [C/Fe] and [N/Fe] abundances as a function of mass.}
            \label{fig:cnmass}%
    \end{figure*}

\begin{figure}[t]
   \centering
        \includegraphics[scale=0.25]{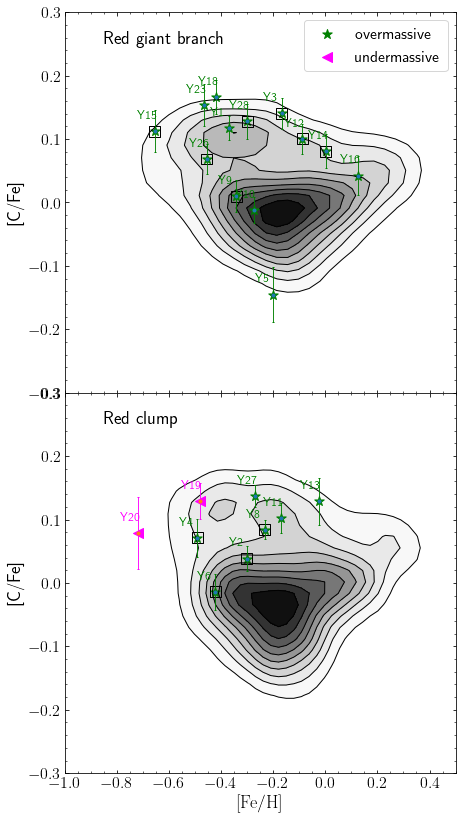}
   \includegraphics[scale=0.25]{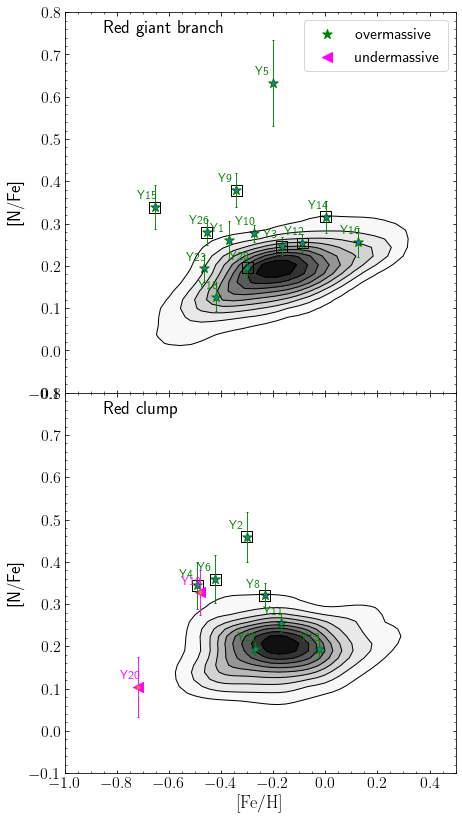}
   \caption{{Same as Fig.~\ref{fig:cnmass}, but for [C/Fe]  and [N/Fe] as a function of metallicity.}}
            \label{fig:c_n_mass}%
    \end{figure}

\subsection{The role of carbon and nitrogen}

Carbon and nitrogen abundances in red giants have extensively been studied as signatures  
{of specific nucleosynthesis processes, and therefore of specific evolutionary phases}.
Since  
  {red giants with masses larger than $\sim1$~M$_\odot$ have exhausted hydrogen  through the CNO cycle operating in their cores during the main sequence}, the abundances of C, N, and O have changed in their interiors. Once red giants experience the first dredge-up, {CNO-processed layers are} brought to the surface, altering the photospheric abundances \citep{iben1967, Smiljanic09}. 

In particular, after the first dredge-up the relative abundance ratio [C/N]  changes by an amount that depends on the stellar mass  \citep{iben1967}. This implies that it is possible to use [C/N] abundances of large samples of giant stars to infer their mass distribution.  \cite{masseron-2015} showed how this fundamental effect could be used to put constraints on the evolution of the thin and thick disk using [C/N] of red giants from APOGEE. That paper opened up a new way to estimate masses and ages in red giants using spectra \citep{martig-2016, ness2016}, and to perform studies of Galaxy evolution when ages are not known with accuracy \citep{hasselquist-2019, jofre-2021}.  However, the exact relation
{between} age and [C/N]  {also depends on} the stellar chemical composition, evolutionary stage and other stellar-evolution processes that are not well constrained yet
\citep{Salaris-2015, lagarde-2019, shetrone2019}.

{In particular, the} stars of our sample 
which do not follow the expected [C/N]-age relation 
might {have experienced mass-transfer}. 
Because of extra-mixing mechanisms, a scatter 
{is admittedly expected in the [C/N]-age relation }
\citep{martig-2016, lagarde-2019} but nonstandard mixing due to mass transfer 
is not considered in such studies. Outliers of the relation, therefore, can help us to identify post-mass-transfer objects.

In the left panels of Fig.~\ref{fig:cnmass} we plot the [C/N] abundances as derived by APOGEE as a function of the mass as derived by APOKASC-3 for stars along the RGB and at the RC.
For reference we also plot the contours of the thick-disk abundances. 
The plots show how badly the majority of the over- and undermassive stars do not follow the disk population, which supports the idea that these stars had an evolution with binary interaction.

The middle and right panels of Fig.~\ref{fig:cnmass} show the [C/Fe] and [N/Fe] abundances as a function of mass.  Overmassive stars have [C/Fe] abundances that are systematically higher than the thick disk. Regarding the [N/Fe] abundances, while some stars have abundances that agree well with the thick disk, a considerable amount is N-rich. In particular,   {Y5} and Y19 have a [N/Fe] abundance that is  very high given their mass.  

In the left-hand panels of Fig.~\ref{fig:c_n_mass} we plot the carbon abundances as a function of metallicity, and our stars lie inside the disk populations although at the higher end. Finally, the right-hand panels show the nitrogen abundances. Here a large number of overmassive and undermassive stars fall off the disk population, pointing toward a different process happening inside the stars.  

There is extensive literature studying the production of nitrogen in stars \citep{Palacios-2016}. In addition to play a role in the CNO cycle, N can be produced as part of the nucleosynthesis happening in AGB stars. It is also a signature of extra mixing because of rotation.  It is therefore interesting that so many of our stars have enhanced nitrogen abundances.

\section{Discussion}
\label{sect:discussion}

In this section we discuss in more detail the nature of the overmassive and undermassive stars given the results obtained above about their chemical abundances, kinematics, masses and binarity. We 
{discuss separately}
overmassive and undermassive stars, 
as summarized in Table~\ref{Tab:starsbyclass_omum}.

\begin{table*}[t]
\caption{\label{Tab:starsbyclass_omum}
Stars classified according to Eq.~\ref{eq:alpha2} for their chemistry and according to the contours of the mass distributions of   {Fig.~\ref{Fig:APO3mass}}.}
   \centering
    \begin{tabular}{|c|c|l|}
    \hline
    class & $N$ & stars \\
    \hline
   overmassive & 20 & Y1, Y2, Y3, Y4, 
Y5, Y6,   Y8, Y9,  Y10, Y11, Y12, Y13, Y14, Y15, Y16, Y18, Y23, Y26, Y27, Y28 \\
    undermassive  & 2 & Y19, Y20 \\
    \hline
    \end{tabular}
    \end{table*}

\subsection{Overmassive stars}
These stars correspond to Y1 (90), Y2 (90), Y3 (90), Y4 (90), 
Y5 (90), Y6 (90),   Y8 (90), Y9 (90),  Y10 (90), Y11 (90), Y12 (90), Y13 (90), Y14 (90), Y15 (90), Y16 (90), Y18 (90), Y23 (80), Y26 (90), Y27 (80), Y28 (90). Numbers between parentheses indicate the contour level of Fig.~\ref{Fig:APO3mass}, which was used to select them.

One very encouraging result from our analysis is the [C/N]-mass relation plotted in Fig.~\ref{fig:cnmass}, because of its striking similarity with the Fig.~5 of \cite{Izzard-2018}, which was made using population synthesis models which included binaries. Firstly, {the models of} \cite{Izzard-2018} 
{show} that many overmassive stars do not follow the decreasing trend of [C/N] with mass; 
{we reach similar conclusions based on observations}. 
Secondly,  their study followed the evolution of the binaries by considering mass transfer and merging events. In particular, they investigated the properties of the binaries after 8 Gyr of evolution (the age of the thick disk) in the stellar population and colored the [C/N]-mass relation according to the final binary frequency. While these exact frequencies were dependent on several parameters of the population synthesis models  (e.g. separation, eccentricities, and mass ratio of the initial binaries, fraction of initial binaries with respect to initial single stars in the model, etc), \cite{Izzard-2018} could conclude that after 8~Gyr of evolution, stars with end masses above about 1.4 M$_\odot$ are products of mergers and {therefore are single at the end of the simulation}.  Y1, Y16, Y10, Y5, and Y11 have masses above that threshold, and are found to be single in our analysis, {supporting the idea that} they are merger products. \cite{Izzard-2018} further found that stars with end masses between about 1.2 and 1.4 M$_\odot$ are products of mass transfer which remain as binaries. Y18, Y3, Y28, Y23, Y12, and Y8 fall in that mass range,  and are binaries.

\cite{Izzard-2018} did not distinguish between RC and RGB stars in their analysis, but we have shown in Fig.~\ref{Fig:APO3mass} that the evolutionary stage might be important for defining a mass threshold between products of mass transfer via Roche-Lobe overflow or via mergers. We note that Y13 and Y27 have high levels of [C/N] compared to the bulk of RC stars, and that in Fig.~5 of \cite{Izzard-2018} a band of single stars for [C/N] around zero is present. The   {single-star} nature of Y13 and Y27 in that sense may be accounted for.   {Y2 and Y6 are on the other hand binaries and have high masses. We have seen how the mass threshold between normal and overmassive stars needs to consider the evolutionary stage; therefore the comparison between our results and those of \cite{Izzard-2018} is not straightforward. }

It is interesting that many of the stars fall well within the [C/N]-{mass} relation (see Fig.~\ref{fig:cnmass}), suggesting they are "truly young" \citep{Hekker-2019}. These are Y4, Y9, Y14, Y15, and Y25. We find that they are  binaries, and that  Y4 and Y15 have a  relative enhancement of [N/Fe], which is an indicator of either extra-mixing, pollution caused by transfer of material synthetized by an AGB star,  or rotation \citep{Palacios-2016}.  Most of the other abundances are consistent with classical thick disk population.  In particular, there are  several overmassive stars with  possible enhancements in [P/Fe].  
Only Y25 does not show evidence of  abundances behaving differently from the thick disk.  

Furthermore, there are several overmassive stars with relatively low Cr abundances and both undermassive stars with very low Cr abundances. 
We do not have an explanation for such a low Cr abundance which seems to affect mostly the undermassive stars and not other giant stars of APOKASC. 

While we have concluded that the binary frequency in the group of overmassive stars is not significantly different from that of the total group of bulk stars  (see  Table~\ref{Tab:SBfrequency2}), it is worth discussing the binary frequency a bit further. The  result of Table~\ref{Tab:SBfrequency2} is obtained by considering all overmassive stars together, namely by considering the entire mass range and not separating the sample by evolutionary stages. 
For YAR stars however,  the binary frequency appears to vary as a function of mass when separating the stars according to their evolutionary stage (Sect.~\ref{Sect:binarity-vs-stellar_properties}). In particular, among the red giant stars, the most massive ones appeared to be single. This might be explained by making a difference between a merger (in which the total mass of the new star is equivalent to the sum of masses of the two individual stars before the merger) and mass transfer via Roche-Lobe overflow (in which the total mass of the new star only increases by a fraction of the primary mass and its final mass is thus smaller than the total mass of the individual stars.) 

\subsubsection{Individual cases}

\noindent {Y1: KIC 9821622.} This is a RGB star, with no signs of binarity after 13 HERMES observations spanning 1080 days. The average of the HERMES RVs agrees with the sole RV APOGEE measurement.
Its enhanced $\alpha$ abundances suggest this star belongs to the {thick disk}. Y1 was classified as binary in Paper I but that was because of a bad measurement in one of the initial runs due to the high crowding in the Kepler field. This star has a  slight enhancement of O and Ca with respect to other stars, and a significant enhancement in K. [C/N] is also significantly enhanced.  \cite{jofree-2015} and \cite{Yong-2016} performed a detailed chemical analysis using high-resolution optical spectra and this star was found to be Li-rich. According to these studies, neutron-capture elements such as Ba, La, Y, and Eu are not enhanced in this star. \cite{Yong-2016}, however, found a mild IR excess in Y1. The anomalies in the different chemical abundances point toward nonstandard evolution, in which this star merged with a companion, 
producing a single star. This might have happened some time ago because its rotational velocity is rather slow \citep[$v\sin i$ of 1.0 km/s, ][] {jofree-2015}. {It is worth noting that its dynamics might  be consistent with the thick or the thin disk, since it lies in an overlapping region in the dynamical planes.}\\

\noindent {Y2: KIC 4143460.} {With new RV measurements and the combination of different RV sources we were able to conclude that this star is a binary.}
In Paper I it was not identified as binary, and the APOGEE or Gaia measurements alone also suggest it is single. Here, thanks to 
7 HERMES RV measurements, which cover a time span of 1382 days, it is possible to detect binarity. Y2 does not stand out in kinematics, nor in most of the chemical abundances, except Na and N. Indeed, this is the most Na and N enhanced overmassive star of the sample. The enhancement of N could point toward pollution from an AGB star. However,  \cite{Yong-2016}  included Y2 in their analysis, finding normal Ba, Y, and other neutron-capture elemental abundances, but a mild IR excess. This star could have gained mass through mass transfer. To characterize the orbit of this star and unveil the nature of the companion, as well as constrain further the mass transfer mechanisms, we still need more RV measurements.  \\

\noindent {Y3: KIC 435050.} This star is a long-period binary, since
the ten HERMES measurements spanning 2200 days reveal a clear drift (Fig.~\ref{Fig:binaries}). 
Y3 is depleted in [Na/Fe], slightly depleted in [Cr/Fe] and enhanced in [K/Fe].  This star has been included in the analysis of \cite{Yong-2016}, also reporting a slight depletion in Cr of $\mathrm{[Cr/Fe]} = -0.1$ but \cite{Matsuno-2018} find solar abundances for Cr. The authors also find enhancement in K. \cite{Yong-2016} also found a slight IR excess in this star.  While Y3 shows normal C and N abundances (although C is at the top of the [C/Fe] distribution), its [C/N] ratio is peculiar, being too large given its mass. 
We conclude that Y3  {experienced} mass transfer. \\

\noindent {Y4: KIC 11394905.} This is a red clump star that was quickly identified as binary after only three RV measurements spanning 650 days. 
\cite{Yong-2016} also 
suspected 
its binary nature, but did not detect any IR excess.
  {Gaia DR3 provides a spectroscopic orbit for this binary.} 
 Its dynamical properties are normal. The elements that stand out are P, C, N, and Cr which is why we dedicate some discussion to Y4. From the visual inspection of the line profiles of P in Fig.~\ref{fig:profiles}, {the [P/Fe] abundance might rather be considered as an upper limit}. \cite{Matsuno-2018} determined a very enhanced K abundance of about 1 dex from optical spectra, but in APOGEE Y4 seems to have normal [K/Fe].\\

\noindent {Y5: KIC 9269081.}  
This RGB star 
does not show up as a binary after eight HERMES RV measurements spanning  1382 days and consistent with the five APOGEE RV measurements and with Gaia DR2. Y5 is the most [N/Fe]-enhanced star and most [C/Fe]-depleted star of our sample. This implies a very low [C/N] abundance ratio.  It could be a low-mass weak G-band star candidate \citep{Palacios-2016}. These stars are thought to be created by \textit{ab initio} pollution with material processed through the CNO cycle. These stars have low carbon  but high nitrogen abundances due to extreme pollution and mixing. These stars tend to be Li-rich, but \cite{Matsuno-2018} did not report any Li enhancement in Y5. It however does not show 
significantly
different {abundances from} 
 the disk population, except for phosphorus but the corresponding spectral line, if existing, is very weak (see Fig.~\ref{fig:profiles}). \cite{Matsuno-2018} comments however on the enhancement of Na and K, but on Fig.~\ref{fig:abus}, they do not clearly stand out. We note that \cite{Hawkins-2016} also find a high Na abundance for this star. Since most of these weak G-band stars are not in binary systems,  the nature of the polluter progenitors is poorly constrained. Y5 has relatively hot kinematics 
which is very consistent with the thick disk.  \\

\noindent {Y6: KIC 11823838.} This is a red clump binary star with enough RV measurements to characterize its orbit (see Appendix~\ref{Sect:orbits}). With ten RV measurements it is possible to estimate a
 816-day period. 
 This star presents enhanced [P/Fe] and [N/Fe] ratios, as well as a [C/N] ratio on the verge of the expected trend for normal RC stars. \cite{Matsuno-2018} found a K enhancement of 0.8 dex but the K APOGEE abundance is normal. \cite{Matsuno-2018} did not find enhancement of other neutron-capture elements.  Y6 is the star with the lowest [C/Fe] ratio in the sample, and shares similar properties to Y5, making it another candidate for a low-mass weak G-band star candidate, but Y6 is a binary. Its kinematics is among the hottest of our sample, but still consistent with the thick disk.  \\

\noindent {Y8: KIC 10525475}. This is a red clump binary that was identified from  seven HERMES RV  measurements spanning 1300 days. These measurements are not numerous enough to characterise the orbit. Y8 has low [Na/Fe] and [Cr/Fe] abundances,  and a slightly enhanced [K/Fe] ratio (although this might be a temperature effect). \cite{Matsuno-2018} finds moderate enhancement of K but solar values for Cr.   C and N are normal in APOGEE for this star, but its [C/N] ratio does not obey the [C/N]-mass relation for red clump stars. \\

\noindent {Y9: KIC 9002884}. This is a binary red giant star identified from eight HERMES RV measurements spanning 1300 days; the mean HERMES  and APOGEE RVs disagree. A circular orbit of period 594~d has been derived (Table~\ref{Tab:orbital_elements}). While its [C/N] abundance ratio follows the expected trend with mass, its [N/Fe] abundance is significantly  enhanced. \\

\noindent {Y11: KIC 11445818}. This is a red clump star that does not seem to be binary, as concluded from 22 RV HERMES measurements spanning 2200 days, as well as consistent HERMES, APOGEE, and Gaia RVs. Unlike Y5, Y11 has normal C and N abundances compared to stars of the same metallicity,  but its C abundance is very high compared to other stars of the same mass. Its [C/N]  ratio does not follow the [C/N]-mass relation for red clump stars. Y11 is the most [P/Fe]-enhanced star in our sample. Its line is however still very weak (Fig.~\ref{fig:profiles}),  and could be blended with another metallic line since Y11 is relatively metal-rich and cool compared to the other stars in our sample; {hence the APOGEE P abundance could be an upper limit}. All the other abundances and kinematics agree well with the disk populations. The abundances derived by \cite{Matsuno-2018} from optical spectroscopy  are normal as well.  This star could be a good candidate for a truly young $\alpha-$rich star if its [C/N]-age relation followed the bulk of APOKASC-3, but that is not the case.  \\

\noindent {Y15: KIC 11753104}. This RGB star was identified as binary from seven HERMES RV measurements spanning 1600 days. Its [C/N] abundance ratio is normal, but its [N/Fe] ratio is enhanced, as well as Ca and O. Its other abundances are normal, except for Cr which is a bit low compared to the disk stars.\\

\noindent {Y26: KIC 3662233}. This red giant branch star was found to be a binary from seven HERMES RV measurements spanning 1600 days. Its abundances are all normal, 
except for Na which is depleted. The Na line is however very weak in Y26 (see Fig.~\ref{fig:profiles}).

\subsection{Undermassive stars}

These are Y19 (90\% contour in Fig.~\ref{Fig:APO3mass}) and Y20 (90\% contour). {Both have been quoted as undermassive by \cite{Yu-2018} and \citet{Li-2022}, therefore their low masses were not subjected to a  systematic analysis in APOKASC-3.}  
Both stars have a mass of 0.6~$\mathrm{M}_\odot$ ($\pm 0.14$ and 0.15~$\mathrm{M}_\odot$ respectively; Table~\ref{Tab:sample}), and although it is possible that they are low-mass horizontal branch stars (see Fig.~\ref{fig:Kiel}), some of their abundance ratios and 
{their extremely low mass}
make them deserve further discussion.  {We wish to clarify that the quoted mass uncertainties encapsulate both systematic and random sources of uncertainty, and that the systematic uncertainties are not negligible in this regime. As mentioned above, both stars have been flagged as low mass by \citet{Yu-2018} and \citet{Li-2022}, as well as by up to seven different analysis pipelines within APOKASC, but the exact scale of asteroseismic masses has been disputed \citep{Gaulme-2016}. This scale is particularly challenging to derive for low-metallicity stars \citep[e.g.][]{Epstein2014} and core helium-burning stars \citep{An-2019} where the asteroseismic correction factors are important \citep[e.g.][]{Sharma-2016}. Given that situation, and the fact that our Galaxy's evolutionary history predicts the majority of $\alpha$-rich clump stars to be uniformly old, we expect the mass dispersion at a given metallicity to be a better tracer of the measurement uncertainties in this regime, and to offer a better way  to determine whether stars are offset from the bulk of the population. Among the main population of $\alpha$-rich clump stars (see Fig.~\ref{Fig:APO3mass}), the average mass seems to be metallicity-dependent, but its (almost uniform) spread 
is only about 0.05~$\mathrm{M}_\odot$, which likely represents the majority of the random measurement uncertainties in this regime. Given that 
situation, Y20 and particularly Y19 are not expected to have a significantly lower mass than other $\alpha$-rich clump stars of similar metallicity (see more details in Appendix~\ref{app:lowmass}).}

undermassive stars offer an interesting opportunity to study the poorly understood process of mass loss in red giant stars. Figure~\ref{fig:Kiel} has shown that some models are consistent with the location of Y19 and Y20 on the blue edge of the red clump/horizontal branch. 
These models have an initial mass of  $0.9~\mathrm{M}_\odot$ and a metallicity of -0.5 and reach the blue horizontal branch within the age of the universe  when a RGB mass loss of at least $0.2~\mathrm{M}_\odot$ is considered. However, our stars have masses below $0.7\;\mathrm{M}_\odot$, suggesting that the mass loss should have been higher than $0.2~\mathrm{M}_\odot$ on the RGB for Y19 and Y20. Quantifying the amount of mass loss in individual low-mass stars is still a matter of debate. In 
the detailed study of mass loss in clusters of age 7~Gyr, \cite{miglio-2012} found that the mass loss might be at most $0.1-0.2~\mathrm{M}_\odot$  during the RGB \citep[see also][]{yu-2021}, arguing that 
very low-mass RC stars must originate from a different channel. {\cite{Salaris-2016} analysed metal-poor stars in the globular cluster 47 Tuc and concluded that the mass loss along the RGB must be between 0.17 and 0.23 $\mathrm{M}_\odot$ for stars of  metallicity $-1$~dex, which is lower than the metalllicity of Y19 and Y20.}

There are extremely few stars with $M < 0.7~ \mathrm{M}_\odot$ in the RC in APOKASC (this is why Y19 and Y20 were selected, because they stand beyond the 90\% mass-distribution contour),  
suggesting that stars with masses that low might have experienced unusually high mass loss. While 
models considering larger amounts of mass loss than the canonical ones predict Y19 and Y20 to be on the horizontal branch, it seems odd that in this case APOKASC would include so few such stars. Perhaps there is another mechanism that formed Y19 and Y20, for example stripping mass due to binary interactions.

Very recently, \cite{Li-2022} published the discovery of about 40 red giant stars that have only partially transferred their envelopes; therefore they are not hot subluminous stars of spectral type B (sdB) but simply very low-mass giant stars. Their study is based on APOKASC, and Y19 is 
one of the
low-mass red giants they uncovered. \cite{Li-2022} have ran stellar evolutionary models with a progenitor mass of $1.5~\mathrm{M}_\odot$ that lost different amounts of mass due to binary stripping. They were unable to reproduce masses of  $0.6~\mathrm{M}_\odot$  with standard evolution without binary interaction, even when including mass loss in their models.    {It was found that after losing part of its envelope, the mass-losing star has a structure} essentially identical to that of a star that began its life with such a low mass and did not experience stripping. \cite{Li-2022} claim that it is impossible to decipher how much mass a star has lost based on its current properties. 
However, the $0.6~\mathrm{M}_\odot$ models without mass loss are older than the universe, while the mass-loss models with binary interaction produce realistic ages for the very low-mass stars.

\subsubsection{Individual cases}

\noindent {Y19: KIC 9644558}. After monitoring this star with HERMES for 1600 days and securing six RVs, no evidence of variability was found. The average HERMES, APOGEE, and Gaia velocities are all consistent with each other. Therefore this star is likely single. It is the warmest star of the sample, with a temperature of 5100~K. Its abundances are at the edge of the disk population, and follow the expected trend when displayed as a function of temperature. The exceptions are [Na/Fe], [P/Fe], [K/Fe], and [Cr/Fe], but considering the weakness of the corresponding lines, we should rather consider Na, P, and Cr abundances
as upper limits.  Y19 deviates from the main trend in the [C/N]-mass relation for red clump stars. Given its [C/N] abundance ratio, this star should be more massive than inferred from asteroseismology. Its [C/Fe] ratio is normal for its metallicity and  mass.  Its [N/Fe] abundance is slightly enhanced for its metallicity, and very enhanced for its mass. \cite{Li-2022} have {included this star in their sample of undermassive stars. They used MIST isochrones to estimate the lower-mass limit that a star with Y19 metallicity can have if it lives for 13.8 Gyr and reaches the zero-age-helium-burning phase. That mass corresponds to 0.87~$\mathrm{M}_\odot$ (Li, priv. comm.), which is larger than Y19 derived mass, even including a generous amount of $0.2 ~\mathrm{M}_\odot$ for mass loss (their Fig.~2).  Their analysis considered a mass for this star of $0.57 \pm 0.2 \mathrm{M}_\odot$, lower than the value adopted by us but consistent with the uncertainties.} \cite{Li-2022} {thus} ran models for this star including stripping from binary interaction, but {from our analysis} Y19 is not a binary. \\

\noindent {Y20: KIC 9946773.} This star does not show significant 
RV variations from the six HERMES measurements spanning 1600 days {so that we consider it to be a single star}.  {\cite{Li-2022} did not flag this star as undermassive because its current mass is consistent with a MIST isochrone that includes the generous mass loss of $0.2\;\mathrm{M}_\odot$ at a metallicity of -0.6. This star is indeed less extreme than Y19 and could well have experienced an evolution with mass loss during its RGB phase without the need to advocate binary interaction. Yet, its mass is still systematically below most of red-clump stars in the sample, hinting at Y20 having experienced a larger mass loss than its peers, or it started with an initial mass at the lower edge of the disk population. We note that} this star is more metal-poor than most of the stars in the sample, including most of the APOKASC stars (Fig.~\ref{fig:c_n_mass}), which makes it hard to compare Y20 with other APOKASC stars. Because of its high temperature, Y20 has P and Cr spectral lines too weak to measure the corresponding abundances. The low abundances reported by APOGEE are probably in this case very uncertain. We cannot assess whether the  [C/N] ratio and the C and N abundances of Y20 are normal because we have no control APOKASC RC stars at these metallicities.  Nevertheless, Na and O are enhanced, suggesting that some extra mixing happened in this star.  Y20 has a quite large radial action and could therefore have formed at larger Galactic radii, where the ISM has a lower chemical enrichment history.

\section{Conclusion}\label{sect:conclusion}

In this paper we have extensively studied the nature of a sample of young $\alpha$-rich stars in the APOKASC catalogue. These stars have masses that are higher than typical $\alpha$-rich stars in the Galaxy, and therefore point toward an interesting formation scenario.  We have performed a long-term RV monitoring campaign using the HERMES spectrograph in La Palma, Spain, which has allowed us to assess with enough confidence if the stars are in binary systems or not. From this stellar sample we have further used information about the chemistry from APOGEE DR16, the RVs, and astrometry from Gaia DR2 and {DR3}, and the inner properties such as masses and evolutionary stages from APOKASC-3 to study their nature.

Using the evolutionary stage of the stars we could separate red clump from red giant branch stars and select carefully the overmassive from the normal stars. 
We found that the most massive stars tend to be single, which is in  agreement with the population synthesis models of \cite{Izzard-2018} who predicted that the most massive ones should be the product of mergers while the others could result from either mergers, mass transfer via RLOF or winds allowing for a variety of binaries and single stars in their models. 

The new masses from APOKASC-3 also allowed us to find that two of our initially selected overmassive stars are in fact undermassive, which means their masses are below the turn-off mass of globular clusters of $0.8~\mathrm{M}_{\odot}$. While mass loss in some models might explain masses below  $0.8~\mathrm{M}_{\odot}$ at the red clump, accounting for masses of $0.6~\mathrm{M}_{\odot}$ might require further understanding of mass loss. One of the stars studied here was already reported by \cite{Li-2022} to be a stripped giant from a binary companion, but we find it to be RV constant in this study. 

From the astrometric Gaia data we found that the stars follow the dynamical properties of other thick disk stars, according to previous works in the literature \citep[e.g.][]{Zhang-2021}. From the APOGEE DR16 data we could investigate further the chemical abundances finding that most of the abundances are consistent with thick disk stars. We found however some anomalies in elements that might be sensitive to nonstandard stellar evolution, such as C, N, Na, P, and K.  We further found very low abundances of Cr for some of the stars.  These abundances in any case need to be confirmed with spectra of higher resolution or in a different wavelength range  because the lines in APOGEE are very weak and could be blended. 

Overall from our analysis we conclude that the young $\alpha$-rich stars are most likely product of binary interaction and are not truly young. The variety of possibilities to form them when binary evolution is involved is large and therefore they cannot be expected to show all the same observational properties. In fact, they are quite different from each other. Some are short- or long-period binaries, and others appear to be single. Some have typical chemistry, others have enhanced or depleted abundances of a particular element. Some follow the usual trends of [C/N] with mass, others stand out. Some have an IR excess, others do not \citep{Yong-2016}. Some have high masses, others very low \citep{Li-2022}. They all however seem to be similar in their kinematics, sharing their origin with the thick disk, hence sharing its formation epoch at least 8 Gyr ago. Our detailed study, however, cannot rule out the possibility that some of the stars in our sample (like Y25) might still be truly young.  

It is long known that determining ages of stars is among the most challenging tasks in astrophysics. We need to have very precise and accurate measurements of several observables such as stellar parameters, distances, and oscillation modes in the case of asteroseismology. These measurements need to agree with well-calibrated stellar models which accurately describe stellar evolution, including poorly understood processes such as mass loss, mixing or convection. But even if stellar models and observational measurements are becoming precise and accurate in modern astronomy, there will always be a good chance that some stars are born in pairs and exchange material throughout their lifetimes. In that case, accurate models and measurements will not be sufficient for dating stars, since its current mass does not tell us for how long a given  star has existed.   

From this relatively small sample of 41 red giants we could already see how much can be potentially learned from the evolution of red giant stars when long-term RV monitoring, high-resolution spectroscopy, asteroseismology, and astrometry is available for them. We could grasp the rich variety of signatures that binary evolution might produce. It is exciting to realise how we are entering a new golden epoch to study these processes in a statistical manner. The arrival of Gaia DR3 combined with TESS and PLATO in the near future, and the new spectroscopic surveys such as SDSS-V, WEAVE or 4MOST being soon released will offer immense opportunities. In the near future we will learn to distinguish single from binary stars, and so be able on the one hand to apply our best stellar models to precise measurements of individual stars and properly reconstruct the Milky Way history, and on the other hand, to address the long-standing problems of stellar evolution theory.

\begin{acknowledgements}
The authors thank the referee for the constructive report that improved significantly our article, and Yaguang Li for providing important information about the modeling of the undermassive stars. P.J. acknowledges partial financial support of FONDECYT Regular grant number 1200703. C.A. acknowledges financial support from ESO-Chile Joint Committee. P.J. and C.A. thank the Millenium Nucleus ERIS NCN2021\_017, K. Hawkins with his group at University Texas Austin as well as D. de Brito and S. Vitali from Universidad Diego Portales, for lively discussions throughout the entire process of understanding the nature of the young $\alpha$-rich stars, much of which helped to continue with this paper. \\

Based on observations made with the Mercator Telescope, operated on the island of La Palma by the Flemish Community, at the Spanish {\it Observatorio del Roque de los Muchachos} of the {\it Instituto de Astrof\'\i sica de Canarias}. Based on observations obtained with the HERMES spectrograph, which is supported by the Research Foundation-Flanders (FWO), Belgium, the Research Council of KU Leuven, Belgium, the Fonds National de la Recherche Scientifique (F.R.S.- FNRS), Belgium, the Royal Observatory of Belgium, the Observatoire de Gen\`eve, Switzerland and the Th\"{u}ringer Landessternwarte Tautenburg, Germany.\\

Funding for the Sloan Digital Sky 
Survey IV has been provided by the 
Alfred P. Sloan Foundation, the U.S. 
Department of Energy Office of 
Science, and the Participating 
Institutions. 

SDSS-IV acknowledges support and 
resources from the Center for High 
Performance Computing  at the 
University of Utah. The SDSS 
website is www.sdss.org.

SDSS-IV is managed by the 
Astrophysical Research Consortium 
for the Participating Institutions 
of the SDSS Collaboration including 
the Brazilian Participation Group, 
the Carnegie Institution for Science, 
Carnegie Mellon University, Center for 
Astrophysics | Harvard \& 
Smithsonian, the Chilean Participation 
Group, the French Participation Group, 
Instituto de Astrof\'isica de 
Canarias, The Johns Hopkins 
University, Kavli Institute for the 
Physics and Mathematics of the 
Universe (IPMU) / University of 
Tokyo, the Korean Participation Group, 
Lawrence Berkeley National Laboratory, 
Leibniz Institut f\"ur Astrophysik 
Potsdam (AIP),  Max-Planck-Institut 
f\"ur Astronomie (MPIA Heidelberg), 
Max-Planck-Institut f\"ur 
Astrophysik (MPA Garching), 
Max-Planck-Institut f\"ur 
Extraterrestrische Physik (MPE), 
National Astronomical Observatories of 
China, New Mexico State University, 
New York University, University of 
Notre Dame, Observat\'ario 
Nacional / MCTI, The Ohio State 
University, Pennsylvania State 
University, Shanghai 
Astronomical Observatory, United 
Kingdom Participation Group, 
Universidad Nacional Aut\'onoma 
de M\'exico, University of Arizona, 
University of Colorado Boulder, 
University of Oxford, University of 
Portsmouth, University of Utah, 
University of Virginia, University 
of Washington, University of 
Wisconsin, Vanderbilt University, 
and Yale University.
\end{acknowledgements}

\bibliographystyle{aa} 
\bibliography{biblio} 

\begin{appendix}


\section{Main differences in masses between APOKASC catalogues}\label{mass_comparison}
\begin{figure}[t]
   \centering
        \includegraphics[scale=0.5]{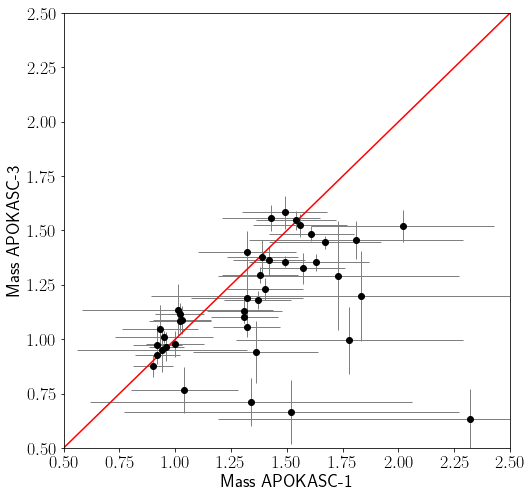}
   \caption{Comparison of masses reported by APOKASC-1 and APOKASC-3.}
            \label{fig:mass12}%
    \end{figure}

{
In Figure~\ref{fig:mass12} we compare the masses reported by APOKASC-1 (our original sample for selecting the stars), and the final masses to be published as part of APOKASC-3. One can note that uncertainties have notably decreased, and that some stars that in APOKASC-1 were reported to have masses above 1~M$_\odot$, in APOKASC-3 have masses below that value. }

{The significant changes in the distribution and categorization of our objects of interest likely stem from the evolution in the analysis that occurred between the first APOKASC data release and the most recent sample. In APOKASC-1 \citep{Pinsonneault-2014}, the analysis focused on a much smaller sample of stars, and they were analyzed using techniques that had been developed for dwarf stars \citep[e.g.][]{Chaplin2014}. In particular, $\nu_{\rm max}$ and $\Delta \nu$ values were inferred from a single analysis pipeline, and then four other pipelines were used for confirmation and outlier rejection. The resulting parameters were then combined with the spectroscopic temperature and metallicity measurements, either raw or calibrated, and compared to the predictions of stellar models \citep[similar to][]{Stello2009, Basu2010, Basu2012, Gai2011, Chaplin2014}. Evolutionary states were only inferred from asteroseismology for a fraction of the stars ($\sim$25\%), and so could not be used as a generic prior on the fits. Results were quoted from a single combination of observables, grid modeling framework, and underlying models, and the variance in the fits was used to define an uncertainty. No further calibrations were applied. }

 {In the intervening time, it became clear that the methods used in APOKASC-1 were not entirely optimal. \citet{Epstein2014} showed that the results were particularly lacking accuracy at low metallicity, and further work suggested that stellar models of red giants tended to be offset from observations as a function of mass and metallicity \citep{Tayar2017, JoyceChaboyer2018b, JoyceChaboyer2018a, Salaris2018}, which likely biased the inferred parameters. As part of the APOKASC-2 analysis \citep{Pinsonneault-2018}, estimates of evolutionary state were substantially improved \citep{Elsworth2019}. In addition, it was realized that different seismic pipelines tended to produce results with constant offsets, and so averaging them, after calibration, could be used to improve the seismic precision. Rather than using a grid of models directly, and inheriting their uncertainties, model-derived corrections to the scaling relations were applied \citep{Pinsonneault-2018} and empirical calibrations were used to put the whole sample onto the mass scale of open clusters. APOKASC-3 (M. Pinsonneault et al., in prep) inherits the same underlying averaging and empirical calibration framework, but instead uses the large ensemble of radii from Gaia that are now available to determine a $\nu_{\rm max}$ - dependent calibration to the scaling relations in order to ensure accuracy as well as precision.}

\section{Complementary information on binarity}

Table~\ref{28stars} collects all information about binarity at hand from the different data sources (see Sect.~\ref{sect:data}). Criteria flagging a star as binary are marked in bold face, and binary stars appear as open squares in  figures throughout the paper.    {Table~\ref{Tab:Gaia_DR2_DR3} provides a comparison of the Gaia DR2 and DR3 RVs for our target stars.} Table~\ref{Tab:RV} provides the individual HERMES RVs for all sample stars. The full table is only available at CDS.
  
\input{tab_summary.tex}

\input{table_RV_DR2_DR3}

\begin{table}[]
    \caption{Individual radial velocities. The first column lists the KIC identifier, the second column the Julian Date, and the third column the barycentric radial velocity. The long-term uncertainty on each RV is 0.07~km~s$^{-1}$. The full table is  available from CDS, Strasbourg.  }
    \label{Tab:RV}
    \centering
    \begin{tabular}{rrc}
    \hline
    \multicolumn{1}{c}{KIC} & \multicolumn{1}{c}{JD} & \multicolumn{1}{c}{RV} \\
        &&\multicolumn{1}{c}{(km~s$^{-1}$)}\\
    \hline
1432587&2457232.6041087&-69.96\\
1432587&2457542.6037500&-69.77\\
1432587&2457564.6391761&-69.68\\
1432587&2457599.5249996&-69.99\\
1432587&2457601.5896846&-69.94\\
1432587&2457940.6157278&-69.73\\
1432587&2458280.6954023&-69.76\\
1432587&2458361.5447534&-69.76\\
\medskip\\
2142095&2457232.5567835&-16.41\\
2142095&2457541.5000000&-16.42\\
2142095&2457564.5794659&-16.37\\
2142095&2457599.5111468&-16.39\\
2142095&2457601.5772863&-16.31\\
2142095&2457866.7037368&-16.15\\
2142095&2458284.5117667&-16.18\\
2142095&2458315.7041689&-16.18\\
\medskip\\
\multicolumn{1}{c}{...}&\multicolumn{1}{c}{...}&\multicolumn{1}{c}{...}\\
\hline
    \end{tabular}

\end{table}

\subsection{Orbits}
\label{Sect:orbits}

\begin{figure}[t]
   \centering
   \vspace*{-4cm}
    \includegraphics[scale=0.45]{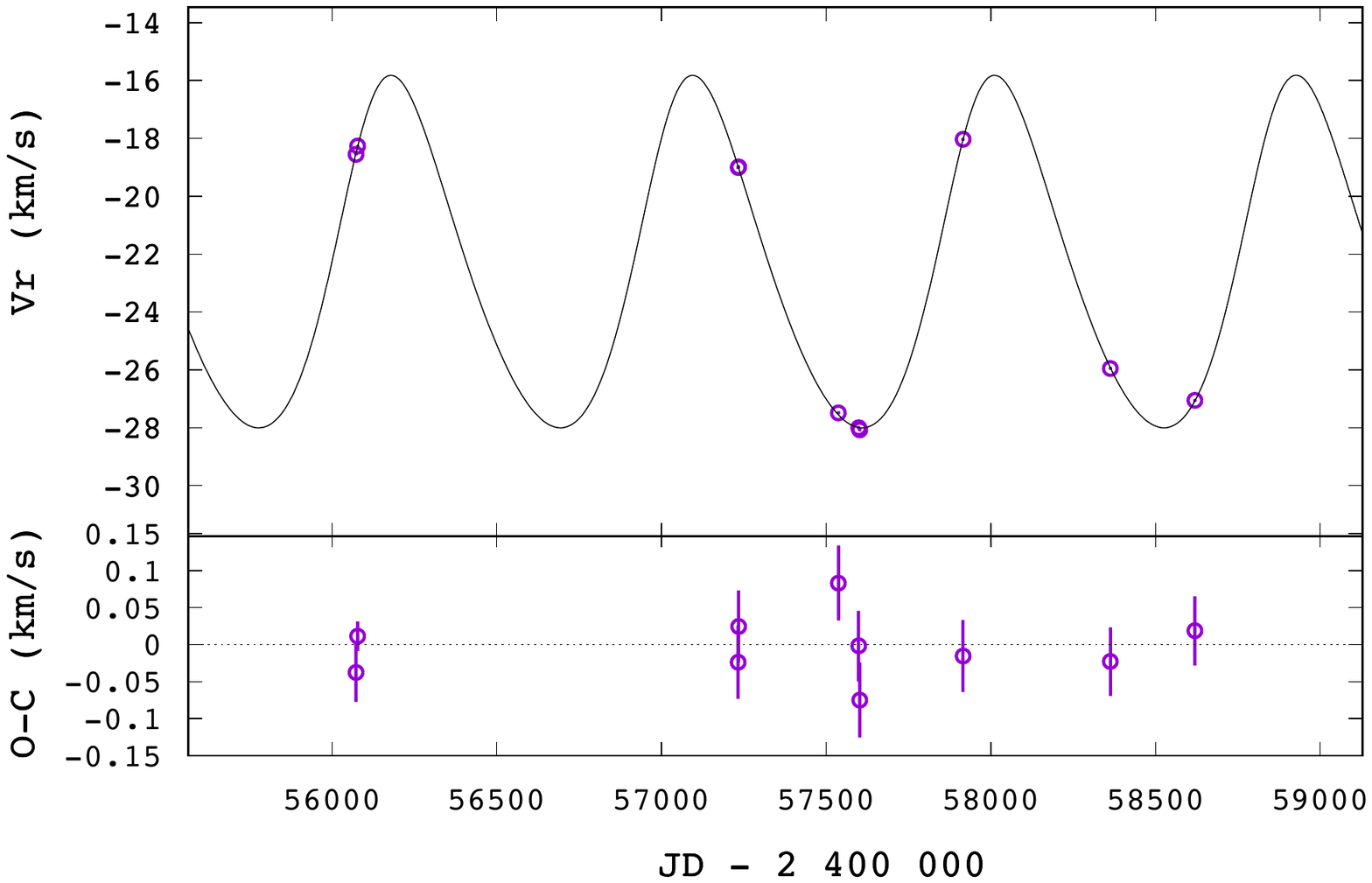}
          \vspace*{-8cm}\\
     \includegraphics[scale=0.45]{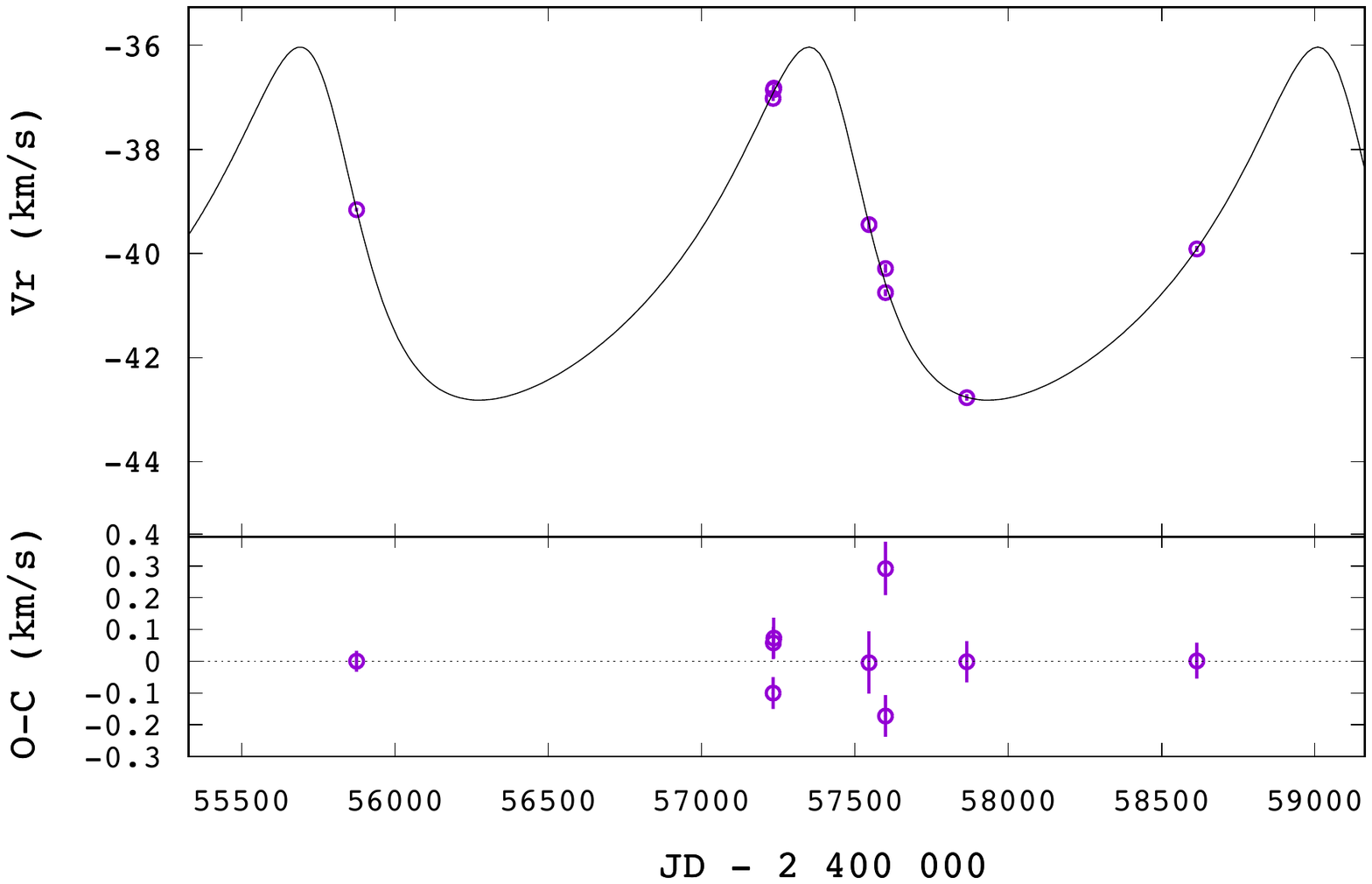}
      \vspace*{-8cm}\\
     \includegraphics[scale=0.45]{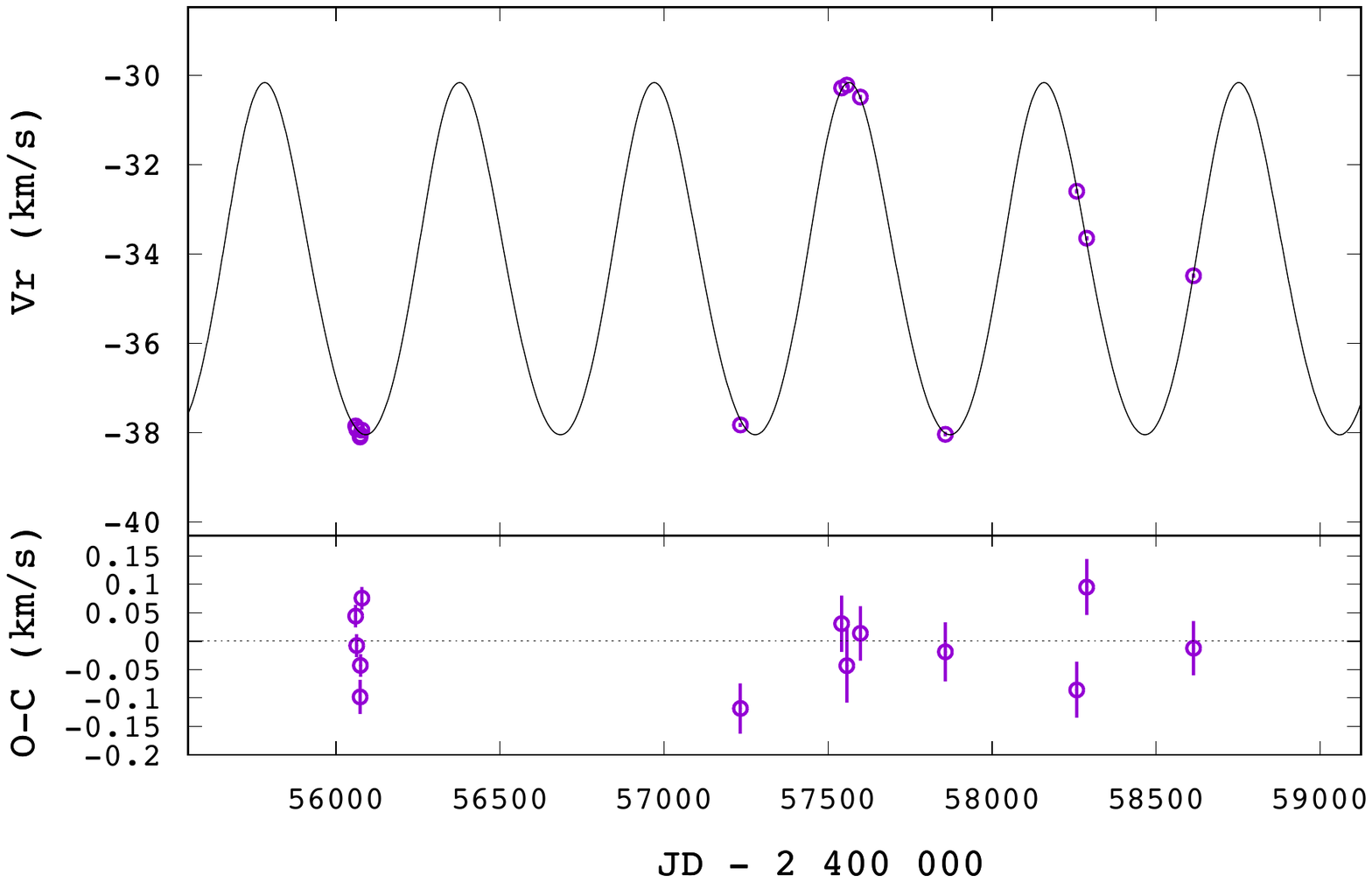}
      \vspace*{-8cm}\\
       \includegraphics[scale=0.45]{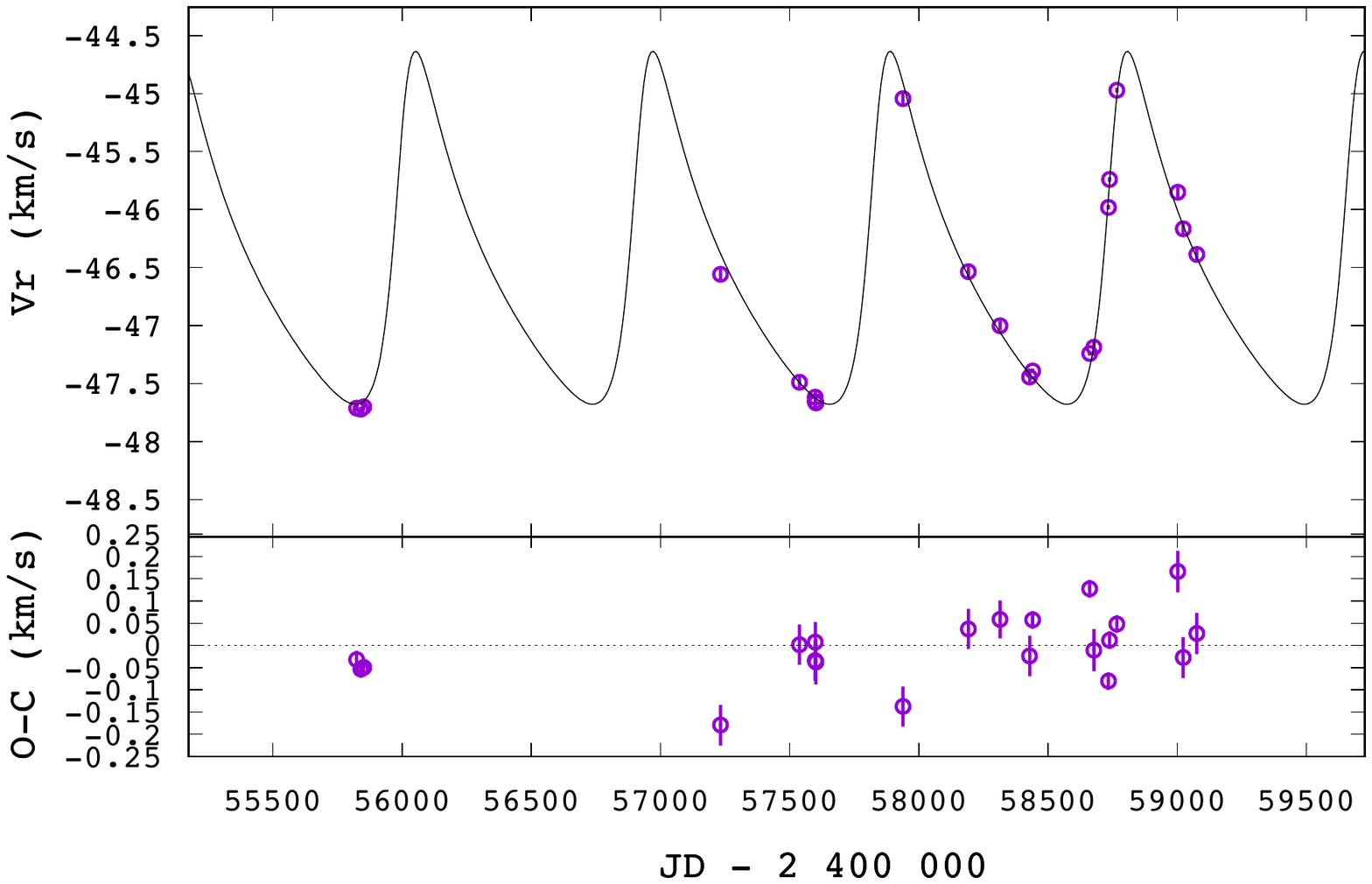}
      \vspace*{-4cm}\\
   \caption{Orbits and O-C residuals of KIC 11823838 (Y6, \textit{Top panel}),  KIC 5512910 (Y7, \textit{Second panel}), KIC 9002884 (Y9, \textit{Third panel}), and KIC 3455760 (Y12, \textit{Bottom panel}).}
 \label{fig:orbits}%
    \end{figure}

We present in Fig.~\ref{fig:orbits} and Table~\ref{Tab:orbital_elements} the orbits of KIC 11823838 (Y6),  KIC 5512910 (Y7), KIC 9002884 (Y9), and KIC 3455760 (Y12), computed along the guidelines described in \citet{Pourbaix1998}.  

\begin{table*}[t]
    \caption{Orbital elements obtained from HERMES or Gaia DR3.   }
    \label{Tab:orbital_elements}
    \begin{tabular}{llllllllll}
    \hline\\
 KIC & 
 $P$& 
   $e$ &
   $\omega$  &
   $T_0$ &
    $V_0$ & 
 $K$ & 
 $f(M)$  &
 $N$ &
 $\sigma (O-C)$
\\    
&
(d) &
&
($^\circ$) &
(JD-2 400 000) &
(km s$^{-1}$) &
 (km s$^{-1}$) &
 ($10^{-2}$ M$_\odot$) &
 &
 (km s$^{-1}$) \\
   \hline\\
 9157260 (O3 - DR3) &
 $675\pm30$ & $0.30\pm0.04$&
 $313\pm11$ &
  $57351\pm18$ &
 $-25.7\pm0.3$ &
$8.64\pm0.41$ &
$3.9\pm1.3$ &
23 &
-
\medskip \\
  11394905 (Y4 - DR3) & 
 $591\pm14$  &
 $0.19\pm0.09$ &
 $170\pm23$ &
 $57600\pm 35$ &
 $-71.6\pm0.2$ &
 $3.71\pm0.32$ &
$0.30\pm0.11$ &
 22 &
 -
\medskip  \\
 11823838 (Y6 - HER) &
 $915.8\pm0.6$ &
 $0.15\pm0.01$ &
 $320\pm3$ &
 $56103\pm8$ &
$-22.61\pm0.02$ &
$6.09\pm0.05$ &
$2.07\pm0.05$ &
 10 &
 0.04
 \\ 
 11823838 (Y6 - DR3) &
 $894\pm14$ &
 $0.26\pm0.04$ &
 $312\pm7$ &
 $57019\pm16$ &
 $-22.50\pm0.16$ &
 $6.62\pm0.21$ &
$2.4\pm0.3$ &
 23 &
 -
\medskip  \\
  5512910 (Y7 - HER) & 
 $1659.5\pm5.4$ &
 $0.36\pm0.03$ &
 $40\pm3$ &
 $59\;093\pm19$ &
  $-40.35\pm0.07$ &
  $3.39\pm0.07$ &
  $0.55\pm0.04$ &
  9 &
  0.13
\medskip  \\
  9002884 (Y9 - HER) &
 $593.8\pm0.8$ &
 $0.079\pm0.005$ &
 $338\pm7$ &
 $57533\pm12$ &
 $-34.40\pm0.02$ &
  $3.94\pm0.04$ &
 $0.37\pm0.01$ &
 13 &
 0.07
  \\
   9002884 (Y9 - DR3) &
 $585\pm16$ &
 $0.034\pm0.064$ &
 $288\pm96$ &
$57451\pm155$&
$-34.41\pm0.16$&
$4.09\pm0.23$ &
$0.41\pm0.03$ &
 21 &
 -
\medskip \\ 
 3455760 (Y12 - HER) &
 $918.1\pm2.8$ &
 $0.460\pm0.006$ &
 $303\pm1$ &
 $58753.2\pm1.7$ &
 $-46.54\pm0.01$ &
$1.52\pm0.01$ &
 $0.0235\pm0.0007$ &
 21 &
 0.08\\
 \hline
  \\
    \end{tabular}
 \tablefoot{$P$ is the orbital period, $e$ is the eccentricity, $\omega$ the argument of periastron, $T_0$ the epoch of periastron passage, $V_0$ the velocity of the centre of mass of the system, $K$ the semi-amplitude of the velocity variations, $f(M)$ the mass function, $N$ the number of observations, and $\sigma (O-C)$ the standard deviation of the O-C ('observed - calculated') residuals.}    \end{table*}

\section{Binary frequencies among specific stellar groups}
\label{Sect:binaries}

\subsection{YAR, AR, and AP}

\begin{table}[t]
\caption{\label{Tab:SBfrequency}
Binary frequencies among chemical classes AP, AR, and YAR.  $p1$ is the one-sided $p$-value for the corresponding 2$\times$2 contingency table (see text). }
   \centering
    \begin{tabular}{l|rr|rr}
  & \multicolumn{3}{c}{$M_{\rm thresh} = 1.3$ M$_\odot$}\\
 \hline\\
 class & $N_{\mathrm{SB}}$ & $N_{\mathrm{nonSB}}$ & \%SB & $p1(\%)$\\
\hline 
AR & 5 & 9 & 35.71\\
YAR & 9 & 7 & 56.25\\
\hline
Tot & 14 & 16 & 46.67\\
$p1(\%)$ & & & & 22.5\\
\hline
AP & 5 & 6 & 45.45\\
AR & 5 & 9 & 35.71\\
\hline
AP + AR & 10 & 15 & 40.00\\
YAR & 9 & 7 & 56.25\\
\hline
Tot & 19 & 22 & 46.34\\
$p1(\%)$ & & && 24.3\\
\hline
YAR + AR & 14 & 16 & 46.67\\
AP & 5 & 6 & 45.45\\
\hline
Tot & 19 & 22 & 46.34\\
$p1(\%)$ & & & &61.3\\
\hline

    \end{tabular}
 \end{table}

Binary frequencies among AP, AR, and YAR stars are listed in Table~\ref{Tab:SBfrequency}. 
To quantify this, the various subpanels of Table~\ref{Tab:SBfrequency} can be seen as $2\times2$ contingency tables $\big(\begin{smallmatrix}
  a & b\\
  c & d
\end{smallmatrix}\big)$, 
where $f_1 - f_2$ can be used to measure the degree of disproportion between the frequencies appearing on the first and second lines, with $f_1 \equiv a/(a+b) =  a/N_1$ and $f_2 \equiv c/(c+d) = c/N_2$.  
If the frequencies listed on lines 1 and 2 correspond to independent quantities, then the probability of occurrence of any such $\big(\begin{smallmatrix}
  a & b\\
  c & d
\end{smallmatrix}\big)$ array with $N_1$ and $N_2$ fixed is expressed by the hyper-geometric distribution $[N_1!\;N_2!\;(a+c)!\;(b+d)]\;/\;(N!\;a!\;b!\;c!\;d!)$, with $N = N_1 + N_2$. The one-sided $p$-value  corresponds to the probability of occurrence of that particular array plus the probabilities associated with all other possible arrays whose degree of disproportion $|f_1 - f_2|$ is equal to or greater than that of the observed array.  For instance, the one-sided $p$-value associated with the array $\big(\begin{smallmatrix}
  2 & 10\\
  5 & 8
\end{smallmatrix}\big)$ is the sum of the probabilities associated with the occurrence of that array and of those having even more extreme degrees of disproportion, namely $\big(\begin{smallmatrix}
  1 & 11\\
  6 & 7
\end{smallmatrix}\big)$ and $\big(\begin{smallmatrix}
  0 & 12\\
  7 & 6
\end{smallmatrix}\big)$. 
 
 To flag significant differences between binary frequencies in our sample, the $p$-value should thus be as small as possible. The $p$-values listed in Table~\ref{Tab:SBfrequency} correspond to the exact Fisher test for a $2\times2$ contingency table \citep[see][]{Jorissen-2016}\footnote{They are derived using the {\tt fisher\_exact} routine from the {\tt scipy.stats}  library of python. }

The first line of the contingency table on top of Table~\ref{Tab:SBfrequency} refers to AR stars and the second one to YAR stars, whereas the first column refers to the number of binary stars and the second column to the number of nonbinary stars. The contingency table in this case corresponds to $\big(\begin{smallmatrix}
  5 & 9\\
  9 & 7
\end{smallmatrix}\big)$, and the corresponding $p-$value of {0.93} reflects the absence of significant difference in binary frequencies between AR and YAR stars. By combining the AP and AR against YAR samples, as done in the middle of Table~\ref{Tab:SBfrequency}, an almost equally poorly significant result is obtained with a $p-$value of  {0.91}. Finally, the lower part of Table~\ref{Tab:SBfrequency} shows the contingency table of all $\alpha-$rich (AR and YAR) against the AP sample. The $p-$value here is {0.61}. 

In summary, the largest difference in the frequencies of binaries and nonbinaries is found when separating the samples on {YAR and AR categories, in which 36\% of the AR stars are binaries compared to 56\% of YAR stars, but the difference is not significant given the $p-$value of 23\%.}

\subsection{Uncertainties due to the adopted definitions of groups}\label{sect:mass_error}

Our classification of stars as AP, AR or YAR described in Sect.~\ref{sect:data} contains some degree of arbitrariness associated with the adopted mass and [$\alpha$/Fe] thresholds.
In this section we therefore study the impact of these thresholds on the derived binary frequencies.

We could for instance have set the mass threshold for YAR stars at $M >1.4$~M$_\odot$ as in Paper~I and \cite{Martig-2015} instead of $M >1.3$~M$_\odot$ adopted throughout this work.  
On the other hand, in Paper~I, the O stars were selected on the condition $M <1.2$~M$_\odot$ \citep[corresponding to stars with ages above 6 Gyr which is the threshold used by][]{Zhang-2021}.  

If one considers  YAR stars as having masses above 1.4~M$_\odot$ and AR stars as having masses below 1.2~M$_\odot$, the binary frequencies for YAR and AR stars are {37.5}\% (4/9) and {25\%, respectively, slightly different from those listed in Table~\ref{Tab:SBfrequency}, but with a high  $p$-value of 62\%.}  If we adopt instead a conservative cut  at $M < 1.0$~M$_\odot$ for AR stars (old thick disk), keeping the YAR threshold at $M > 1.4$~M$_\odot$, then the binary frequencies are 33\%  for YAR stars and  {15\%} for AR stars (corresponding to a one-sided $p$-value of 60\% but the sample of AR stars is small).

The second source of arbitrariness is the chemical threshold for separating AR from  AP stars.  
We define AR and AP stars following Eq.~\ref{eq:alpha2} but could have adopted our initial criterion (Eq.~\ref{eq:alpha}). In that case, we obtain 61\%  of YAR being binaries, and  35\% of AR being binaries.

Another important aspect to consider is the uncertainty on the masses and chemical abundances. By running a Monte-Carlo simulation over the uncertainties on the masses and chemical abundances, we can estimate the uncertainty in the binary fraction of each category by counting the AP, AR, YAR stars in each simulation. To do so, we assigned to each star a random value for the mass and abundances drawn from a normal distribution centered on the value given in Table~\ref{Tab:sample} and with a standard deviation equal to the corresponding uncertainty on the mass, [$\alpha$/Fe] or [Fe/H]. These values are on average 0.1~M$_\odot$, 0.008 dex and 0.008 dex, respectively.  Counting the binaries in each stellar group in 1000 realisations, we find a mean binary frequency of  $0.52\pm 0.04$ for YAR stars,   $0.40 \pm 0.04$ for AR stars, and $0.44 \pm  0.02$ for AP stars, 
indicating that mass and abundance uncertainties are not the major source of uncertainty on the binary fractions.

Although the binary frequencies do depend  on the cuts adopted for defining the different populations, these variations never reach the level necessary to make the AP, AR, YAR binary frequencies differ significantly from one another, as may be judged from the above $p-$values which never reached below 22\%.

If anything, we may conclude that the frequency of binaries among YAR stars seems slightly larger than among AR stars, but not at all in a statistically-significant manner, whereas the binary frequency among AR and simply AP seem identical, but we cannot rule out the possibility that all AR stars are old $\alpha-$rich stars that have evolved without binary interaction.

We comment that these frequencies cannot be directly compared to other multiplicity frequencies as a function of stellar parameters found in other studies such as \citet{badenes18} or \citet{mazzola2020},  because our sample is still very small and it does not intend to be representative of the red giant population in the disk nor has been corrected for selection biases. Here we have attempted to see if YAR stars have a higher tendency to be in binary systems compared to other lower-mass stars in the disk, and our analysis shows this is not the case.

\subsection{Overmassive, undermassive, and bulk}

Table~\ref{Tab:SBfrequency2} shows the $2\times2$ contingency tables for the new groups defined in Sect.~\ref{sect:new_groups} following the same line of thought as Table~\ref{Tab:SBfrequency}. 
Again, there is no significant difference in the binary frequencies for the different groups considered here.   Furthermore, it is difficult to compare the binary frequencies among overmassive  and  undermassive stars, since there are only two undermassive stars. 
From here we conclude that regardless of how the samples are defined (YAR, AR, and AP or overmassive, undermassive, and bulk), the binary frequencies are not significantly different.  
We remark that our analysis considers the different above categories as a whole, that is, there is no attempt to study possible variations of the binary frequency within each category as a function of other stellar properties.

\begin{table}[t!]
\caption{\label{Tab:SBfrequency2}
Binary frequencies among the classes "overmassive", "undermassive", and "bulk".  $p1$ is the one-sided $p$-value for the corresponding 2$\times$2 contingency table. }
   \centering
    \begin{tabular}{l|rr|rr}
  & \multicolumn{3}{c}{$M_{\rm thresh} =80$\% contours }\\
 \hline\\
 Class    & $N_{\rm SB}$ & $N_{\rm nonSB}$ & \% SB & $p1(\%)$\\
\hline 
overmassive & 11 & 9 & 55.00\\
bulk & 8 & 11 & 42.11\\
\hline
Tot & 19 & 20 & 48.72\\
$p1(\%)$ & & & & 31.4\\
\hline
undermassive & 0 & 2 & 0.00\\
bulk & 8 & 11 & 42.11\\
\hline
over + undermassive & 11 & 11 & 50.00\\
bulk & 8 & 11 & 42.11\\
\hline
Tot & 19 & 22 & 46.34\\
$p1(\%)$ & & & & 42.5\\
\hline
overmassive& 11 & 9 & 55.00\\
undermassive& 0 & 2 & 0.00\\
\hline
Tot & 11 & 11 & 50.00\\
$p1(\%)$ & & & & 23.8\\
\hline
   \end{tabular}
 \end{table}

\section{Discussion about mass uncertainties of Y19 and Y20}\label{app:lowmass}

\begin{figure}[t]
   \centering
        \includegraphics[scale=0.35]{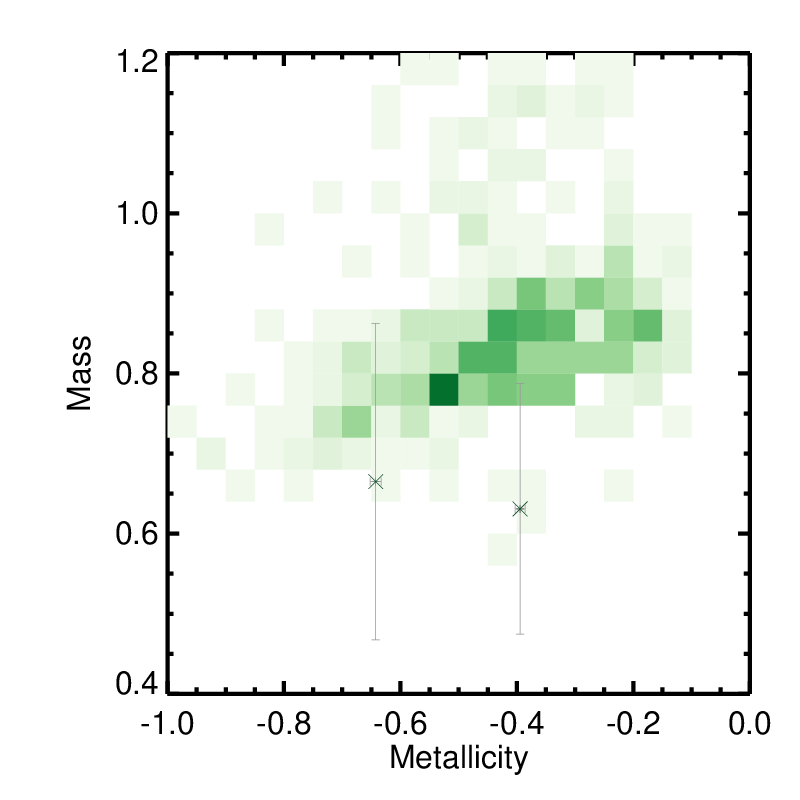}
   \includegraphics[scale=0.5]{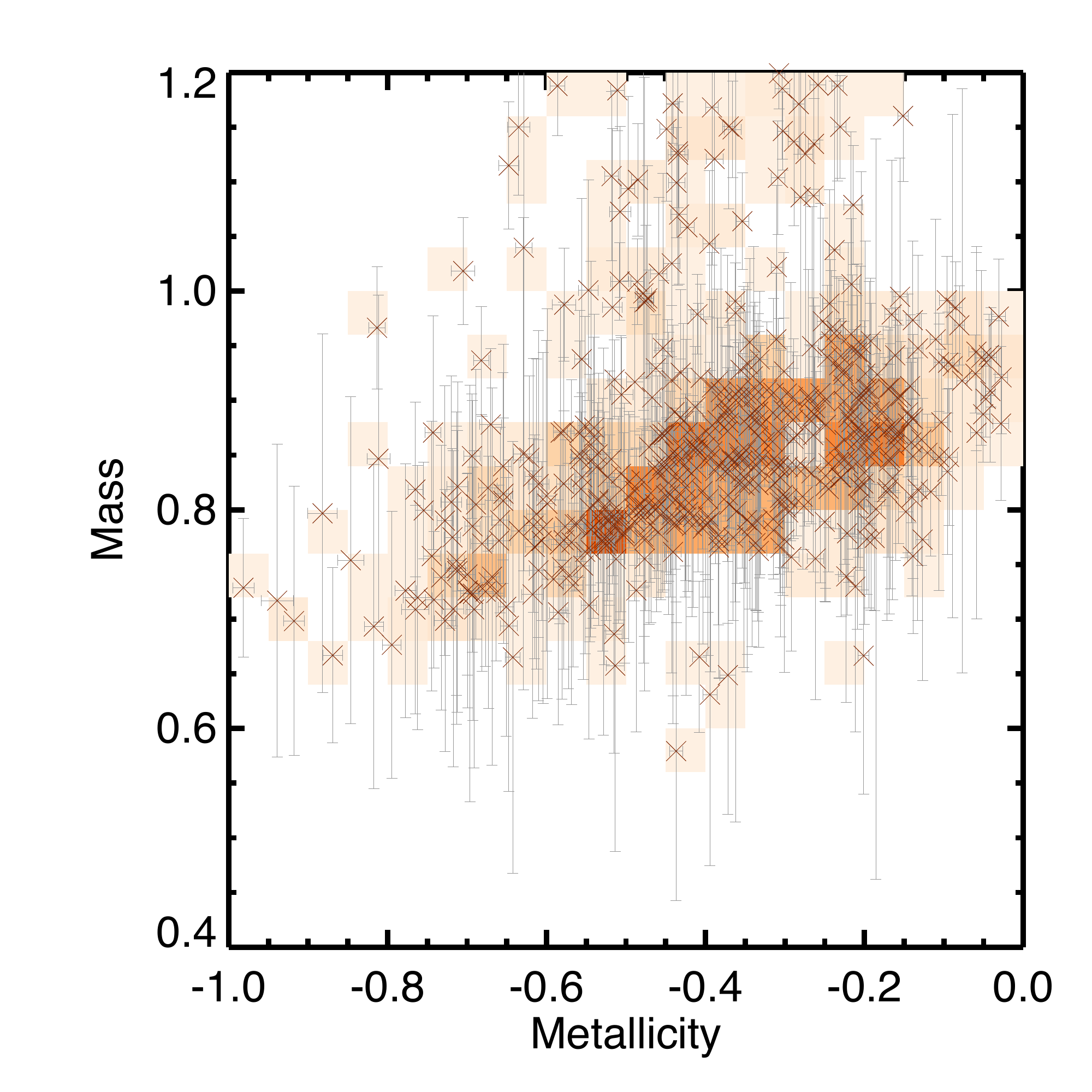}

   \caption{Upper panel: Mass and metallicity density distribution of red clump $\alpha-$enhanced stars in APOKASC-3 alongside with Y19 and Y20 with crosses. Lower panel: Same distribution as above but now random uncertainties are plotted with orange bands and systematic uncertainties with gray error bars.
   }
            \label{fig:uncertainty_ensemble}%
    \end{figure}

{Both stars were processed by up to seven pipelines in the APOKASC analysis, so their resulting masses are robust. The measurement errors and spread between the pipelines are a little bit larger, reflecting a slightly less precise estimate than usual, but those errors are properly propagated into the mass uncertainty. In addition, the fact that it also came out low when analyzed by for example \cite{Yu-2018} and \cite{Li-2022} points toward an uncertainty that is not a simple systematic error in APOKASC-3 measurements.  

Because these stars are in the clump and have low masses, the corrections to the $\delta \nu$ {the small frequency interval}, which depends on models) vary more than usual from pipeline to pipeline. That is expected given that we are in a parameter range where such corrections have had less calibration and some of the modeling choices matter more. Still these stars always come out as having low masses, therefore there is no obvious seismic  indication that their calculated masses could be incorrect.  

Given that the uncertainties on the masses include both random and systematic contributions, we compare Y19 and Y20 to the ensemble of other stars.

The upper panel of Fig.~\ref{fig:uncertainty_ensemble} shows the mass and metallicity distribution of the red clump sample for the $\alpha-$enhanced stars, similar to Fig.~\ref{Fig:APO3mass}. Here we only show Y19 and Y20 as crosses. From the error bars one might think they could well be normal stars whose uncertainties are large. The lower panel in that figure, however, shows the distribution of uncertainties, distinguishing random from systematic. While systematic uncertainties are larger for lower masses, which is explained by the discussion above, the random uncertainties are comparable for all stars. 

Assuming that the large error bars mean that they could be part of the normal ensemble thus implies that these stars are not affected by the same systematic effects than all other stars, which is unlikely. }

\end{appendix}
\end{document}

%% file: table_ids_jul2022.tex
\renewcommand{\tabcolsep}{4pt}
\begin{table*}
\caption{Various identifiers of the program stars (KIC, APOGEE, and Gaia DR3) followed by the Gaia $G$ and JASS $K_s$ magnitudes. The atmospheric parameters, masses and  abundances are taken from APOGEE DR16, and APOKASC-3. The column labeled 'Evol' lists the evolutionary stage of the stars, {as derived from APOKASC-3}, where  1 represents the red giant branch and 2 the red clump.}\label{Tab:sample}
\small{
\begin{tabular}{|c|c|c|c|cc|ccccc|ccc|}
\hline 
Star & KIC ID & APOGEE ID & Gaia DR3 ID & $G$  & $K_s$ & $T_{\rm eff}$ & $\log g$ & [Fe/H] & [$\alpha$/Fe] & [C/N] & $M$ & $\sigma_M$ & Evol \\
 &  &  &   & (mag) & (mag) & (K) &  &  &  &  & (M$_\odot$) & (M$_\odot$) &  \\
\hline 
O1 & 10586902 & 2M19031115+4753540 & 2131506438378495872 & 12.00 & 9.78 & 4682 & 2.56 & -0.35 & 0.09 & 0.06 & 1.01 & 0.03 & 1 \\
O2 & 2142095 & 2M19055465+3735053 & 2099076519018313344 & 11.50 & 9.26 & 4898 & 2.38 & -0.37 & 0.04 & -0.08 & 1.12 & 0.05 & 2 \\
O3 & 9157260 & 2M19313671+4532500 & 2126539467621880320 & 11.74 & 9.68 & 4876 & 3.28 & -0.14 & 0.09 & 0.10 & 0.97 & 0.09 & 1 \\
O4 & 4844527 & 2M19370398+3954132 & 2076511722922992896 & 13.76 & 11.44 & 4920 & 2.38 & -0.38 & 0.21 & -0.04 & 0.77 & 0.11 & 2 \\
O5 & 8611114 & 2M19023483+4444086 & 2106349960231621760 & 10.62 & 8.48 & 4864 & 2.43 & -0.20 & 0.01 & 0.19 & 0.98 & 0.06 & 2 \\
O6 & 10463137 & 2M19151828+4736422 & 2130913148779139840 & 12.08 & 10.01 & 4948 & 2.38 & -0.42 & 0.08 & 0.03 & 0.88 & 0.05 & 2 \\
O7 & 11870991 & 2M19411280+5009279 & 2134976565796276096 & 11.82 & 9.72 & 4907 & 2.41 & -0.38 & 0.07 & -0.04 & 0.93 & 0.04 & 2 \\
O8 & 7594865 & 2M19084355+4317586 & 2105616031923137152 & 11.91 & 9.73 & 4822 & 2.40 & -0.25 & 0.05 & -0.06 & 1.09 & 0.06 & 2 \\
O9 & 1432587 & 2M19254985+3701028 & 2051748212801991296 & 11.46 & 8.70 & 4289 & 1.63 & -0.27 & 0.07 & -0.12 & 1.05 & 0.11 & 1 \\
O10 & 3658136 & 2M19382435+3847454 & 2052200588812290560 & 11.75 & 8.87 & 4406 & 1.80 & -0.15 & 0.04 & -0.23 & 1.14 & 0.12 & 1 \\
O11 & 10880958 & 2M19551232+4817344 & 2086860120203064960 & 10.57 & 8.21 & 4779 & 2.43 & -0.08 & 0.01 & -0.11 & 1.08 & 0.05 & 2 \\
O12 & 9143924 & 2M19070280+4530112 & 2106267904878928896 & 12.12 & 9.56 & 4438 & 1.86 & -0.11 & 0.03 & -0.19 & 0.95 & 0.10 & 1 \\
O13 & 9605294 & 2M19513344+4617498 & 2079527099200861440 & 11.25 & 8.86 & 4721 & 2.39 & -0.03 & 0.04 & 0.06 & 0.97 & 0.07 & 2 \\
Y1 & 9821622 & 2M19083615+4641212 & 2130439469722896000 & 11.98 & 9.84 & 4819 & 2.71 & -0.35 & 0.23 & 0.03 & 1.45 & 0.03 & 1 \\
Y2 & 4143460 & 2M19101154+3914584 & 2099658950942122624 & 11.67 & 9.31 & 4822 & 2.51 & -0.27 & 0.20 & -0.26 & 1.52 & 0.05 & 2 \\
Y3 & 4350501 & 2M19081716+3924583 & 2100420946858839168 & 11.69 & 9.31 & 4829 & 3.05 & -0.11 & 0.14 & 0.02 & 1.33 & 0.07 & 1 \\
Y4 & 11394905 & 2M19093999+4913392 & 2131256887898248576 & 11.35 & 9.24 & 4910 & 2.50 & -0.46 & 0.17 & -0.19 & 1.29 & 0.04 & 2 \\
Y5 & 9269081 & 2M19032243+4547495 & 2106500829544203264 & 12.03 & 9.89 & 4820 & 2.30 & -0.18 & 0.16 & -0.55 & 1.52 & 0.07 & 1 \\
Y6 & 11823838 & 2M19455292+5002304 & 2135277767562458624 & 11.01 & 8.83 & 4943 & 2.53 & -0.39 & 0.15 & -0.22 & 1.55 & 0.04 & 2 \\
Y7 & 5512910 & 2M18553092+4042447 & 2103628703312376448 & 12.99 & 10.74 & 4946 & 2.49 & -0.36 & 0.13 & 0.10 & 1.35 & 0.04 & 2 \\
Y8 & 10525475 & 2M19102133+4743193 & 2130894220860474752 & 10.73 & 8.55 & 4781 & 2.50 & -0.20 & 0.18 & -0.13 & 1.38 & 0.08 & 2 \\
Y9 & 9002884 & 2M18540578+4520474 & 2106992311244648448 & 11.53 & 8.68 & 4174 & 1.56 & -0.30 & 0.18 & -0.34 & 1.46 & 0.09 & 1 \\
Y10 & 9761625 & 2M19093801+4635253 & 2130434517620507136 & 11.47 & 8.89 & 4426 & 1.86 & -0.25 & 0.16 & -0.26 & 1.58 & 0.08 & 1 \\
Y11 & 11445818 & 2M19052620+4921373 & 2132045164717799808 & 12.10 & 9.88 & 4751 & 2.49 & -0.12 & 0.11 & -0.06 & 1.56 & 0.06 & 2 \\
Y12 & 3455760 & 2M19374569+3835356 & 2052173380209489408 & 10.91 & 8.41 & 4611 & 2.58 & -0.06 & 0.12 & -0.15 & 1.36 & 0.03 & 1 \\
Y13 & 3833399 & 2M19024305+3854594 & 2100234510918046592 & 9.22 & 6.97 & 4663 & 2.47 & 0.05 & 0.08 & 0.06 & 1.36 & 0.07 & 2 \\
Y14 & 8547669 & 2M19052572+4437508 & 2106312473757027200 & 12.18 & 9.68 & 4474 & 2.34 & 0.03 & 0.10 & -0.26 & 1.23 & 0.05 & 1 \\
Y15 & 11753104 & 2M19025410+4957320 & 2132141994758193024 & 12.48 & 10.27 & 4750 & 2.43 & -0.62 & 0.24 & -0.07 & 1.13 & 0.06 & 1 \\
Y16 & 11413138 & 2M19460251+4913014 & 2086818952931779072 & 11.50 & 9.06 & 4594 & 2.57 & 0.17 & 0.05 & -0.18 & 1.49 & 0.03 & 1 \\
Y17 & 11824403 & 2M19464438+5002378 & 2135279962285745280 & 11.66 & 8.59 & 4067 & 1.38 & -0.43 & 0.22 & -0.00 & 0.99 & 0.15 & 1 \\
Y18 & 12066292 & 2M19355053+5030152 & 2135223066858975616 & 11.52 & 8.58 & 4141 & 1.57 & -0.38 & 0.22 & 0.08 & 1.40 & 0.10 & 1 \\
Y19 & 9644558 & 2M19194578+4619275 & 2127706388756329088 & 11.42 & 9.37 & 5099 & 2.37 & -0.39 & 0.19 & -0.06 & 0.63 & 0.14 & 2 \\
Y20 & 9946773 & 2M19193356+4648258 & 2127816099398726912 & 11.46 & 9.46 & 5012 & 2.38 & -0.64 & 0.26 & 0.15 & 0.67 & 0.15 & 2 \\
Y21 & 9390558 & 2M18592488+4556131 & 2107185760867863296 & 12.45 & 10.20 & 4708 & 2.37 & -0.20 & 0.15 & 0.17 & 0.94 & 0.14 & 2 \\
Y22 & 10554179 & 2M19534542+4742439 & 2086447013067213824 & 10.05 & 8.09 & 5080 & 2.40 & -0.65 & 0.19 & 0.18 & 0.71 & 0.11 & 2 \\
Y23 & 9474021 & 2M19411424+4601483 & 2080027303965885312 & 10.86 & 7.69 & 4008 & 1.21 & -0.42 & 0.23 & -0.02 & 1.20 & 0.21 & 1 \\
Y24 & 7670489 & 2M19083152+4321542 & 2105623041309762048 & 11.73 & 9.38 & 4564 & 2.44 & -0.26 & 0.16 & -0.15 & 1.05 & 0.03 & 1 \\
Y25 & 5938430 & 2M18521894+4116496 & 2104372729087003648 & 12.49 & 10.06 & 4619 & 2.67 & -0.09 & 0.17 & -0.06 & 1.10 & 0.03 & 1 \\
Y26 & 3528656 & 2M19050492+3839005 & 2100001689331818368 & 11.34 & 8.84 & 4561 & 2.21 & -0.40 & 0.18 & -0.20 & 1.18 & 0.04 & 1 \\
Y27 & 2570715 & 2M19211713+3748039 & 2051105857492905600 & 12.68 & 10.28 & 4713 & 2.34 & -0.23 & 0.17 & 0.06 & 1.19 & 0.18 & 2 \\
Y28 & 3662233 & 2M19415752+3845044 & 2073158173792306816 & 11.76 & 8.65 & 4076 & 1.49 & -0.27 & 0.20 & -0.03 & 1.29 & 0.25 & 1 \\
\hline 
\end{tabular}
}

\end{table*}%

%% file: tab_summary.tex
\renewcommand{\tabcolsep}{2pt}
\begin{table*}[t!]
\caption{Results of the HERMES and APOGEE RV monitoring are displayed in columns 3 to 8 and 9 to 11, respectively, and are statistically combined in columns 12 to 14. 
 Columns $\overline{RV}$ and $\sigma$ list the mean RV and the corresponding standard deviation. 
 The probability $Prob$ of the star being in a binary system, from the HERMES data alone (and adopting an uncertainty of 0.07~km~s$^{-1}$) is listed in column 5 (in fact, $Prob$ is the 
 probability integral of the $\chi^2$ distribution from 0 to the observed $\chi^2$ value for $N - 1$ 
 degrees of freedom) are tabulated as well. $N$ denotes  the total number of HERMES or APOGEE RV measurements 
 for each star and $\Delta t$ is the number of days between the first and the last observation. 
 The value $F2$ in column 13 corresponds to the $F2$-statistics of the combined HERMES-APOGEE RVs, assuming a 0.4~km~s$^{-1}$  offset between their zero points and a minimum uncertainty on the RVs of  {0.09}~km~s$^{-1}$ (see text).
 Columns 15 to 18 correspond to the Gaia DR2 RV measurements, which cover the time span JD 2456863.5 to 2457531.5 (2014 July 25 to 2016 May 23), just prior to the HERMES monitoring thus. 
 The column labelled $\epsilon$ lists the expected uncertainty on the Gaia DR2 RV (see text). The column labelled $\Delta RV$ under the heading Gaia DR2 corresponds to $RV_{\rm HERMES} - RV_{\rm DR2}$.  $\Delta RV$ under the heading APOGEE corresponds to $RV_ \mathrm{HERMES} - RV_\mathrm{APOGEE}$, after applying the zero-point offset of $-0.40$~km~s$^{-1}$ to the APOGEE velocities. Values in bold face identify binary signatures. The last column summarizes the situation regarding binarity, collecting the diagnostics from HERMES, Gaia DR2, and APOGEE. When the only binary signature is an offset between Gaia DR2 and HERMES, the star is flagged as 'SB?'.  
 } \label{28stars}
{ \small
\begin{tabular}{|cr|rclrrc|rrrr|rrr|rrrr|c|}
\hline
&&\multicolumn{6}{c}{HERMES}&\multicolumn{6}{c}{APOGEE}&\multicolumn{5}{c}{Gaia DR2}&HER \\
& & & & & & & & & & &&\multicolumn{3}{c}{HER} &&&&&\& APO\\
& & & & & & & & & & &&\multicolumn{3}{c}{\&APO} &&&&&\& DR2\\
\cline{3-8}\cline{9-15}\cline{16-19}\cline{20-20}
Star & class &  \multicolumn{1}{c}{$\overline{RV}\pm\sigma$}&  $F2$ &$Prob$ & $N$  &$\Delta t$  & SB && \multicolumn{1}{c}{$\overline{RV}\pm\sigma$} & $N$ & \multicolumn{1}{c}{$\Delta RV$} &  \multicolumn{1}{c}{$\sigma$} & $F2$ & SB & \multicolumn{1}{c}{$\overline{RV}\pm\sigma$} & \multicolumn{1}{c}{$\epsilon$} &  \multicolumn{1}{c}{$\Delta RV$} & SB & SB\\
 & & \multicolumn{1}{c}{(km/s)} &   & &&(d)& && \multicolumn{1}{c}{(km/s)}  && \multicolumn{1}{c}{(km/s)} & \multicolumn{1}{c}{(km/s)} & & &\multicolumn{1}{c}{(km/s)} & \multicolumn{1}{c}{(km/s)} & \multicolumn{1}{c}{(km/s)}  &&\\
\hline 
O1       & AP  & $5.56\pm2.33$  &{  41}&1.0&7&1390&SB  & & $8.23$\phantom{$\pm0.00$}  & 1 & {  -2.67}& 2.36 & {  45}&SB    & $1.86\pm1.39$&1.04 & {  3.70} & SB                         &SB\tablefootmark{e}\\
O2       & AP  & $-16.30 \pm0.11$ &1.2&0.8762&8&1083&&      &$-16.26\pm0.06$  & 3 & -0.04&  0.10 & 0.77                    &      & $-15.62\pm0.47$& 0.78 &-0.68&                                                           &\\
O3       & AR  & $-29.27 \pm3.39$ &{  53}&1.0&7&1392&SB& &$-22.13$\phantom{$\pm0.00$}  & 1 & {  -7.14}& 4.03 & {  66}&SB   & $-30.39\pm2.40$& 0.87 &1.12 &SB\tablefootmark{a}             &SB\tablefootmark{c}\\
O4       & AR  & $-150.89 \pm0.23$ &2.6&0.9953&3&708&&&$-151.01\pm0.03$  & 3 & 0.12 & 0.16 & 2.2 &                   & - & - & - & -                                               &\\
O5       & AP  & $5.84 \pm0.14$ &2.1&0.9807&7&1082&&&$6.23\pm0.16$  & 5 & -0.39 & 0.24 &{  6.9}                 &SB   &$6.15\pm0.23$& 0.42 &-0.31&                                         & SB \\
O6       & AP  & $-30.86 \pm0.12$ &1.4&0.9162&7&1391&&      &$-30.78$\phantom{$\pm0.00$}  & 1 & -0.08& 0.12 & 1.3 &             & $-30.06\pm0.47$ & 1.41  &-0.80&                                    &\\ 
O7       & AP  &  $2.00 \pm0.07$ &-0.6&0.2840&6&1390&&       & $1.98\pm0.07$  & 2 & 0.02& 0.06 & -0.92 &                          & $3.18\pm$ 1.00 & 1.37  &  -1.18 &                                  &\\
O8       & AP  & $-46.47 \pm0.10$&0.7&0.7537&9&1392&&       &$-46.75\pm0.11$  & 3 & 0.28& 0.16 &  {  3.5}&SB                   & $-47.73\pm0.58$ & 1.11  & 1.26 &                                   & SB\\ 
O9       & AP  & $-69.82 \pm0.12$&1.3&0.9073&8&1128&&       &$-69.77\pm0.06$  & 4 & -0.05 & 0.10 &  0.92 &                        & $-69.45\pm0.27$& 0.64 & -0.37&                                     &\\  
O10      & AP  & $6.99 \pm0.09$&0.2&0.5624&7&1393&&         &$6.96 \pm${  0.25} & 8 & 0.03&  0.19 & {  5.3}&   SB           & $ 7.55\pm0.84$ & 0.81 &  -0.56&                                    & SB\\
O11      & AP  & $-5.98 \pm0.07$&-0.6&0.2677&7&1392& &      &$-5.93\pm0.07$  & 3 & -0.05& 0.07 & -0.82 &                          & $-5.41\pm0.29$& 0.50 &-0.57&                                       &\\
O12      & AP  & $29.21\pm0.12$ &1.3&0.9021&7&1391&  &      &$29.09\pm0.10$  & 5 & 0.12&  0.13 & 2.1&                   & $29.97\pm0.40$& 0.99 &-0.76&                                       &\\
O13      & AP  & $-41.42\pm0.09$ &0.2&0.5625&6&1389&    &   &$-41.01\pm0.08$  & 3 & -0.41& 0.22 &  {  5.2}&SB                  & $-41.19\pm0.52$& 0.80 & -0.23&                                     &SB\\
Y1 & YAR & $-5.78\pm0.10$ &0.6 &0.7089&13&1080&           &&$-5.75$\phantom{$\pm0.00$} & 1 & -0.03 & 0.09 & 0.49 &               & $ -4.80\pm0.41$& 0.94  & -0.98 &                           &\\  
Y2 & YAR & $6.30 \pm0.20$ &{  4.0}&1.0&7&1382&SB        &&$6.11\pm0.15$    & 3 & 0.19 & 0.20 & {  4.9} & SB                    & $  7.02\pm0.63$& 1.00  & -0.72&                            &SB\\
Y3 & YAR & $-83.48 \pm0.13$ &2.0&0.9795&10&2208&  &&$-83.78\pm0.05$  & 3 & 0.30& 0.18 & {  4.5} & SB                   & $-82.95\pm0.49$& 0.82  & -0.53&                           &SB\\
Y4 & YAR & $-70.62 \pm3.30$ &{  30}&1.0&3&648&SB       &&$-75.73$\phantom{$\pm0.00$}  & 1 & {  4.31} & 3.71 & {  41} &SB   & $-70.54\pm1.48$& 0.82  & -0.08 &                                        &SB\tablefootmark{c}\\
Y5 & YAR & $-86.11 \pm0.10$ &0.7&0.7430&8&1382&           &&$-86.10\pm0.05$  & 5 & -0.01& 0.08 &-0.17 &                          & $-85.98\pm0.60$& 1.12  &-0.13&                             &\\ 
Y6 & YAR & $-24.07 \pm4.52$ &{  71}&1.0&8&1386&SB      &&$-18.41\pm0.20$   & 2 & {  -5.66}& 4.65 & {  83} &SB               & $-19.71\pm${  1.84}&0.61& {  -4.36}&SB                          &SB\tablefootmark{b,c}\\
Y7 & YAR & $-39.23 \pm2.16$ &{  41}&1.0&8&1382&SB       &&$-39.16$\phantom{$\pm0.00$}   & 1 &-0.07 & 2.02 & {  43} & SB       & $-36.65\pm$ 1.51& 2.39  & -2.58 &                                      &SB\tablefootmark{b}\\
Y8 & YAR & $-44.36 \pm1.22$ &{  24}&1.0&7&1387&SB        && $-39.98$\phantom{$\pm0.00$}& 1 &{  -4.38} &1.92 & {  38} &SB   & $-43.58\pm0.75$& 0.58  &-0.78&                                               & SB\tablefootmark{d}\\
Y9       & YAR  & $-33.45 \pm3.19$&{  55}&1.0&8&1381&SB&    &$-37.98\pm0.10$    & 5 & {  4.43} & 3.34 & {  76}           &SB  & $-31.94\pm0.93$& 0.55 & {  -1.51} & SB              &SB\tablefootmark{b,c}\\ 
Y10& YAR & $-56.97 \pm0.12$ & 1.6&0.9454&10&2213&        &&$-56.94$\phantom{$\pm0.08$}  & 1 & -0.03&0.12 & 1.5 &               & $-56.59\pm0.25$& 0.62  & -0.38&                                        & \\
Y11& YAR & $-26.73 \pm0.08$ &-0.3&0.3809&11&2208&           &&$-26.75$\phantom{$\pm0.00$}  & 1 & 0.02& 0.08 & -0.44 &              & $-25.95\pm0.70$& 1.18  &-0.78&                                        &\\
Y12& YAR & $-46.81\pm0.81$  &{  25}&1.0&13&1842&SB       &&$-46.80\pm${ 1.08}  & 8 & -0.01 &0.90 & {  35} & SB             & $-45.74\pm0.38$& 0.70  & -1.07 &                                   &SB \tablefootmark{b}\\   
Y13      & YAR  & $-62.15\pm0.09$ &-0.1&0.4566&14&1488&   &   &$-62.30\pm0.06$  & 3 & 0.15& 0.10 &  0.79 &                          & $-61.73\pm0.15$& 0.37  &-0.42& & \\
Y14      & AR  & $-45.46\pm0.08$ &-1.6&0.0063&2&309&      & &$-44.97\pm0.11$    & 5 & {  -0.49} &0.26 & {  5.6}          &SB  & $-44.90\pm0.41$& 1.22 & -0.56&                              & SB                                  \\
Y15      & AR  & $-54.79 \pm0.21$&{  4.1}&1.0&7&1604&SB& &$-55.68$\phantom{$\pm0.00$}  & 1 & {  0.89} & 0.37 & {  9.2}&SB & $-55.57\pm0.57$& 1.52  &  0.78  &                         &SB\\
Y16      & YAR  & $-15.15 \pm0.03$&-2.0&0.0223&5&1603&     &&$-15.25\pm0.00$   & 2 & 0.10& 0.05 & -1.3 &                          & $-14.63\pm0.27$& 0.83  &-0.52&                                     &\\
Y17      & AR  & $-67.70 \pm0.08$&0.1&0.5427&4&1603& &       &$-67.71\pm0.06$  & 2 & 0.01 & 0.07 & -0.43&                        & $-66.45\pm0.57$& 0.80  &-1.25 &                           & \\
Y18&       YAR  & $  2.17 \pm0.15$&2.1&0.9812&6&1604& &&$2.18\pm0.03$    & 2 & -0.01 & 0.13  & 1.7 &                        & $  3.55\pm0.77$& 0.77  &-1.38 &                           &\\
Y19&       AR  & $-41.10 \pm0.06$&-0.9&0.1893&6&1604&     & &$-41.03$\phantom{$\pm0.08$}  & 1 & -0.07 & 0.06 & -0.90 &          & $-40.28\pm0.45$& 0.82  &-0.82&                              &\\  
Y20&       AR  & $-30.63 \pm0.15$&2.0&0.9768&6&1603& &&$-30.87$\phantom{$\pm0.08$} & 1 & 0.24 & 0.16 & 2.8 &    & $-31.05\pm0.65$& 0.66 & 0.42&                               &\\
Y21&       AR  & $-30.71 \pm0.07$&-0.4&0.3427&7&1602&     &  &$-30.89\pm0.12$  & 5 & 0.18 & 0.13 & 2.4                &  & $-29.91\pm0.75$& 1.40 &-0.80&                                &\\
Y22&       AR  & $ 75.64 \pm0.05$&-1.2&0.1128&6&1604&     & & $75.89\pm0.08$  & 3 & -0.25& 0.14 & 2.3                & & $ 76.37\pm0.46$& 0.47  &-0.73&                              &\\
Y23&       AR  & $-121.84 \pm0.07$&-0.4&0.3349&5&1605&  &    &$-122.02$\phantom{$\pm0.00$}  & 1 & 0.18& 0.10 & 0.57 &             & $-121.06\pm0.40$& 0.47  &-0.78&                            &\\
Y24&       AR  & $-51.08 \pm0.07$&-0.3&0.3895&5&1603&      &  & $-51.31\pm0.04$  & 2 & 0.23& 0.13 & 1.8                     &    & $-49.72\pm0.40$& 1.01  &-1.36 &                           &\\
Y25&       AR  & $-49.41 \pm0.52$&{  13}&1.0&9&1604&SB      &&$-43.35$\phantom{$\pm0.00$}  & 1&{  -6.06} &1.98 & {  44}&SB& $-43.23\pm$ 1.32& 1.37  & {  -6.18}&SB                 &SB\tablefootmark{d}\\
Y26&       AR  & $-87.78 \pm3.94$&{  59}&1.0&7&1604&SB    &&$-84.22\pm0.25$ & 3 & {  -3.56}& 3.65 & {  70} &SB            & $-93.54\pm0.55$& 0.67  & {  5.76} & SB                 &SB\tablefootmark{d}\\
Y27&       AR  & $-3.60  \pm0.09$&0.3&0.6345&6&1602&        &&$-3.65\pm0.04$ & 3 & 0.05& 0.08 & -0.13 &                          & $ -2.15\pm$0.94& 1.94  & -1.45 &                         & \\ 
Y28&       YAR  & $ 9.06  \pm1.03$&{  20}&1.0&6&1606&SB     &&$11.15\pm0.14$ & 3 & {  -2.09} & 1.32 & {  31} &SB           & $  9.40\pm0.52$&0.85  &-0.34&                                & SB\\
\hline
\end{tabular}
}
\tablefoot{
\tablefoottext{a}{The Gaia eDR3 RUWE ('reduced unit-weight error') parameter is 4.01 for this star, further indicative of its binary nature.}
\tablefoottext{b}{Spectroscopic orbit available from HERMES data (see Appendix~\ref{Sect:orbits}).}
\tablefoottext{c}{Spectroscopic orbit available from Gaia DR3 (see Appendix~\ref{Sect:orbits}).
\tablefoottext{d}{First-degree RV trend in Gaia DR3}
\tablefoottext{e}{Second-degree RV trend in Gaia DR3}}
}
\end{table*}%

%% file: table_RV_DR2_DR3.tex
\begin{table*}[t!]
\caption{{Gaia DR2 and DR3 radial velocities.  
RUWE is the 'reduced unit weight error' from Gaia DR3. 'renormalized\_gof' is the renormalized goodness-of-fit of Gaia DR3 RVs and 'Prob' is the $p$-value of the renormalised DR3 $\chi^2$ (see Gaia DR3 documentation). The SB assignment in the last column has been copied from Table~\ref{28stars} whereas the SB DR3 assignment is based on Prob being smaller than 0.01. The column 'NSS Rem.' lists the assignment made by the 'non-single star' (NSS) team \citep{Arenou2022} to any one of the subtypes 'first-degree RV trend' (1D trend), 'second-degree RV trend' (2D trend), or 'SB orbit' (SBO), as listed in Table~\ref{Tab:orbital_elements}.}   
\label{Tab:Gaia_DR2_DR3}}
\begin{tabular}{rrrrrcrrrrrrr}
\hline
Star & \multicolumn{1}{c}{$\overline{RV}_{\rm DR2}\pm\sigma$}&$\epsilon_{\mathrm{DR2}}$ &  
\multicolumn{1}{c}{$\overline{RV}_{\rm DR3}\pm\sigma$} & RUWE & renormalized\_gof & Prob & \multicolumn{2}{c}{SB} & NSS Rem.\\
\cline{8-9}
 &    \multicolumn{1}{c}{(km/s)} &   \multicolumn{1}{c}{(km/s)} & \multicolumn{1}{c}{(km/s)} & &&& DR3 & all\\
\hline 
O1    & $1.86\pm1.39$  &1.04   &$3.49	\pm1.13$& 1.06&11.45&0.00& SB & SB & 2D trend\\  
O2    & $-15.62\pm0.47$& 0.78  &$-15.95	\pm0.24$& 0.93&-1.36&0.90& \\   
O3    & $-30.39\pm2.40$& 0.87  &$-30.42	\pm1.36$& 4.01&22.88&0.00& SB & SB\\
O4    & -              & -     &$-150.05\pm1.76$& 0.99&  -  &    &     \\
O5    & $6.15\pm0.23$  & 0.42  &$5.98	\pm0.18$& 0.97& 0.28&0.23&    & SB\\
O6    & $-30.06\pm0.47$ &1.41  &$-31.05	\pm0.45$& 0.91&-0.91&0.81& \\ 
O7    & $  3.18\pm1.00$ &1.37  &$2.15	\pm0.37$& 0.95&-0.59&0.63& \\
O8    & $-47.73\pm0.58$ & 1.11 &$-47.66	\pm0.55$& 0.85& 1.21&0.06&    & SB\\ 
O9    & $-69.45\pm0.27$& 0.64  &$-69.64	\pm0.25$& 1.05& 0.39&0.31& \\  
O10   & $ 7.55\pm0.84$ & 0.81  &$7.14	\pm0.23$& 0.87&-0.37&0.56&    & SB\\
O11   & $-5.41\pm0.29$& 0.50   &$-5.65	\pm0.17$& 0.94& 0.52&0.11& \\
O12   & $29.97\pm0.40$& 0.99   &$29.67	\pm0.30$& 0.87& 0.06&0.46& \\
O13   & $-41.19\pm0.52$& 0.80  &$-40.79	\pm0.24$& 0.90&-0.18&0.45&    & SB\\
Y1    & $ -4.80\pm0.41$& 0.94  &$-5.46	\pm0.32$& 0.95&-0.51&0.59& \\  
Y2    & $  7.02\pm0.63$& 1.00  &$6.13	\pm0.37$& 0.88& 0.41&0.31&    & SB\\
Y3    & $-82.95\pm0.49$& 0.82  &$-83.04	\pm0.28$& 1.06& 0.53&0.19&    & SB\\
Y4    & $-70.54\pm1.48$& 0.82  &$-72.05	\pm0.75$& 1.29& 9.74&0.00& SB & SB & SBO\\
Y5    & $-85.98\pm0.60$& 1.12  &$-85.69	\pm0.45$& 0.99& 0.15&0.53& \\ 
Y6 & $-19.71\pm${ 1.84}&0.61&$-23.56	\pm1.31$& 1.47&25.84&0.00& SB & SB & SBO\\
Y7    & $-36.65\pm$ 1.51& 2.39 &$-38.62	\pm1.24$& 1.19& 0.33&0.27&    & SB\\
Y8    & $-43.58\pm0.75$& 0.58  &$-43.79	\pm0.45$& 1.05& 8.31&0.00& SB & SB & 1D trend\\
Y9    & $-31.94\pm0.93$& 0.55  &$-33.81	\pm0.80$& 1.10&16.15&0.00& SB & SB & SBO\\ 
Y10   & $-56.59\pm0.25$& 0.62  &$-56.94	\pm0.23$& 0.82& 0.39&0.23& \\
Y11   & $-25.95\pm0.70$& 1.18  &$-26.90	\pm0.39$& 0.95&-1.06&0.83& \\
Y12   & $-45.74\pm0.38$& 0.70  &$-46.52	\pm0.31$& 1.03& 3.90&0.00& SB & SB\\   
Y13   & $-61.73\pm0.15$& 0.37  &$-61.90	\pm0.13$& 1.07&-1.18&0.67& \\
Y14   & $-44.90\pm0.41$& 1.22  &$-45.01	\pm0.54$& 0.99& 0.38&0.99&    & SB \\
Y15   & $-55.57\pm0.57$& 1.52  &$-55.68	\pm0.43$& 0.94&-1.30&0.91&    & SB\\
Y16   & $-14.63\pm0.27$& 0.83  &$-15.13	\pm0.23$& 1.00&-0.30&0.52& \\
Y17   & $-66.45\pm0.57$& 0.80  &$-67.25	\pm0.23$& 0.86& 0.38&0.36& \\
Y18   & $  3.55\pm0.77$& 0.77  &$2.39	\pm0.37$& 0.87& 1.62&0.26& \\
Y19   & $-40.28\pm0.45$& 0.82  &$-40.45	\pm0.32$& 0.89& 0.17&0.32& \\  
Y20   & $-31.05\pm0.65$& 0.66  &$-30.89	\pm0.33$& 0.82& 1.03&0.08&\\
Y21   & $-29.91\pm0.75$& 1.40  &$-30.35	\pm0.77$& 1.01& 4.43&0.00& SB & \\
Y22   & $ 76.37\pm0.46$& 0.47  &$75.93	\pm0.25$& 0.94& 0.88&0.11& \\
Y23   & $-121.06\pm0.40$& 0.47 &$-121.46\pm0.16$& 0.99&-1.12&0.79& \\
Y24   & $-49.72\pm0.40$& 1.01  &$-51.00	\pm0.33$& 0.98&-0.61&0.68& \\
Y25   & $-43.23\pm$ 1.32& 1.37 &$-46.48	\pm1.07$& 1.83& 8.09&0.00& SB & SB& 1D trend\\
Y26   & $-93.54\pm0.55$& 0.67  &$-92.53	\pm0.37$& 0.90& 4.68&0.00& SB & SB& 1D trend\\
Y27   & $ -2.15\pm$0.94& 1.94  &$-3.10	\pm0.72$& 1.06& 1.18&0.04& \\ 
Y28   & $  9.40\pm0.52$&0.85   &$8.99	\pm0.38$& 0.93& 1.61&0.08&    & SB\\
\hline
\end{tabular}
\end{table*}